\documentclass[12pt, draftclsnofoot, onecolumn]{IEEEtran}
\usepackage{epsfig,graphicx,subfigure,psfrag,amsmath,cases,bm}
\usepackage{latexsym,amssymb,amsmath,epsfig,subfigure,algorithm,mathtools}
\usepackage{algorithmic}
\usepackage{color}
\usepackage{url}
\usepackage{scrtime}

\author{\IEEEauthorblockN{Yan Sun, Derrick Wing Kwan Ng, Jun Zhu, and Robert Schober \\
\thanks{Yan Sun and Robert Schober are  with the Institute for Digital Communications, Friedrich-Alexander-University Erlangen-N\"urnberg (FAU), Germany (email:\{yan.sun, robert.schober\}@fau.de). Derrick Wing Kwan Ng is with the School of Electrical Engineering and Telecommunications, the University of New South Wales, Australia (email: w.k.ng@unsw.edu.au).
Jun Zhu is with the Department of Electrical and Computer Engineering, University of British Columbia, Canada (email: zhujun@ece.ubc.ca).  This paper has been presented in part at IEEE ICC 2017 \cite{sun2017MISONOMA} and IEEE SPAWC 2017 \cite{sun2017OFDMFD}.
}}\vspace*{-0mm}
}

\title{Robust and Secure Resource Allocation for Full-Duplex MISO Multicarrier NOMA Systems\vspace*{-2mm}}

\newtheorem{Thm}{Theorem}
\newtheorem{Lem}{Lemma}

\newtheorem{T-Prob}{Transformed Problem}
\newtheorem{Prop}{Proposition}

\DeclareMathOperator{\Tr}{Tr}
\DeclareMathOperator{\Rank}{Rank}
\DeclareMathOperator{\maxo}{maximize}
\DeclareMathOperator{\mino}{minimize}

\DeclareMathOperator{\Diag}{\mathrm{Diag}}
\DeclareMathOperator{\rem}{\mathrm{rem}}

 \newcommand{\qed}{\hfill \ensuremath{\blacksquare}}

\newcommand{\abs}[1]{\lvert#1\rvert}
\newcommand{\norm}[1]{\lVert#1\rVert}
\begin{document}
\maketitle \vspace*{-18mm}
\begin{abstract}\vspace*{-2mm}
In this paper, we study the resource allocation algorithm design for multiple-input single-output (MISO) multicarrier non-orthogonal multiple access (MC-NOMA) systems, in which a full-duplex base station serves multiple half-duplex uplink and downlink users on the same subcarrier simultaneously.
The resource allocation is optimized for maximization of the weighted system throughput while the information leakage is constrained and artificial noise is injected to guarantee secure communication in the presence of multiple potential eavesdroppers.
To this end, we formulate a robust non-convex optimization problem taking into account the imperfect channel state information (CSI) of the eavesdropping channels and the quality-of-service (QoS) requirements of the legitimate users.
Despite the non-convexity of the optimization problem, we solve it optimally by applying monotonic optimization which yields the optimal beamforming, artificial noise design, subcarrier allocation, and power allocation policy.
The optimal resource allocation policy serves as a performance benchmark since the corresponding monotonic optimization based algorithm entails a high computational complexity. Hence, we also develop a low-complexity suboptimal resource allocation algorithm which converges to a locally optimal solution. Our simulation results reveal that the performance of the suboptimal algorithm closely approaches that of the optimal algorithm. Besides, the proposed optimal MISO NOMA system can not only ensure downlink and uplink communication security simultaneously but also provides a significant system secrecy rate improvement compared to traditional MISO orthogonal multiple access (OMA) systems and two other baseline schemes. 
\end{abstract}\vspace*{-2mm}
\begin{IEEEkeywords}\vspace*{-3mm}Non-orthogonal multiple access (NOMA), multicarrier systems, full-duplex radio, physical layer security, imperfect channel state information, non-convex and combinatorial optimization.
\end{IEEEkeywords}

\renewcommand{\baselinestretch}{1.22}
\vspace*{-2mm}
\section{Introduction}
Over the past two decades, multicarrier (MC) techniques have been widely adopted in wireless
communication standards and their design has been extensively studied, since they provide a high flexibility in resource allocation and are able to exploit multiuser diversity.
For example, the authors of \cite{Cui2009Distrib} studied the power and subcarrier allocation design for maximization of the weighted sum rate of multiuser MC relay systems.
However, traditional MC systems adopt orthogonal multiple access (OMA) serving at most one user on one subcarrier.
Therefore, resource allocation strategies designed for traditional MC systems underutilize the spectral resources, since each subcarrier is allocated exclusively to one user to avoid multiuser interference (MUI).

To overcome this shortcoming, non-orthogonal multiple access (NOMA) has been recently proposed to improve spectral efficiency and to provide fairness in resource allocation by multiplexing multiple users on the same time-frequency resource \cite{Ding17surveyNOMA}--\nocite{book:Key5GWong,Zhu17OptimalDLNOMA,Chen16Quasi,ding2015general,Lei2016NOMAJournal,Wei17optimal,sun2016optimal}\cite{Zhang17subchannel}.
In particular, NOMA exploits the power domain for multiple access and harnesses MUI via superposition coding at the transmitter and successive interference cancellation (SIC) at the receiver.
In \cite{Zhu17OptimalDLNOMA}, the authors investigated the optimal power allocation design for maximization of the system throughput in single-input single-output (SISO) single-carrier (SC) NOMA systems. It is shown in \cite{Zhu17OptimalDLNOMA} that SISO NOMA achieves a higher spectral efficiency compared to conventional SISO OMA.
The authors of \cite{Chen16Quasi} proposed a suboptimal precoding design for minimization of the transmit power in multiple-input single-output (MISO) NOMA systems.
In \cite{ding2015general}, multiple-input multiple-output (MIMO) SC-NOMA systems are shown to achieve a substantially higher spectral efficiency compared to traditional  MIMO SC-OMA systems by exploiting the degrees of freedom (DoF) offered in both the spatial domain and the power domain.
On the other hand, the application of NOMA to improve the fairness and spectrum utilization in MC systems was studied in \cite{Lei2016NOMAJournal}\nocite{Wei17optimal,sun2016optimal}--\cite{Zhang17subchannel}.
In \cite{Lei2016NOMAJournal}, the authors developed a suboptimal subcarrier and power allocation algorithm  for maximization of the weighted system throughput in single-antenna MC-NOMA systems.
Optimal subcarrier and power allocation algorithms for minimization of the total transmit power and maximization of the weighted system throughput in MC-NOMA systems were proposed in \cite{Wei17optimal} and \cite{sun2016optimal}, respectively.
The authors of \cite{Zhang17subchannel} developed a game theory based joint subcarrier and power allocation algorithm for maximization of the average system throughput of MC-NOMA relaying networks.
However, in \cite{Zhu17OptimalDLNOMA}\nocite{Chen16Quasi,ding2015general,Lei2016NOMAJournal,Wei17optimal,sun2016optimal}--\cite{Zhang17subchannel}, the spectral resources are not fully utilized even if NOMA is employed, since the base station (BS) operates in the half-duplex (HD) mode and uses orthogonal time or frequency resources for uplink (UL) and downlink (DL) transmission which may lead to a significant loss in spectral efficiency.

Full-duplex (FD) transceivers allow simultaneous DL and UL transmission in the same frequency band \cite{FDRad}\nocite{ng2015power}--\cite{Chang17optimalFD}, at the expense of introducing strong self-interference (SI).
In particular, the optimal beamforming and power allocation design for maximization of the minimum signal-to-interference-plus-noise ratio (SINR) of the DL and UL users in FD systems was studied in \cite{Chang17optimalFD}.
Motivated by the potential benefits of FD transmission, the integration of FD and NOMA was advocated in \cite{sun2016optimalJournal}\nocite{Zhong16FDNOMARelay}--\cite{Zhang17coopFDNOMA}.
In \cite{sun2016optimalJournal}, the authors investigated the optimal power and subcarrier allocation algorithm design for maximization of the weighted system throughput in FD MC-NOMA systems.
The authors of  \cite{Zhong16FDNOMARelay} studied the outage probability and the ergodic sum rate of a NOMA-based FD relaying system.
In \cite{Zhang17coopFDNOMA}, the optimal power allocation minimizing the outage probability of NOMA-based FD relaying systems was investigated.
However, in FD NOMA systems, the communication is more susceptible to eavesdropping compared to conventional HD OMA systems, since the simultaneous DL and UL transmissions and the multiplexing of multiple DL and UL users on each subcarrier increases the potential for information leakage. Nevertheless, the existing designs of FD NOMA systems in \cite{sun2016optimalJournal}\nocite{Zhong16FDNOMARelay}--\cite{Zhang17coopFDNOMA} cannot guarantee communication security.

In practical systems, secrecy is a critical concern for the design of wireless communication protocols due to the broadcast nature of the wireless medium \cite{Chen2015Multi,Zhu2014Secure}. The conventional approach for securing communications is to perform cryptographic encryption at the application layer. However, new powerful computing technologies (e.g. quantum computing) weaken the effectiveness of this approach since they provide tremendous processing power for deciphering the encryption.
Physical layer security is an emerging technique  that promises to overcome these challenges \cite{Wu16Secure}\nocite{Zhu2016Linear,li2013spatially,Kwan14Robust}--\cite{Sun16FDSecurity}.
Particularly, BSs equipped with multiple antennas can steer their beamforming vectors and inject artificial noise (AN) to impair the information reception of potential eavesdroppers.
In \cite{li2013spatially}, joint transmit signal and AN covariance matrix optimization was studied for secrecy rate maximization.
Taking into account imperfect channel state information (CSI), the authors of \cite{Kwan14Robust} developed a robust resource allocation algorithm to guarantee DL communication security in multiuser communication systems.
The authors of \cite{Sun16FDSecurity} studied the tradeoff between the total DL transmit power consumption and the total UL transmit power consumption in secure FD multiuser systems.
Recently, secure communication in NOMA systems has been investigated in \cite{Zhang16secrecy}\nocite{Liu17Enhancing}--\cite{Ding17SESecure}.
In \cite{Zhang16secrecy}, a power allocation strategy for maximization of the system secrecy rate in SISO NOMA systems was studied.
In \cite{Liu17Enhancing}, the authors considered the secrecy outage probability of MISO NOMA systems, where AN was generated at the BS to deliberately degrade the channels of the eavesdroppers.
In \cite{Ding17SESecure}, the power allocation and the information-bearing beamforming vector were designed to limit the eavesdropping capacities of the eavesdroppers.
However, the above works \cite{Zhang16secrecy}\nocite{Liu17Enhancing}--\cite{Ding17SESecure} focused on securing the DL in SC-NOMA systems employing HD BSs. Hence, the schemes proposed in \cite{Zhang16secrecy}\nocite{Liu17Enhancing}--\cite{Ding17SESecure} cannot guarantee communication security in FD MISO MC-NOMA systems since the coupling between the SIC decoding order, the subcarrier allocation, the
DL transmit beamforming, the DL AN injection, and the UL power allocation significantly complicates the resource allocation algorithm design.
Besides, \cite{Zhang16secrecy}\nocite{Liu17Enhancing}--\cite{Ding17SESecure} optimistically assumed that the CSI of the potential eavesdroppers (idle users) is available at the BS.
However, if some idle users in the system misbehave and eavesdrop the active users' information signals, perfect CSI of these eavesdroppers may not be available at the BS since they may not transmit reference signals during the idle period.
In fact, the design of a robust and secure resource allocation policy for FD MISO MC-NOMA systems has not been investigated in the literature yet and is more difficult to obtain than the optimal power and subcarrier allocation for FD SISO MC-NOMA studied in \cite{sun2016optimalJournal}.
In particular, the methodology used for modelling the user pairing for FD SISO MC-NOMA in \cite{sun2016optimalJournal} cannot be applied for FD MISO MC-NOMA, where the user pairing does not only depend on the users channel conditions but also on the transmit beamforming vectors and the injected AN. Besides, the transmit beamforming vector design for FD MISO MC-NOMA systems leads to a rank constrained optimization problem which is more challenging to solve than the power allocation problem for FD SISO MC-NOMA systems in \cite{sun2016optimalJournal}.
Moreover, the imperfect CSI leads to an infinite number of constraints, which makes the resource allocation optimization problem formulated in this paper difficult to tackle.
Furthermore, in modern communication systems, meeting the quality-of-service (QoS) requirements of the users is crucial. However, \cite{sun2016optimalJournal} did not take the QoS requirements of the users into account for resource allocation.
In fact, the robust and secure resource allocation algorithm design for FD MISO MC-NOMA systems with QoS constraints is still an open problem.

In this paper, we address the above issues. To this end, the robust and secure resource allocation algorithm design for FD MC-NOMA systems is formulated as a non-convex optimization problem. The proposed optimization problem formulation aims to maximize the weighted system throughput while taking into account the imperfectness of the CSI of the eavesdroppers' channels and the QoS constraints of the legitimate users. Although the considered problem is non-convex and difficult to tackle, we solve it optimally by employing monotonic optimization theory \cite{tuy2000monotonic,zhang2013monotonic}. Besides, we also develop a low-complexity suboptimal scheme which is shown to achieve a close-to-optimal performance.

\vspace*{-6mm}
\section{Notation and System Model}
In this section, we present the notation and the FD MISO MC-NOMA system and channel models.

\vspace*{-4mm}
\subsection{Notation}%
We use boldface lower and upper case letters to denote vectors and matrices, respectively. $\mathbf{X}^H$, $\det(\mathbf{X})$, $\Rank(\mathbf{X})$, and $\Tr(\mathbf{X})$  denote the Hermitian transpose, determinant, rank, and trace of matrix $\mathbf{X}$, respectively; $\mathbf{X}^{-1}$ represents the inverse of matrix $\mathbf{X}$; $\mathbf{X}\preceq \mathbf{0}$ and $\mathbf{X}\succeq \mathbf{0}$ indicate that $\mathbf{X}$ is a negative  semidefinite and a positive semidefinite matrix, respectively; $\mathbf{I}_N$ is the $N\times N$ identity matrix; $\mathbb{C}^{N\times M}$ denotes the set of all $N\times M$ matrices with complex entries; $\mathbb{H}^N$ denotes the set of all $N\times N$ Hermitian matrices;
$\mathbb{R}^{N\times 1}_{\mathrm{+}}$ denotes the set of all $N\times 1$ vectors with non-negative real entries; $\mathbb{Z}^{N\times 1}$ denotes the set of all $N\times 1$ vectors with integer entries;
$\abs{\cdot}$, $\norm{\cdot}_{\mathrm{F}}$, and $\norm{\cdot}$ denote the absolute value of a complex scalar, the Frobenius matrix norm, and the Euclidean vector norm, respectively; ${\cal E}\{\cdot\}$ denotes statistical expectation;
$\Diag(\mathbf{X})$ returns a diagonal matrix having the main diagonal elements of $\mathbf{X}$ on its main diagonal;
$\lceil x \rceil$ denotes the minimum integer which is larger than or equal to $x$; $\rem(x,y)$ denotes the remainder of $x/y$; $\Re(\cdot)$ extracts the real part of a complex-valued input;  $\lambda_{\mathrm{max}}(\mathbf{X})$ denotes the maximum eigenvalue of matrix $\mathbf{X}$; $[x]^+$ stands for $\mathrm{max}\{0,x\}$; the circularly symmetric complex Gaussian distributions with mean $w$, variance $\sigma^2$ and mean vector $\mathbf{a}$, covariance matrix $\mathbf{X}$ are denoted by ${\cal CN}(w,\sigma^2)$ and ${\cal CN}(\mathbf{a},\mathbf{X})$, respectively; and $\sim$ stands for ``distributed as". $\nabla_{\mathbf{x}} f(\mathbf{x})$ denotes the gradient vector of a function $f(\mathbf{x})$ whose components are the partial derivatives of $f(\mathbf{x})$.

\vspace*{-2mm}
\subsection{FD MISO MC-NOMA System}
\begin{figure}
\centering\vspace*{-6mm}
\includegraphics[width=3.5in]{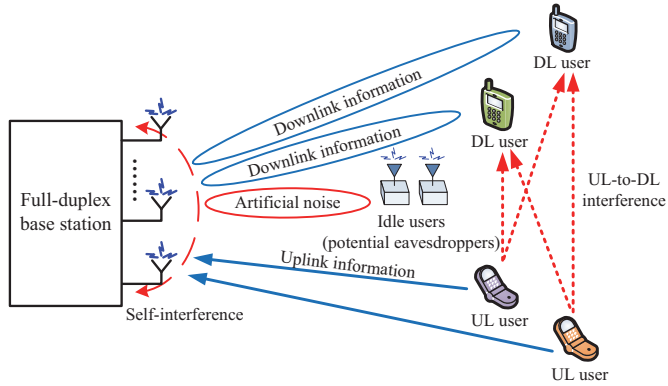} \vspace*{-4mm}
\caption{FD MISO MC-NOMA system with one FD BS, $K=2$ HD DL users, $J=2$ HD UL users, and $M=2$ HD idle users (potential eavesdroppers).}
\label{fig:system_model}\vspace*{-5mm}
\end{figure}

The considered FD MISO MC-NOMA system comprises  $K$  DL users, $J$ UL users, $M$ idle users, and an FD BS, cf. Figure \ref{fig:system_model}. We assume that the FD BS is equipped with $N_{\mathrm{T}} \hspace*{-1.5mm} > \hspace*{-1.5mm} 1$  antennas and $N_{\mathrm{T}} \hspace*{-1mm} > \hspace*{-1mm} M$ to facilitate secure communication. The FD BS enables simultaneous UL reception and DL transmission in the same frequency band\footnote{Transmitting and receiving signals simultaneously with the same antenna has been demonstrated for circulator based FD radio prototypes \cite{FDRad}.}. The DL, UL, and idle users have portable communication devices which operate in the HD mode and are equipped with a single antenna to ensure low hardware complexity.
The available frequency band is divided into ${N_{\mathrm{F}}}$ orthogonal subcarriers.
Besides, we assume that at most two UL users and two DL users are scheduled on each subcarrier to ensure low hardware complexity and low processing delay\footnote{We note that hardware complexity and processing delay scale with the number of users multiplexed on the same subcarrier due to the required successive decoding and cancellation of other users' signals \cite{book:Key5GWong}.}, and to limit the interference on each subcarrier\footnote{Multiplexing more UL and DL users on the same subcarrier leads to more severe UL-to-DL co-channel interference and more severe MUI which cause a larger performance loss for individual users.}.
In order to enable NOMA, we assume that each of the DL users and the FD BS are equipped with successive interference cancellers for multiuser detection.
The idle users are legitimate users that are not scheduled in the current time slot and may deliberately eavesdrop the information signals intended for the DL and UL users.
As a result, the idle users are treated as potential eavesdroppers which have to be taken into account for resource allocation algorithm design to guarantee communication security.

\vspace*{-3mm}
\subsection{Channel Model}
In each scheduling time slot, the FD BS transmits two independent signal streams simultaneously to two selected DL users on each subcarrier. In particular, assuming that DL users $m, n \in \{1,\ldots,K\}$ and UL users $r, t \in \{1,\ldots,J\}$ are scheduled on subcarrier $i\in \{1,\ldots,N_{\mathrm{F}}\}$ in a given time slot, the FD BS transmits signal stream $\mathbf{w}_m^{[i]} x_{\mathrm{DL}_m}^{[i]}$ to  DL user $m$ on subcarrier $i$, where $x_{\mathrm{DL}_m}^{[i]}\in\mathbb{C}$ and $\mathbf{w}_m^{[i]} \in\mathbb{C}^{N_\mathrm{T}\times1}$ are the information bearing symbol and the corresponding beamforming vector for DL user $m$ on subcarrier $i$, respectively. Without loss of generality, we assume ${\cal E}\{\abs{x_{\mathrm{DL}_m}^{[i]}}^2\}=1,\forall m\in\{1,\ldots,K\}$. Besides, UL user $r$ transmits signal $P_r^{[i]}x_{\mathrm{UL}_r}^{[i]}$ on subcarrier $i$ to the FD BS, where
$x_{\mathrm{UL}_r}^{[i]}$, ${\cal E}\{\abs{x_{\mathrm{UL}_r}^{[i]}}^2\}=1$, and $P_r^{[i]}$ denote the transmitted data symbol and the corresponding transmit power, respectively.
Since the signal intended for the desired user and the FD BS may be intercepted by the idle users (potential eavesdroppers), in order to ensure secure communications, the FD BS injects AN to interfere the reception of the idle users.
In particular, the transmit signal vector on subcarrier $i$ at the FD BS, $\mathbf{x}^{[i]}\in\mathbb{C}^{N_{\mathrm{T}}\times1}$, comprising data and AN, is given by $\mathbf{x}^{[i]}= \mathbf{w}_m^{[i]} x_{\mathrm{DL}_m}^{[i]} + \mathbf{w}_n^{[i]} x_{\mathrm{DL}_n}^{[i]} + \mathbf{z}^{[i]}$, where $\mathbf{z}^{[i]} \in\mathbb{C}^{N_\mathrm{T}\times1}$ represents the AN vector on subcarrier $i$ generated by the FD BS to degrade the channels of the potential eavesdroppers.
$\mathbf{z}^{[i]}$ is modeled as a complex Gaussian random vector with $\mathbf{z}^{[i]}\sim{\cal CN}(\mathbf{0},\mathbf{Z}^{[i]})$,
where $\mathbf{Z}^{[i]} \in\mathbb{H}^{N_\mathrm{T}}$, $\mathbf{Z}^{[i]} \succeq \mathbf{0}$, denotes the covariance matrix of the AN. Therefore, the received signals at DL user $m\hspace*{-1mm}\in\hspace*{-1mm}\{1,\hspace*{-0.5mm}\ldots,\hspace*{-0.5mm}K\}$, DL user $n\hspace*{-1mm}\in\hspace*{-1mm}\{1,\hspace*{-0.5mm}\ldots,\hspace*{-0.5mm}K\}$, and the FD BS on subcarrier $i$ are given by \vspace*{-2mm}
\begin{eqnarray}
\label{eqn:dl_user_rcv_signal1}
\hspace*{-4mm} y_{\mathrm{DL}_m}^{[i]}\hspace*{-3.5mm} &=&\hspace*{-3mm}\mathbf{h}_m^{[i]H}\mathbf{w}_m^{[i]} x_{\mathrm{DL}_m}^{[i]} \hspace*{-0.5mm} + \hspace*{-0.5mm} \underbrace{\mathbf{h}_m^{[i]H}\mathbf{w}_n^{[i]} x_{\mathrm{DL}_n}^{[i]}}_{\substack{\text{multiuser}\\\text{interference}}} \hspace*{-0.5mm} + \hspace*{-0.5mm}\underbrace{\mathbf{h}_m^{[i]H} \mathbf{z}^{[i]}}_{\substack{\text{artificial}\\\text{noise}}}\hspace*{-0mm} + \hspace*{-0mm}\underbrace{ \sqrt{P_r^{[i]}}f_{r,m}^{[i]} x_{\mathrm{UL}_r}^{[i]}\hspace*{-0.5mm} + \hspace*{-0.5mm} \sqrt{P_t^{[i]}}f_{t,m}^{[i]} x_{\mathrm{UL}_t}^{[i]}}_{\substack{\text{UL-to-DL}\\\text{co-channel interference}}}  \hspace*{-0.5mm} + \hspace*{1mm} n_{\mathrm{DL}_m}^{[i]},
 \\[-2mm]
\label{eqn:dl_user_rcv_signal2}
\hspace*{-4mm} y_{\mathrm{DL}_n}^{[i]}\hspace*{-3.5mm} &=&\hspace*{-3mm}\mathbf{h}_n^{[i]H}\mathbf{w}_n^{[i]} x_{\mathrm{DL}_n}^{[i]} \hspace*{-0.8mm} + \hspace*{-0.8mm} \underbrace{\mathbf{h}_n^{[i]H} \hspace*{-0.5mm} \mathbf{w}_m^{[i]} x_{\mathrm{DL}_m}^{[i]}}_{\substack{\text{multiuser}\\\text{interference}}} \hspace*{-0.8mm} + \hspace*{-0.8mm}\underbrace{\mathbf{h}_n^{[i]H} \mathbf{z}^{[i]}}_{\substack{\text{artificial}\\\text{noise}}}\hspace*{-0.8mm} + \hspace*{-0.8mm}\underbrace{ \sqrt{P_r^{[i]}}f_{r,n}^{[i]} x_{\mathrm{UL}_r}^{[i]}\hspace*{-0.8mm} + \hspace*{-0.8mm} \sqrt{P_t^{[i]}}f_{t,n}^{[i]} x_{\mathrm{UL}_t}^{[i]}}_{\substack{\text{UL-to-DL}\\\text{co-channel interference}}}\hspace*{-0.5mm} + \hspace*{0.5mm} n_{\mathrm{DL}_n}^{[i]}, \text{and}
\\[-2mm]
\label{eqn:ul_rcv_signal}
\hspace*{-4mm} \mathbf{y}_{\mathrm{BS}}^{[i]}\hspace*{-3.5mm}&=&\hspace*{-3mm} \sqrt{P_r^{[i]}} \mathbf{g}_r^{[i]} x_{\mathrm{UL}_r}^{[i]}\hspace*{-0.5mm} + \hspace*{-0.5mm} \sqrt{P_t^{[i]}} \mathbf{g}_t^{[i]} x_{\mathrm{UL}_t}^{[i]}\hspace*{-0.5mm} + \hspace*{-0.5mm} \underbrace{\mathbf{H}_{\mathrm{SI}}^{[i]}  (\mathbf{w}_m^{[i]} x_{\mathrm{DL}_m}^{[i]} + \mathbf{w}_n^{[i]} x_{\mathrm{DL}_n}^{[i]})}_{\substack{\text{self-interference}}} \hspace*{-0.5mm} + \hspace*{-0.5mm} \underbrace{\mathbf{H}_{\mathrm{SI}}^{[i]} \mathbf{z}^{[i]}}_{\substack{\text{artificial noise}}} \hspace*{-1mm} + \hspace*{1mm}\mathbf{n}_{\mathrm{BS}}^{[i]},
\end{eqnarray}
respectively. Here, the channel vector between the FD BS and DL user $m$ on subcarrier $i$ is denoted by $\mathbf{h}_m^{[i]}\in\mathbb{C}^{N_{\mathrm{T}}\times1}$, and the channel gain between UL user $r$ and DL user $m$ on subcarrier $i$  is denoted by $f_{r,m}^{[i]} \in\mathbb{C}$.  $\mathbf{g}_r^{[i]}\in\mathbb{C}^{N_{\mathrm{T}}\times1}$ denotes the channel between UL user $r$ and the FD BS on subcarrier $i$. Matrix $\mathbf{H}_{\mathrm{SI}}^{[i]} \in{\mathbb{C}^{N_{\mathrm{T}}\times N_{\mathrm{T}}}}$ represents the SI channel of the FD BS on subcarrier $i$. Variables $\mathbf{h}_m^{[i]}$, $f_{r,m}^{[i]}$, $\mathbf{g}_r^{[i]}$, and $\mathbf{H}_{\mathrm{SI}}^{[i]}$ capture the joint effect of path loss and small scale fading.  $\mathbf{n}_{\mathrm{BS}}^{[i]}\sim{\cal CN}(\mathbf{0},\sigma_{\mathrm{BS}}^2\mathbf{I}_{N_{\mathrm{T}}})$ and $n_{\mathrm{DL}_m}^{[i]}\sim{\cal CN}(0,\sigma_{m}^2)$ represent the additive white Gaussian noise (AWGN) at the FD BS and DL user $m$, respectively, where $\sigma_{\mathrm{BS}}^2$ and $\sigma_{m}^2$ denote the corresponding noise powers.

Moreover, for secure communication design, we make the worst-case assumption that the $M$ potential eavesdroppers fully cooperate with each other to form an equivalent super-eavesdropper equipped with $M$ antennas. Thus, the received signal at the equivalent multiple-antenna eavesdropper on subcarrier $i$ is given by \vspace*{-1mm}
\begin{eqnarray}
\label{eqn:eve_rcv_signal}\hspace*{-4mm}\mathbf{y}_{\mathrm{E}}^{[i]}\hspace*{-2mm} &=&\hspace*{-2mm} \mathbf{L}^{[i]H} (\mathbf{w}_m^{[i]} x_{\mathrm{DL}_m}^{[i]} + \mathbf{w}_n^{[i]} x_{\mathrm{DL}_n}^{[i]} ) \hspace*{-0.5mm}+\hspace*{-0.5mm} \sqrt{P_r^{[i]}}\mathbf{e}_{r}^{[i]} x_{\mathrm{UL}_r}^{[i]}\hspace*{-0.5mm} + \hspace*{-0.5mm} \sqrt{P_t^{[i]}}\mathbf{e}_{t}^{[i]} x_{\mathrm{UL}_t}^{[i]}\hspace*{-0.5mm} +\hspace*{-2mm} \underbrace{\mathbf{L}^{[i]H} \mathbf{z}^{[i]}}_{\mbox{artificial noise}}\hspace*{-1mm}  + \hspace*{1mm}\mathbf{n}_{\mathrm{E}}^{[i]},
\end{eqnarray}
where matrix $\mathbf{L}^{[i]} \in{\mathbb{C}^{N_{\mathrm{T}}\times M}}$ represents the channel between the FD BS and the equivalent eavesdropper. Vector $\mathbf{e}_{r}^{[i]}\in\mathbb{C}^{M\times 1}$ models the channel between UL user $r$ and the equivalent eavesdropper on subcarrier $i$. $\mathbf{L}^{[i]}$ and $\mathbf{e}_{r}^{[i]}$ take the joint effect of small scale fading and path loss into account. Finally, $\mathbf{n}_{\mathrm{E}}^{[i]}\sim{\cal CN}(\mathbf{0},\sigma_{\mathrm{E}}^2\mathbf{I}_{M})$ represents the AWGN at the equivalent eavesdropper, where $\sigma_{\mathrm{E}}^2$ denotes the corresponding noise power.

\vspace*{-1mm}
\section{Resource Allocation Problem Formulation}
In this section, we first define the performance metric adopted for the considered FD MC-NOMA system. Then, we present the CSI model employed for resource allocation algorithm design. Finally, we formulate the resource allocation design as a non-convex optimization problem.

\vspace*{-1mm}
\subsection{Weighted System Throughput and Secrecy Rate}
In the considered FD MISO MC-NOMA system, we assume that the FD BS can provide communication service simultaneously to at most two UL users and two DL users on each subcarrier. To mitigate the MUI caused by multiplexing multiple users on the same subcarrier, SIC is performed at the DL users and the FD BS.
Specifically, assuming that DL users $m$, $n$ and UL users $r$, $t$ are multiplexed on subcarrier $i$, DL user $n$ first decodes the message of DL user $m$ and performs SIC to cancel the interference caused by DL user $m$, before attempting to decode its own signal\footnote{We assume that a DL user can cancel only the signals of other DL users by performing SIC but treats the signals of UL users as noise.}. Besides, DL user $m$ directly decodes its own signal while treating the signals of DL user $n$ and the UL users as noise. Thus, the achievable rates (bits/s/Hz) of DL users $m$ and $n$ on subcarrier $i$ are given by \vspace*{-2mm}
\begin{eqnarray} \label{rate-dl-m}
R_{m,n,r,t}^{[i]\mathrm{DL}_m}=&&\hspace*{-6mm} \log_2 \Bigg(  1  +  \frac{\abs{\mathbf{h}_{m}^{[i]H} \mathbf{w}_{m}^{[i]} }^2 } {\abs{\mathbf{h}_{m}^{[i]H} \mathbf{w}_{n}^{[i]} }^2 + \Tr(\mathbf{H}_m^{[i]} \mathbf{Z}^{[i]}) + P_r^{[i]} \abs{f_{r,m}^{[i]}}^2 + P_t^{[i]} \abs{f_{t,m}^{[i]}}^2 + \sigma_{m}^2}  \Bigg) \,\,\, \text{and}  \\[-1mm]
\label{rate-dl-n}
R_{m,n,r,t}^{[i]\mathrm{DL}_n}=&&\hspace*{-6mm} \log_2 \Bigg(  1  +  \frac{\abs{\mathbf{h}_{n}^{[i]H} \mathbf{w}_{n}^{[i]} }^2 } {\Tr(\mathbf{H}_n^{[i]} \mathbf{Z}^{[i]}) + P_r^{[i]} \abs{f_{r,n}^{[i]}}^2 + P_t^{[i]} \abs{f_{t,n}^{[i]}}^2 + \sigma_{n}^2}  \Bigg),
\end{eqnarray}
respectively, where $\mathbf{H}_m^{[i]}=\mathbf{h}_m^{[i]} \mathbf{h}_m^{[i]H}$, $m\in\{1,\ldots,K\}$. For UL reception, we assume that the FD BS performs SIC by first decoding the signal of UL user $r$ and removing it from the received signal before decoding the signal of UL user $t$. Hence, the achievable rates (bits/s/Hz) of UL users $r$ and $t$ on subcarrier $i$ are given by \vspace*{-2mm}
\begin{eqnarray}
\label{rate-ul-r}
\hspace*{-3mm} R_{m,n,r,t}^{[i]\mathrm{UL}_r}\hspace*{-1mm}=&& \hspace*{-7mm} \log_2  \hspace*{-1mm} \Bigg(\hspace*{-1.5mm} 1 \hspace*{-1mm}  + \hspace*{-1mm}  \frac{P_r^{[i]}\abs{\mathbf{g}_r^{[i]H} \mathbf{v}_r^{[i]}}^2}{P_t^{[i]}\abs{\mathbf{g}_t^{[i]H} \hspace*{-1mm} \mathbf{v}_r^{[i]}}^2 \hspace*{-1mm}  + \hspace*{-1mm} \Tr\hspace*{-1mm}\Big(\hspace*{-0.5mm}\rho\mathbf{V}_r^{[i]} \hspace*{-1mm} \Diag \hspace*{-1mm} \big( \mathbf{H}_{\mathrm{SI}}^{[i]} (\mathbf{w}_m^{[i]} \mathbf{w}_m^{[i]H} \hspace*{-2mm} + \hspace*{-1mm} \mathbf{w}_n^{[i]} \mathbf{w}_n^{[i]H} \hspace*{-2.5mm} + \hspace*{-1.5mm} \mathbf{Z}^{[i]})\mathbf{H}_{\mathrm{SI}}^{[i]H}\big)\hspace*{-1mm} \Big) \hspace*{-1mm}  + \hspace*{-1mm} \sigma_{\mathrm{BS}}^2 \norm{\mathbf{v}_r^{[i]}}^2} \hspace*{-1.5mm} \Bigg),\notag\\[-5mm]
\\[-2mm]
\label{rate-ul-t}
\hspace*{-3mm} R_{m,n,r,t}^{[i]\mathrm{UL}_t}=&& \hspace*{-7mm} \log_2 \hspace*{-1mm} \Bigg( \hspace*{-1.5mm} 1 \hspace*{-1mm} + \hspace*{-1mm} \frac{P_t^{[i]}\abs{\mathbf{g}_t^{[i]} \mathbf{v}_t^{[i]}}^2} {\Tr\hspace*{-1mm}\Big(\hspace*{-0.5mm}\rho\mathbf{V}_t^{[i]} \hspace*{-0.5mm} \Diag \hspace*{-1mm} \big( \mathbf{H}_{\mathrm{SI}}^{[i]} (\mathbf{w}_m^{[i]} \mathbf{w}_m^{[i]H} \hspace*{-1mm} + \hspace*{-1mm} \mathbf{w}_n^{[i]} \mathbf{w}_n^{[i]H} \hspace*{-1mm} + \hspace*{-1mm} \mathbf{Z}^{[i]})\mathbf{H}_{\mathrm{SI}}^{[i]H}\big)\hspace*{-1mm} \Big) + \sigma_{\mathrm{BS}}^2 \norm{\mathbf{v}_t^{[i]}}^2} \Bigg),
\end{eqnarray}
respectively, where  variable $\mathbf{v}_r^{[i]} \in \mathbb{C}^{N_{\mathrm{T}} \times 1}$ denotes the receive beamforming vector adopted at the FD BS for detecting the information received from UL user $r$ on subcarrier $i$ and we define $\mathbf{V}_r^{[i]} = \mathbf{v}_r^{[i]} \mathbf{v}_r^{[i]H}, r \in\{1,\ldots,J\}$, for notational simplicity.
In this paper, maximum ratio combining (MRC) is adopted at the FD BS for UL signal reception \cite{tse2005fundamentals}, i.e., $\mathbf{v}_r^{[i]}=\mathbf{h}_r^{[i]}/\norm{\mathbf{h}_r^{[i]}}$. As a result, an MRC-SIC multiuser detector is employed in the UL \cite{tse2005fundamentals}. In practice, MRC-SIC achieves full diversity \cite{tse2005fundamentals} and leads to a high performance if the number of receive antennas is sufficiently large \cite{JR:TWC_large_antennas,Ngo_Energy13}. In addition, the use of an MRC-SIC multiuser detector facilitates the design of a computationally tractable and efficient resource allocation algorithm. Besides, $\rho$, $0 <\rho \ll 1$, in \eqref{rate-ul-r} and \eqref{rate-ul-t} reflects the noisiness of the SI cancellation at the FD BS where the variance of the residual SI is proportional to the power received at an antenna \cite{JR:FD_model}.

Therefore, the weighted system throughput on subcarrier $i$ is given by \vspace*{-2mm}
\begin{eqnarray} \label{throughput_i}
U_{m,n,r,t}^{[i]}(\mathbf{s},\mathbf{W},\mathbf{p},\mathbf{Z}) = s_{m,n,r,t}^{[i]}\Big[  w_m  R_{m,n,r,t}^{[i]\mathrm{DL}_m} + w_n  R_{m,n,r,t}^{[i]\mathrm{DL}_n} + \mu_r  R_{m,n,r,t}^{[i]\mathrm{UL}_r} + \mu_t R_{m,n,r,t}^{[i]\mathrm{UL}_t} \Big],
\end{eqnarray}
where $s_{m,n,r,t}^{[i]} \in\{0,1\}$ indicates the subcarrier allocation policy. Specifically, $s_{m,n,r,t}^{[i]}=1$ if DL users $m$, $n$ and UL users $r$, $t$ are scheduled on subcarrier $i$ and $s_{m,n,r,t}^{[i]}=0$ if another scheduling policy is executed.
The priorities of DL user $m$ and UL user $r$ in resource allocation are reflected by the positive constants $0  \le  w_m \le 1$  and $0  \le  \mu_r \le 1$ which are defined in the upper layers to enforce fairness.
To facilitate the presentation, we introduce $\mathbf{s}\in\mathbb{Z}^{{N_{\mathrm{F}}}K^2 J^2 \times 1}$, $\mathbf{W}\in\mathbb{C}^{{N_{\mathrm{F}}}K \times N_{\mathrm{T}}}$, $\mathbf{p}\in\mathbb{R}_{\mathrm{+}}^{{N_{\mathrm{F}}}J \times 1}$, and $\mathbf{Z}\in\mathbb{C}^{{N_{\mathrm{F}}} M \times N_{\mathrm{T}}}$ as the collections of the optimization variables $s_{m,n,r,t}^{[i]}$, $\mathbf{w}_m^{[i]}$, $P_r^{[i]}$, and $\mathbf{Z}^{[i]}$ for $\forall i,m,n,r,t$, respectively.

NOMA requires successful SIC at the receivers.
In particular, for a given subcarrier $i$, DL user $n$ can only successfully decode and remove the signal of DL user $m$ by SIC when the following inequality holds \cite{sun2016optimalJournal, tse2005fundamentals, ImpactPairNOMA}:\vspace*{-2mm}
\begin{eqnarray}\label{rate-inequality}
\hspace*{-6mm}J_{m,n,r,t}^{[i]}(\mathbf{W},\mathbf{p},\mathbf{Z})\hspace*{-0mm} \hspace*{-2mm}&\triangleq&  \hspace*{-2mm} \log_2 \hspace*{-0mm} \Big(1 \hspace*{-0mm} + \hspace*{-0mm} \frac{\abs{\mathbf{h}_{m}^{[i]H} \mathbf{w}_{m}^{[i]} }^2 } {\abs{\mathbf{h}_{m}^{[i]H} \mathbf{w}_{n}^{[i]} }^2 + \Tr(\mathbf{H}_m^{[i]} \mathbf{Z}^{[i]}) + P_r^{[i]} \abs{f_{r,m}^{[i]}}^2 + P_t^{[i]} \abs{f_{t,m}^{[i]}}^2 + \sigma_{m}^2}\Big) \notag \\[-2mm]
\hspace*{-2mm}&-& \hspace*{-2mm}
\log_2 \hspace*{-0mm} \Big(1 \hspace*{-0.5mm} + \hspace*{-0.5mm} \frac{\abs{\mathbf{h}_{n}^{[i]H} \mathbf{w}_{m}^{[i]} }^2 } {\abs{\mathbf{h}_{n}^{[i]H} \mathbf{w}_{n}^{[i]} }^2 \hspace*{-0.5mm} + \hspace*{-0.5mm} \Tr(\mathbf{H}_n^{[i]} \mathbf{Z}^{[i]}) \hspace*{-0.5mm} + \hspace*{-0.5mm} P_r^{[i]} \abs{f_{r,n}^{[i]}}^2 + P_t^{[i]} \abs{f_{t,n}^{[i]}}^2 + \sigma_{n}^2}  \hspace*{-0mm} \Big) \le 0.
\end{eqnarray}
Eq. \eqref{rate-inequality} indicates that successful SIC at DL user $n$ is possible only if the SINR of the signal of user $m$ at DL user $n$ is higher than or equal to the corresponding SINR at DL user $m$. For the UL reception, since the FD BS is the intended receiver of the signals from both UL users, the FD BS is able to perform successful SIC in arbitrary order.

Next, for guaranteeing communication security in the considered system, we design the resource allocation under a worst-case assumption. In particular, we assume that the equivalent eavesdropper can cancel the MUI and the UL (DL) user interference before decoding the information of the desired DL (UL) user on each subcarrier.
Thus, under this assumption, the capacities of the eavesdropper in eavesdropping DL user $m$ and UL user $r$ on subcarrier $i$ are given by\vspace*{-2mm}
\begin{eqnarray}
\label{eqn:DL-EVE-rate}C_{\mathrm{DL}_m}^{[i]}\hspace*{-1mm}&=&\hspace*{-1mm} \log_2\det(\mathbf{I}_{M} \hspace*{-1mm} +  \hspace*{-1mm} (\mathbf{X}^{[i]})^{-1}\mathbf{L}^{[i]H} \mathbf{w}_m^{[i]} \mathbf{w}_m^{[i]H} \mathbf{L}^{[i]}) \,\,\, \text{and} \\[-2mm]
C_{\mathrm{UL}_r}^{[i]}\hspace*{-1mm}&=&\hspace*{-1mm} \log_2\det(\mathbf{I}_{M} \hspace*{-1mm} + \hspace*{-1mm} P_r^{[i]} (\mathbf{X}^{[i]})^{-1}\mathbf{e}_{r}^{[i]}\mathbf{e}_{r}^{[i]H}),
\end{eqnarray}
respectively, where $\mathbf{X}^{[i]}=\mathbf{L}^{[i]H} \mathbf{Z}^{[i]} \mathbf{L}^{[i]} +\sigma_{\mathrm{E}}^2\mathbf{I}_{M}$ denotes the AN-plus-channel-noise covariance matrix of the equivalent eavesdropper on subcarrier $i$. Therefore, the achievable secrecy rates between the FD BS and  DL user $l$ and UL user $h$ on subcarrier $i$ are given by $R_{\mathrm{DL}_l}^{[i]\mathrm{Sec}}\hspace*{-1.5mm}=\hspace*{-1.5mm}\Big[R_{\mathrm{DL}_l}^{[i]} - C_{\mathrm{DL}_l}^{[i]}\Big]^+$  and
$R_{\mathrm{UL}_h}^{[i]\mathrm{Sec}}\hspace*{-1.5mm}=\hspace*{-1.5mm}\Big[R_{\mathrm{UL}_h}^{[i]} - C_{\mathrm{UL}_h}^{[i]}\Big]^+$, respectively,
where\vspace*{-2mm}
\begin{eqnarray}
R_{\mathrm{DL}_l}^{[i]}\hspace*{-1.5mm}&=&\hspace*{-1.5mm} \sum_{r=1}^{J} \sum_{t=1}^{J}\Big( \sum_{n=1}^{K}  s_{l,n,r,t}^{[i]} R_{l,n,r,t}^{[i]\mathrm{DL}_l} +  \sum_{m=1}^{K}  s_{m,l,r,t}^{[i]} R_{m,l,r,t}^{[i]\mathrm{DL}_l}\Big)  \,\,\, \text{and}\\[-2mm]
R_{\mathrm{UL}_h}^{[i]}\hspace*{-1.5mm}&=&\hspace*{-1.5mm} \sum_{m=1}^{K}\sum_{n=1}^{K} \Big( \sum_{t=1}^{J}  s_{m,n,h,t}^{[i]} R_{m,n,h,t}^{[i]\mathrm{UL}_h} +  \sum_{r=1}^{J}  s_{m,n,r,h}^{[i]} R_{m,n,r,h}^{[i]\mathrm{UL}_h}\Big),
\end{eqnarray}
denote the achievable rates (bits/s/Hz) of DL user $l$ and UL user $h$ on subcarrier $i$, respectively.

\vspace*{-3mm}
\subsection{Channel State Information}
In this paper, we assume that all wireless channels vary slowly over time. Thus, the perfect CSI of the DL and UL transmission links and the UL-to-DL co-channel interference channels is available at the FD BS via handshaking between the FD BS and the DL and UL users at the beginning of each scheduling slot.
On the other hand, the idle users constituting the equivalent eavesdropper only perform handshaking with the FD BS occasionally due to their idle and unscheduled status.
Thus, only imperfect CSI of the equivalent eavesdropper's channels is available at the FD BS. To capture the impact of the imperfect CSI, the link between the FD BS and the equivalent eavesdropper on subcarrier $i$, i.e., $\mathbf{L}^{[i]}$, and the CSI of the link between UL user $r$  and the equivalent eavesdropper on subcarrier $i$, i.e., $\mathbf{e}_{r}^{[i]}$, are modeled as \cite{jiaheng13Robust}: \vspace*{-3mm}
\begin{eqnarray}
\label{eqn:BS-EVE-channel}\mathbf{L}^{[i]}\hspace*{-2mm}&=&\hspace*{-2mm}\mathbf{\hat L}^{[i]} + \Delta\mathbf{L}^{[i]},\,\,\,
\label{eqn:BS-EVE-set}\mathbf{\Omega}_{\mathrm{DL}}^{[i]} \triangleq \Big\{\mathbf{L}^{[i]}\in \mathbb{C}^{N_{\mathrm{T}}\times M}  :\norm{\Delta\mathbf{L}^{[i]}}_{\mathrm{F}} \le \varepsilon_{\mathrm{DL}}^{[i]}\Big\} \,\,\, \text{   and}\\[-2mm]
\label{eqn:UL-EVE-channel}\mathbf{e}_{r}^{[i]}\hspace*{-2mm}&=&\hspace*{-2mm}\mathbf{\hat e}_{r}^{[i]} + \Delta\mathbf{e}_{r}^{[i]},\,  \,\,\,
\label{eqn:UL-EVE-set}\mathbf{\Omega}_{\mathrm{UL}_{r}}^{[i]} \triangleq \Big\{\mathbf{e}_{r}^{[i]} \in \mathbb{C}^{M\times 1}  :\norm{\Delta\mathbf{e}_{r}^{[i]}} \le \varepsilon_{\mathrm{UL}_{r}}^{[i]}\Big\},
\end{eqnarray}
respectively, where $\mathbf{\hat L}^{[i]}$ and $\mathbf{\hat e}_{r}^{[i]}$ are the CSI estimates available at the FD BS, and $\Delta\mathbf{L}^{[i]}$ and $\Delta\mathbf{e}_{r}^{[i]}$ model the respective channel uncertainties. The continuous sets $\mathbf{\Omega}_{\mathrm{DL}}^{[i]}$ and $\mathbf{\Omega}_{\mathrm{UL}_{r}}^{[i]}$ contain all possible channel uncertainties with bounded magnitudes $\varepsilon_{\mathrm{DL}}^{[i]}$ and $\varepsilon_{\mathrm{UL}_{r}}^{[i]}$, respectively.

\vspace*{-3mm}
\subsection{Optimization Problem Formulation}
The system design objective is the maximization of the weighted system throughput under secrecy and QoS constraints. The resource allocation policy is obtained by solving the following optimization problem:\vspace*{-4mm}
\begin{eqnarray} \label{pro}
&&\hspace*{-7mm}\underset{\mathbf{s},\mathbf{W},\mathbf{p},\mathbf{Z}}{\maxo}\,\, \,\,  \sum_{i=1}^{N_{\mathrm{F}}}\sum_{m=1}^{K} \sum_{n=1}^{K} \sum_{r=1}^{J} \sum_{t=1}^{J} U_{m,n,r,t}^{[i]}(\mathbf{s},\mathbf{W},\mathbf{p},\mathbf{Z}) \\  [-1.5mm]
\notag\mathrm{s.t.}
&&\hspace*{-5mm}\mbox{C1: } s_{m,n,r,t}^{[i]} J_{m,n,r,t}^{[i]}(\mathbf{W},\mathbf{p},\mathbf{Z}) \le 0, \forall i,m,n,r,t, \\[-1.5mm]
&&\hspace*{-5mm}\mbox{C2: }\overset{N_{\mathrm{F}}}{\underset{i=1}{\sum}} \Big(\overset{K}{\underset{m=1}{\sum}} \overset{K}{\underset{n=1}{\sum}} \overset{J}{\underset{r=1}{\sum}} \overset{J}{\underset{t=1}{\sum}} s_{m,n,r,t}^{[i]} (\norm{\mathbf{w}_{m}^{[i]}}^2 + \norm{\mathbf{w}_{n}^{[i]}}^2) + \Tr(\mathbf{Z}^{[i]})\Big) \le P_{\mathrm{max}}^{\mathrm{DL}}, \notag\\ [-1.5mm]
&&\hspace*{-5mm}\mbox{C3: }\sum_{i=1}^{N_{\mathrm{F}}} R_{\mathrm{DL}_l}^{[i]} \hspace*{-1mm} \ge \hspace*{-1mm} R_{\mathrm{req}_l}^{\mathrm{DL}},  \forall l \hspace*{-0.5mm} \in \hspace*{-0.5mm} \{1,\hspace*{-0.5mm}\ldots\hspace*{-0.5mm},K\}, \quad \,\hspace*{9.7mm}
\mbox{C4: }\sum_{i=1}^{N_{\mathrm{F}}} R_{\mathrm{UL}_h}^{[i]} \hspace*{-1mm} \ge \hspace*{-1mm} R_{\mathrm{req}_h}^{\mathrm{UL}},  \forall h \hspace*{-0.5mm} \in \hspace*{-0.5mm} \{1,\hspace*{-0.5mm}\ldots\hspace*{-0.5mm},J\}, \notag \\[-1.5mm]
&&\hspace*{-5mm}\mbox{C5: }\hspace*{-1mm} \underset{\Delta\mathbf{L}^{[i]}\in \mathbf{\Omega}_{\mathrm{DL}}^{[i]}}{\max} \hspace*{-2mm} C_{\mathrm{DL}_l}^{[i]} \hspace*{-1mm} \le \hspace*{-1mm} R_{\mathrm{tol}_{l}}^{\mathrm{DL}[i]}, \forall l \hspace*{-0.5mm} \in \hspace*{-0.5mm} \{1,\hspace*{-0.5mm}\ldots\hspace*{-0.5mm},K\},  \hspace*{8.5mm}
\mbox{C6: }\hspace*{-1mm} \underset{\Delta \mathbf{e}_{r}^{[i]} \in \mathbf{\Omega}_{\mathrm{UL}_{r}}^{[i]}}{\underset{\Delta\mathbf{L}^{[i]}\in \mathbf{\Omega}_{\mathrm{DL}}^{[i]},}{\max}} \hspace*{-2mm} C_{\mathrm{UL}_h}^{[i]} \hspace*{-1mm} \le \hspace*{-1mm} R_{\mathrm{tol}_{h}}^{\mathrm{UL}[i]},  \forall h \hspace*{-0.5mm} \in \hspace*{-0.5mm} \{1,\hspace*{-0.5mm}\ldots\hspace*{-0.5mm},J\}, \notag\\[-4mm]
&&\hspace*{-5mm}\mbox{C7: }\overset{N_{\mathrm{F}}}{\underset{i=1}{\sum}} \overset{K}{\underset{m=1}{\sum}} \overset{K}{\underset{n=1}{\sum}} \Big(\overset{J}{\underset{t=1}{\sum}} s_{m,n,h,t}^{[i]} P_h^{[i]} + \overset{J}{\underset{r=1}{\sum}} s_{m,n,r,h}^{[i]} P_h^{[i]} \Big) \le P_{\mathrm{max}_h}^{\mathrm{UL}}, \forall h \hspace*{-0.5mm} \in \hspace*{-0.5mm} \{1,\hspace*{-0.5mm}\ldots\hspace*{-0.5mm},J\}, \notag \\ [-1mm]
&&\hspace*{-5mm} \mbox{C8: } P_r^{[i]} \ge 0, \forall i,r \hspace*{50mm} \mbox{C9: } s_{m,n,r,t}^{[i]} \in \{0,1\}, \forall i,m,n,r,t, \notag \\ [-2mm]
&&\hspace*{-5mm} \mbox{C10: } \overset{K}{\underset{m=1}{\sum}} \overset{K}{\underset{n=1}{\sum}} \overset{J}{\underset{r=1}{\sum}} \overset{J}{\underset{t=1}{\sum}} s_{m,n,r,t}^{[i]} \le 1, \forall i, \hspace*{20mm} \mbox{C11: } \mathbf{Z}^{[i]} \succeq \mathbf{0}, \,\mathbf{Z}^{[i]} \in\mathbb{H}^{N_\mathrm{T}}, \, \forall i. \notag
\end{eqnarray}
Constraint C1 ensures the success of SIC at user $n$ if $s_{m,n,r,t}^{[i]} = 1$.
The constant $P_{\mathrm{max}}^{\mathrm{DL}}$ in constraint C2 is the maximum transmit power of the FD BS. Constraints C3 and C4 impose minimum required constant throughputs $R_{\mathrm{req}_l}^{\mathrm{DL}}$ and $R_{\mathrm{req}_h}^{\mathrm{UL}}$ for DL user $l$ and UL user $h$, respectively.
$R_{\mathrm{tol}_{l}}^{\mathrm{DL}[i]}$ and $R_{\mathrm{tol}_{h}}^{\mathrm{UL}[i]}$, in C5 and C6, respectively, are pre-defined system parameters representing the maximum tolerable data rate at the equivalent eavesdropper for decoding the information of DL user $l$ and UL user $h$ on subcarrier $i$, respectively.
In fact, communication security in DL and UL is guaranteed by constraints C5 and C6 for given CSI uncertainty sets $\mathbf{\Omega}_{\mathrm{DL}}^{[i]}$ and $\mathbf{\Omega}_{\mathrm{UL}_{r}}^{[i]}$.
If the above optimization problem is feasible, the proposed problem formulation guarantees that the secrecy rate of DL user $l$ is bounded below by $R_{\mathrm{DL}_l}^{\mathrm{Sec}}\hspace*{-1mm}\ge\hspace*{-1mm} R_{\mathrm{req}_l}^{\mathrm{DL}}- \sum_{i=1}^{N_{\mathrm{F}}} R_{\mathrm{tol}_{l}}^{\mathrm{DL}[i]}$
and the secrecy rate of UL user $h$ is bounded below by
$R_{\mathrm{UL}_h}^{\mathrm{Sec}}\hspace*{-1mm}\ge\hspace*{-1mm} R_{\mathrm{req}_h}^{\mathrm{UL}}- \sum_{i=1}^{N_{\mathrm{F}}} R_{\mathrm{tol}_{h}}^{\mathrm{UL}[i]}$.
We note that the secrecy constraints in C5 and C6 provide flexibility in controlling the security level of communication for different services\footnote{For example, low data rate services (e.g. E-mail) have lower information leakage tolerance than high data rate services (e.g. video streaming).}.
Constraint C7 limits the maximum transmit power of UL user $r$ to $P_{\mathrm{max}_r}^{\mathrm{UL}}$. Constraint C8 ensures that the power of UL user $r$ is non-negative.
Constraints C9 and C10 are imposed since each subcarrier can be allocated to at most two UL and two DL users.
Constraint C11 ensures that  after optimization the resulting $\mathbf{Z}^{[i]}$ is a Hermitian positive semidefinite matrix.

The problem in \eqref{pro} is a mixed non-convex and combinatorial optimization problem which is very difficult to solve. In particular, the binary selection constraint in C9, the non-convex constraints C1--C7, and the non-convex objective function are the main obstacles for the systematic design of a resource allocation algorithm. In particular, constraints C5 and C6 involve infinite numbers of inequality constraints due to the continuity of the corresponding CSI uncertainty sets. Nevertheless, despite these challenges, in the next section, we will develop optimal and suboptimal solutions to problem \eqref{pro}.

\vspace*{-5mm}
\section{Solution of the Optimization Problem}
In this section, we first solve the problem in \eqref{pro} optimally by applying monotonic optimization theory \cite{tuy2000monotonic,zhang2013monotonic} leading to an iterative resource allocation algorithm. In particular, a non-convex optimization problem is solved by semidefinite programming (SDP) relaxation in each iteration. Then, a suboptimal solution based on sequential convex approximation is proposed to reduce computational complexity while achieving close-to-optimal performance.

\vspace*{-4mm}
\subsection{Optimal Resource Allocation Scheme}

\subsubsection{Monotonic Optimization Framework}

Let us define $\mathbf{W}_{m}^{[i]}=\mathbf{w}_m^{[i]} \mathbf{w}_m^{[i]H}$, $\mathbf{W}_{m}^{[i]}\in\mathbb{H}^{N_{\mathrm{T}}}$. Also, we define new variables $\tilde{\mathbf{W}}_{m,n,r,t}^{[i]\mathrm{DL}_m}=s_{m,n,r,t}^{[i]} \mathbf{W}_{m}^{[i]}$ and $\tilde{P}_{m,n,r,t}^{[i]\mathrm{UL}_r}=s_{m,n,r,t}^{[i]}  P_r^{[i]}$. Then, the weighted throughput on subcarrier $i$ in  \eqref{throughput_i} can be rewritten equivalently as: \vspace*{-3mm}
\begin{eqnarray} \label{throughput-i-eqv}
U_{m,n,r,t}^{[i]}(\mathbf{s},\mathbf{W},\mathbf{p},\mathbf{Z}) = w_m  \tilde{R}_{m,n,r,t}^{[i]\mathrm{DL}_m} + w_n  \tilde{R}_{m,n,r,t}^{[i]\mathrm{DL}_n} + \mu_r  \tilde{R}_{m,n,r,t}^{[i]\mathrm{UL}_r} + \mu_t \tilde{R}_{m,n,r,t}^{[i]\mathrm{UL}_t},
\end{eqnarray}
where \vspace*{-4mm}
\begin{eqnarray} \label{eqv-rate-dl-m}
\tilde{R}_{m,n,r,t}^{[i]\mathrm{DL}_m}=&&\hspace*{-6mm}  \log_2 \hspace*{-0.5mm} \Big(  1 \hspace*{-1mm} + \hspace*{-1mm} \frac{ \Tr(\mathbf{H}_{m}^{[i]} \tilde{\mathbf{W}}_{m,n,r,t}^{[i]\mathrm{DL}_m} ) } {\Tr(\mathbf{H}_{m}^{[i]} (\tilde{\mathbf{W}}_{m,n,r,t}^{[i]\mathrm{DL}_n} \hspace*{-0.5mm} + \hspace*{-0.5mm} \mathbf{Z}^{[i]})) \hspace*{-1mm} + \hspace*{-1mm} \tilde{\mathbf{p}}_{m,n,r,t}^{[i]} \mathbf{f}_{r,t,m}^{[i]}   \hspace*{-0.5mm} + \hspace*{-0.5mm} \sigma_{m}^2}  \Big)
\triangleq \log_2 (u_{m,n,r,t}^{[i]}), \\[-2mm]
\label{eqv-rate-dl-n}
\tilde{R}_{m,n,r,t}^{[i]\mathrm{DL}_n}=&&\hspace*{-6mm}  \log_2 \Big(  1 \hspace*{-1mm} + \hspace*{-1mm} \frac{ \Tr(\mathbf{H}_{n}^{[i]} \tilde{\mathbf{W}}_{m,n,r,t}^{[i]\mathrm{DL}_n} ) } { \tilde{\mathbf{p}}_{m,n,r,t}^{[i]} \mathbf{f}_{r,t,n}^{[i]} \hspace*{-1mm} + \hspace*{-0.5mm}\Tr(\mathbf{H}_n^{[i]} \mathbf{Z}^{[i]}) \hspace*{-0.5mm} + \hspace*{-0.5mm} \sigma_{n}^2}  \Big)
\triangleq  \log_2 (v_{m,n,r,t}^{[i]}),
\end{eqnarray}
respectively. Here, we  define $\tilde{\mathbf{p}}_{m,n,r,t}^{[i]}=\Big[\tilde{P}_{m,n,r,t}^{[i]\mathrm{UL}_r},\tilde{P}_{m,n,r,t}^{[i]\mathrm{UL}_t}\Big]$ and $\mathbf{f}_{r,t,m}^{[i]}=\Big[\abs{f_{r,m}^{[i]}}^2, \abs{f_{t,m}^{[i]}}^2\Big]^T$ for simplicity of notation.
In addition, $\tilde{R}_{m,n,r,t}^{[i]\mathrm{UL}_r}$ and $\tilde{R}_{m,n,r,t}^{[i]\mathrm{UL}_t}$ in \eqref{throughput-i-eqv} are given by  \vspace*{-2mm}
\begin{eqnarray}
\label{eqv-rate-ul-r}
\tilde{R}_{m,n,r,t}^{[i]\mathrm{UL}_r}\hspace*{-0.5mm} =&& \hspace*{-6mm}  \log_2 \hspace*{-0.5mm} \Big( \hspace*{-0.5mm}1 \hspace*{-1mm} + \hspace*{-1mm} \frac{ \tilde{P}_{m,n,r,t}^{[i]\mathrm{UL}_r} \Tr(\mathbf{G}_r^{[i]} \mathbf{V}_r^{[i]})}{\tilde{P}_{m,n,r,t}^{[i]\mathrm{UL}_t} \hspace*{-0.5mm} \Tr(\mathbf{G}_t^{[i]} \mathbf{V}_r^{[i]}) \hspace*{-1mm} + \hspace*{-0.8mm} \Tr\hspace*{-0.5mm}\big(\hspace*{-0mm}\rho\mathbf{V}_r^{[i]} \hspace*{-0.5mm} \Diag \hspace*{-0.5mm} \big(\mathbf{S}_{\mathrm{SI}_{m,n,r,t}}^{[i]} \big)\big) \hspace*{-0.8mm} + \hspace*{-0.8mm} \sigma_{\mathrm{BS}}^2 \hspace*{-0.5mm} \Tr( \hspace*{-0.5mm} \mathbf{V}_r^{[i]} \hspace*{-0.5mm} )}  \Big) \hspace*{-0.5mm} \triangleq \hspace*{-0.5mm}  \log_2 (\zeta_{m,n,r,t}^{[i]}), \notag \\[-5mm]
\\[-2mm]
\label{eqv-rate-ul-t}
\hspace*{-3mm} \tilde{R}_{m,n,r,t}^{[i]\mathrm{UL}_t}\hspace*{-0.5mm}=&& \hspace*{-6mm} \log_2 \hspace*{-0.5mm} \Big( \hspace*{-0.5mm}1 \hspace*{-1mm} + \hspace*{-1mm} \frac{ \tilde{P}_{m,n,r,t}^{[i]\mathrm{UL}_t} \Tr(\mathbf{G}_t^{[i]} \mathbf{V}_t^{[i]})} {\Tr\hspace*{-0.5mm}\big(\hspace*{-0mm}\rho\mathbf{V}_t^{[i]} \hspace*{-0.5mm} \Diag \hspace*{-0.5mm} \big(\mathbf{S}_{\mathrm{SI}_{m,n,r,t}}^{[i]} \big)\big) + \hspace*{-0.5mm} \sigma_{\mathrm{BS}}^2 \hspace*{-0.5mm} \Tr( \hspace*{-0.5mm} \mathbf{V}_t^{[i]} \hspace*{-0.5mm} )}  \Big)
 \triangleq  \hspace*{-0mm} \log_2 (\xi_{m,n,r,t}^{[i]}),
\end{eqnarray}
respectively, where $\mathbf{G}_r^{[i]} \hspace*{-1mm} = \hspace*{-1mm} \mathbf{g}_r^{[i]} \hspace*{-0.5mm} \mathbf{g}_r^{[i]H}\hspace*{-1mm},$ $r \hspace*{-1mm} \in \hspace*{-1mm} \{\hspace*{-0.5mm} 1,\hspace*{-0.5mm}\ldots,\hspace*{-0.5mm}J \hspace*{-0.5mm} \}$, and
$\mathbf{S}_{\mathrm{SI}_{m,n,r,t}}^{[i]} \hspace*{-2mm} = \hspace*{-1mm} \mathbf{H}_{\mathrm{SI}}^{[i]}  (\tilde{\mathbf{W}}_{m,n,r,t}^{[i]\mathrm{DL}_m} \hspace*{-0.5mm} + \hspace*{-0.5mm} \tilde{\mathbf{W}}_{m,n,r,t}^{[i]\mathrm{DL}_n} \hspace*{-0.5mm}+ \hspace*{-1mm} \mathbf{Z}^{[i]})\mathbf{H}_{\mathrm{SI}}^{[i]H}\hspace*{-1mm}.$

Next, we note that constraint C1 is the difference of two logarithmic functions which is not a monotonic function. To facilitate the use of monotonic optimization, we introduce the following equivalent transformation of constraint C1: \vspace*{-2mm}
\begin{eqnarray}
&&\hspace*{-12mm}\mbox{C1a: }
\hspace*{-1mm} \log_2 \hspace*{-0.5mm} \Big(  1 \hspace*{-1mm} + \hspace*{-1mm} \frac{ \Tr(\mathbf{H}_{m}^{[i]} \tilde{\mathbf{W}}_{m,n,r,t}^{[i]\mathrm{DL}_m} ) } {\Tr(\mathbf{H}_{m}^{[i]} (\tilde{\mathbf{W}}_{m,n,r,t}^{[i]\mathrm{DL}_n} \hspace*{-0.5mm} +\hspace*{-0.5mm} \mathbf{Z}^{[i]})) \hspace*{-1mm} + \hspace*{-1mm} \tilde{\mathbf{p}}_{m,n,r,t}^{[i]} \mathbf{f}_{r,t,m}^{[i]}   \hspace*{-1mm}  + \hspace*{-0.5mm} \sigma_{m}^2}  \Big) \hspace*{-0.5mm} + \hspace*{-0.5mm} \varsigma_{m,n,r,t}^{[i]}  \hspace*{-0.5mm} \le \hspace*{-0.5mm} \log_2 \hspace*{-0.5mm} \Big (\hspace*{-0.5mm} 1 \hspace*{-0.5mm} + \hspace*{-0.5mm} \frac{P_{\mathrm{max}}^{\mathrm{DL}}}{\sigma_{m}^2} \hspace*{-0.5mm} \Big), \\[-2mm]
&&\hspace*{-12mm}\mbox{C1b: }
\hspace*{-1.3mm} \log_2 \hspace*{-0.5mm} \Big(  1 \hspace*{-1mm} + \hspace*{-1mm} \frac{ \Tr(\mathbf{H}_{n}^{[i]} \tilde{\mathbf{W}}_{m,n,r,t}^{[i]\mathrm{DL}_m} ) } {\Tr(\mathbf{H}_{n}^{[i]} (\tilde{\mathbf{W}}_{m,n,r,t}^{[i]\mathrm{DL}_n} + \mathbf{Z}^{[i]})) \hspace*{-1mm} + \hspace*{-1mm} \tilde{\mathbf{p}}_{m,n,r,t}^{[i]} \mathbf{f}_{r,t,n}^{[i]}   \hspace*{-1mm} + \hspace*{-0.5mm} \sigma_{n}^2}  \Big) \hspace*{-1mm} + \hspace*{-1mm} \varsigma_{m,n,r,t}^{[i]}   \hspace*{-0.5mm} \ge \hspace*{-0.5mm} \log_2 \hspace*{-0.5mm} \Big ( \hspace*{-0.5mm} 1 \hspace*{-0.5mm} + \hspace*{-0.5mm} \frac{P_{\mathrm{max}}^{\mathrm{DL}}}{\sigma_{m}^2} \hspace*{-1mm} \Big),
\end{eqnarray}
where $ \varsigma_{m,n,r,t}^{[i]} \ge 0$ is a new  scalar slack optimization variable. With the aforementioned definitions, constraint C1a can be rewritten as:\vspace*{-2mm}
\begin{eqnarray}
&&\hspace*{-12mm} \overline{\mbox{C1}}\mbox{a: } \log_2 \hspace*{-0mm} ( u_{m,n,r,t}^{[i]} ) \hspace*{-0mm} + \hspace*{-0mm} \varsigma_{m,n,r,t}^{[i]}  \hspace*{-0mm} \le \hspace*{-0mm} \log_2 \hspace*{-0mm} \Big (\hspace*{-0mm} 1 \hspace*{-0mm} + \hspace*{-0mm} \frac{P_{\mathrm{max}}^{\mathrm{DL}}}{\sigma_{m}^2} \hspace*{-0mm} \Big).
\end{eqnarray}
We note that the left hand sides of constraints  $\overline{\mbox{C1}}\mbox{a}$ and $\mbox{C1b}$ are monotonically increasing functions as required for monotonic optimization.

Then, the original problem in \eqref{pro} can be rewritten in the following equivalent form:\vspace*{-2mm}
\begin{eqnarray} \label{eqv-pro}
\hspace*{-1mm}&&\hspace*{-12mm}\underset{\mathbf{s},\tilde{\mathbf{W}},\tilde{\mathbf{p}},\mathbf{Z},\mathbf{t}}{\maxo} \sum_{i=1}^{N_{\mathrm{F}}} \hspace*{-0.7mm} \sum_{m=1}^{K} \hspace*{-0.7mm} \sum_{n=1}^{K} \hspace*{-0.7mm} \sum_{r=1}^{J} \hspace*{-0.7mm} \sum_{t=1}^{J} \log_2(u_{m,n,r,t}^{[i]})^{w_m} \hspace*{-0.7mm} + \hspace*{-0.7mm} \log_2(v_{m,n,r,t}^{[i]})^{w_n} \hspace*{-0.7mm} + \hspace*{-0.7mm} \log_2(\zeta_{m,n,r,t}^{[i]})^{\mu_r} \hspace*{-0.7mm} + \hspace*{-0.7mm} \log_2(\xi_{m,n,r,t}^{[i]})^{\mu_t} \notag \\[-2mm]
\hspace*{-1mm}\mathrm{s.t.} \hspace*{0mm}
&&\hspace*{-5mm}\overline{\mbox{C1}}\mbox{a}, \mbox{C1b}, \mbox{C9}, \mbox{C10}, \mbox{C11}, \mbox{ C2: }\overset{N_{\mathrm{F}}}{\underset{i=1}{\sum}} \Big(\overset{K}{\underset{m=1}{\sum}} \overset{K}{\underset{n=1}{\sum}} \overset{J}{\underset{r=1}{\sum}} \overset{J}{\underset{t=1}{\sum}} (\tilde{\mathbf{W}}_{m,n,r,t}^{[i]\mathrm{DL}_m} \hspace*{-0mm} + \hspace*{-0mm} \tilde{\mathbf{W}}_{m,n,r,t}^{[i]\mathrm{DL}_n}) \hspace*{-0mm} + \hspace*{-0mm} \mathbf{Z}^{[i]} \Big) \le P_{\mathrm{max}}^{\mathrm{DL}}, \notag\\ [-2mm]
&&\hspace*{-5mm}\mbox{C3: }\sum_{i=1}^{N_{\mathrm{F}}} \sum_{r=1}^{J} \sum_{t=1}^{J} \Big( \sum_{n=1}^{K}   \tilde{R}_{l,n,r,t}^{[i]\mathrm{DL}_l} +  \sum_{m=1}^{K}  \tilde{R}_{m,l,r,t}^{[i]\mathrm{DL}_l}\Big) \ge R_{\mathrm{req}_l}^{\mathrm{DL}},  \forall l \in \{1,\ldots,K\}, \notag\\[-2mm]
&&\hspace*{-5mm}\mbox{C4: }\sum_{i=1}^{N_{\mathrm{F}}} \sum_{m=1}^{K} \sum_{n=1}^{K} \Big( \sum_{t=1}^{J}  \tilde{R}_{m,n,h,t}^{[i]\mathrm{UL}_h} +  \sum_{r=1}^{J}  \tilde{R}_{m,n,r,h}^{[i]\mathrm{UL}_h}\Big) \ge R_{\mathrm{req}_h}^{\mathrm{UL}},  \forall h \in \{1,\ldots,J\}, \notag \\[-2mm]
&&\hspace*{-5mm}\mbox{C5: }\underset{\Delta\mathbf{L}^{[i]}\in \mathbf{\Omega}_{\mathrm{DL}}^{[i]}}{\max} \log_2\det(\mathbf{I}_{M}+ (\mathbf{X}^{[i]})^{-1}\mathbf{L}^{[i]H} \tilde{\mathbf{W}}_{m,n,r,t}^{[i]\mathrm{DL}_m} \mathbf{L}^{[i]}) \hspace*{-0mm} \le \hspace*{-0mm} R_{\mathrm{tol}_{m}}^{[i]\mathrm{DL}}, \,\,\,  \forall i,m,n,r,t, \notag \\[-2mm]
&&\hspace*{-5mm}\mbox{C6: }\underset{\Delta\mathbf{L}^{[i]}\in \mathbf{\Omega}_{\mathrm{DL}}^{[i]}, \Delta \mathbf{e}_{r}^{[i]} \in \mathbf{\Omega}_{\mathrm{UL}_{r}}^{[i]}}{\max} \log_2\det(\mathbf{I}_{M}+ \tilde{P}_{m,n,r,t}^{[i]\mathrm{UL}_r} (\mathbf{X}^{[i]})^{-1}\mathbf{e}_{r}^{[i]}\mathbf{e}_{r}^{[i]H}) \hspace*{-0mm}\le \hspace*{-0mm} R_{\mathrm{tol}_{r}}^{[i]\mathrm{UL}},  \,\,\, \forall i,m,n,r,t,\notag\\[-2mm]
&&\hspace*{-5mm}\mbox{C7: }\overset{N_{\mathrm{F}}}{\underset{i=1}{\sum}} \overset{K}{\underset{m=1}{\sum}} \overset{K}{\underset{n=1}{\sum}} \Big(\overset{J}{\underset{t=1}{\sum}} \tilde{P}_{m,n,h,t}^{[i]\mathrm{UL}_h} + \overset{J}{\underset{r=1}{\sum}} \tilde{P}_{m,n,r,h}^{[i]\mathrm{UL}_h} \Big) \le P_{\mathrm{max}_h}^{\mathrm{UL}}, \forall h \hspace*{-0.5mm} \in \hspace*{-0.5mm} \{1,\hspace*{-0.5mm}\ldots\hspace*{-0.5mm},J\},
\quad \, \mbox{C8: } \tilde{P}_{m,n,r,t}^{[i]\mathrm{UL}_r} \ge 0, \forall i,r, \notag \\ [-2mm]
&&\hspace*{-5mm}  \mbox{C12: } \tilde{\mathbf{W}}_{m,n,r,t}^{[i]\mathrm{DL}_m} \hspace*{-0.8mm} \succeq \hspace*{-0.8mm}  \mathbf{0}, \tilde{\mathbf{W}}_{m,n,r,t}^{[i]\mathrm{DL}_n} \hspace*{-0.8mm} \succeq \hspace*{-0.8mm} \mathbf{0}, \quad \quad  \quad\mbox{C13:} \Rank(\tilde{\mathbf{W}}_{m,n,r,t}^{[i]\mathrm{DL}_m} ) \le 1, \Rank(\tilde{\mathbf{W}}_{m,n,r,t}^{[i]\mathrm{DL}_n} ) \le 1,
\end{eqnarray}
where $\tilde{\mathbf{W}}\in\mathbb{C}^{N_{\mathrm{F}}K^2J^2 \times N_{\mathrm{T}}}$ and $\tilde{\mathbf{p}}\in\mathbb{C}^{N_{\mathrm{F}}K^2J^2 \times 1}$ are the collections of all $\tilde{\mathbf{W}}_{m,n,r,t}^{[i]\mathrm{DL}_m}$ and $\tilde{P}_{m,n,r,t}^{[i]\mathrm{UL}_r}$, respectively.
Constraints C12 and C13 are imposed to guarantee that $\tilde{\mathbf{W}}_{m,n,r,t}^{[i]\mathrm{DL}_m}=s_{m,n,r,t}^{[i]}\mathbf{w}_m^{[i]}\mathbf{w}_m^{[i]H}$ holds after optimization.
We also define
$\mathbf{t}\hspace*{-1mm}=\hspace*{-1mm}[t_1\hspace*{-0.5mm},\hspace*{-0.5mm}\ldots\hspace*{-0.5mm}, \hspace*{-0.5mm}t_{D}]^T\hspace*{-1.8mm}=\hspace*{-1mm}[\varsigma_{1,1,1,1}^1\hspace*{-0.5mm}, \hspace*{-0.5mm}\ldots\hspace*{-0.5mm},\hspace*{-0.5mm}\varsigma_{K,K,J,J}^{N_{\mathrm{F}}}]^T$ to collect all $\varsigma_{m,n,r,t}^{[i]}$ in one vector, where $D=N_{\mathrm{F}}K^2 J^2$.

For notational simplicity, we define functions $f_{d}(\tilde{\mathbf{W}},\hspace*{-0.5mm}\tilde{\mathbf{p}},\hspace*{-0.5mm}\mathbf{Z})$ and $g_{d}(\tilde{\mathbf{W}},\hspace*{-0.5mm}\tilde{\mathbf{p}},\hspace*{-0.5mm}\mathbf{Z})$ as the collections of the numerator and denominator of variables $u_{m,n,r,t}^{[i]}$, $v_{m,n,r,t}^{[i]}$, $\zeta_{m,n,r,t}^{[i]}$, and $\xi_{m,n,r,t}^{[i]}$, respectively: \vspace*{-2mm}
\begin{equation}
\hspace*{-3mm} f_{d}(\tilde{\mathbf{W}},\hspace*{-0.8mm}\tilde{\mathbf{p}},\hspace*{-0.8mm}\mathbf{Z})\\
=
\hspace*{-1mm} \begin{cases}
\hspace*{-1mm} \Tr \hspace*{-0.5mm} \big(\mathbf{H}_{m}^{[i]} (\tilde{\mathbf{W}}_{m,n,r,t}^{[i]\mathrm{DL}_m} \hspace*{-0.5mm} + \hspace*{-0.5mm} \tilde{\mathbf{W}}_{m,n,r,t}^{[i]\mathrm{DL}_n} \hspace*{-0.5mm} + \hspace*{-0.5mm} \mathbf{Z}^{[i]}) \big) \hspace*{-0.5mm} + \hspace*{-0.5mm} \tilde{\mathbf{p}}_{m,n,r,t}^{[i]} \mathbf{f}_{r,t,m}^{[i]}   \hspace*{-0.5mm} +  \hspace*{-0.5mm} \sigma_{m}^2,
& \hspace*{0mm} d \hspace*{-1mm} = \hspace*{-1mm} \Delta, \\[-1.5mm]
\hspace*{-1mm} \Tr(\mathbf{H}_{n}^{[i]} (\tilde{\mathbf{W}}_{m,n,r,t}^{[i]\mathrm{DL}_n} \hspace*{-0.5mm}+\hspace*{-0.5mm} \mathbf{Z}^{[i]}) ) + \tilde{\mathbf{p}}_{m,n,r,t}^{[i]} \mathbf{f}_{r,t,n}^{[i]} \hspace*{-1mm}  + \hspace*{-0.5mm} \sigma_{n}^2,
& \hspace*{0mm} d\hspace*{-1mm} = \hspace*{-1mm} D \hspace*{-1mm} + \hspace*{-1mm} \Delta,\\[-1.5mm]
\hspace*{-1mm} \tilde{\mathbf{p}}_{m,n,r,t}^{[i]} \mathbf{r}_{r,t}^{[i]} \hspace*{-1mm} + \hspace*{-0.5mm} \Tr\hspace*{-0.5mm}\big(\hspace*{-0mm}\rho\mathbf{V}_r^{[i]} \hspace*{-0.5mm} \Diag \hspace*{-0.5mm} \big(\mathbf{S}_{\mathrm{SI}_{m,n,r,t}}^{[i]} \big)\big) \hspace*{-0.5mm} + \hspace*{-0.5mm} \sigma_{\mathrm{BS}}^2 \hspace*{-0.5mm} \Tr( \hspace*{-0.5mm} \mathbf{V}_r^{[i]} \hspace*{-0.5mm} ),
& \hspace*{0mm} d\hspace*{-1mm} = \hspace*{-1mm} 2D \hspace*{-1mm} + \hspace*{-1mm} \Delta,\\[-1.5mm]
\hspace*{-1mm} \tilde{P}_{m,n,r,t}^{[i]\mathrm{UL}_t} \Tr(\mathbf{G}_t^{[i]} \mathbf{V}_t^{[i]}) \hspace*{-0.5mm} + \hspace*{-0.5mm} \Tr\hspace*{-0.5mm}\big(\hspace*{-0mm}\rho\mathbf{V}_t^{[i]} \hspace*{-0.5mm} \Diag \hspace*{-0.5mm} \big(\mathbf{S}_{\mathrm{SI}_{m,n,r,t}}^{[i]} \big)\big) \hspace*{-0.5mm} + \hspace*{-0.5mm} \sigma_{\mathrm{BS}}^2 \hspace*{-0.5mm} \Tr( \hspace*{-0.5mm} \mathbf{V}_t^{[i]} \hspace*{-0.5mm} ),
& \hspace*{0mm} d\hspace*{-1mm} = \hspace*{-1mm} 3D \hspace*{-1mm} + \hspace*{-1mm} \Delta,
\end{cases}
\end{equation}\vspace*{-2mm}
\begin{equation}
\hspace*{-1mm}g_{d}(\tilde{\mathbf{W}},\hspace*{-1mm}\tilde{\mathbf{p}},\hspace*{-1mm}\mathbf{Z})\hspace*{-1mm}=
\hspace*{-1mm} \begin{cases}
\hspace*{-1mm} \Tr \hspace*{-0.5mm} \big(\mathbf{H}_{m}^{[i]}  (\tilde{\mathbf{W}}_{m,n,r,t}^{[i]\mathrm{DL}_n} \hspace*{-0.5mm} + \hspace*{-0.5mm} \mathbf{Z}^{[i]})\big) \hspace*{-0.5mm} + \hspace*{-0.5mm} \tilde{\mathbf{p}}_{m,n,r,t}^{[i]} \mathbf{f}_{r,t,m}^{[i]}   \hspace*{-0.5mm} +  \hspace*{-0.5mm} \sigma_{m}^2,  & \hspace*{0mm} d\hspace*{-1mm} = \hspace*{-1mm} \Delta, \\[-1.5mm]
\hspace*{-1mm} \tilde{\mathbf{p}}_{m,n,r,t}^{[i]} \mathbf{f}_{r,t,n}^{[i]} \hspace*{-1mm} + \hspace*{-0.5mm}\Tr(\mathbf{H}_n^{[i]} \mathbf{Z}^{[i]}) \hspace*{-0.5mm} + \hspace*{-0.5mm} \sigma_{n}^2, & \hspace*{0mm} d\hspace*{-1mm} = \hspace*{-1mm} D\hspace*{-1mm} +\hspace*{-1mm} \Delta,\\[-1.5mm]
\hspace*{-1mm} \tilde{P}_{m,n,r,t}^{[i]\mathrm{UL}_t} \Tr(\mathbf{G}_t^{[i]} \mathbf{V}_r^{[i]}) \hspace*{-1mm} + \hspace*{-0.5mm} \Tr\hspace*{-0.5mm}\big(\hspace*{-0mm}\rho\mathbf{V}_r^{[i]} \hspace*{-0.5mm} \Diag \hspace*{-0.5mm} \big(\mathbf{S}_{\mathrm{SI}_{m,n,r,t}}^{[i]} \big)\big) \hspace*{-0.5mm} + \hspace*{-0.5mm} \sigma_{\mathrm{BS}}^2 \hspace*{-0.5mm} \Tr( \hspace*{-0.5mm} \mathbf{V}_r^{[i]} \hspace*{-0.5mm} ), & \hspace*{0mm} d\hspace*{-1mm} = \hspace*{-1mm} 2D \hspace*{-1mm} + \hspace*{-1mm} \Delta,\\[-1.5mm]
\hspace*{-1mm} \Tr\hspace*{-0.5mm}\big(\hspace*{-0mm}\rho\mathbf{V}_t^{[i]} \hspace*{-0.5mm} \Diag \hspace*{-0.5mm} \big(\mathbf{S}_{\mathrm{SI}_{m,n,r,t}}^{[i]} \big)\big) + \hspace*{-0.5mm} \sigma_{\mathrm{BS}}^2 \hspace*{-0.5mm} \Tr( \hspace*{-0.5mm} \mathbf{V}_t^{[i]} \hspace*{-0.5mm} ), & \hspace*{0mm} d \hspace*{-1mm} = \hspace*{-1mm} 3D \hspace*{-1mm} + \hspace*{-1mm} \Delta,
\end{cases}
\end{equation}
where $\mathbf{r}_{r,t}^{[i]}=[\Tr(\mathbf{G}_r^{[i]} \mathbf{V}_r^{[i]}), \Tr(\mathbf{G}_t^{[i]} \mathbf{V}_r^{[i]})]^T$ and $\Delta=(i-1)K^2 J^2+(m-1)KJ^2+(n-1)J^2+(r-1)J+t$.
We further define
$\mathbf{z}\hspace*{-1mm}=\hspace*{-1mm}[z_1\hspace*{-0.5mm},\hspace*{-0.5mm}\ldots\hspace*{-0.5mm},\hspace*{-0.5mm}z_{4D}]^T$, where $z_\Delta=u_{m,n,r,t}^{[i]}$, $z_{D+\Delta}=v_{m,n,r,t}^{[i]}$, $z_{2D+\Delta}=\zeta_{m,n,r,t}^{[i]}$, and $z_{3D+\Delta}=\xi_{m,n,r,t}^{[i]}$, $\forall i,m,n,r,t$.
Now, we rewrite the original problem in $\eqref{pro}$ as: \vspace*{-3mm}
\begin{eqnarray}\label{MO-pro}
\hspace*{-15mm}\underset{\mathbf{z},\mathbf{t}}{\maxo}\,\, \,\,  \sum_{d=1}^{4D} \log_2(z_d)^{\chi_d}   \quad \quad \quad
\mathrm{s.t.}
\hspace*{5mm} (\mathbf{z},\mathbf{t})\in\mathcal{V},
\end{eqnarray}
where $\hspace*{-0mm}\chi_d\hspace*{-0mm}$ is the corresponding user weight, i.e., $\chi_d\hspace*{-1mm}=\hspace*{-1mm}w_m$, $m=\rem(\lceil d/(KJ^2) \rceil, K)$, $\forall d \in \{1,\ldots,D\}$, $\chi_d=w_n$, $n=\rem(\lceil d/J^2 \rceil, K)$, $\forall d \in \{D+1,\ldots,2D\}$, $\chi_d=\mu_r$, $r=\rem(\lceil d/J \rceil, J)$, $\forall d \in \{2D+1,\ldots,3D\}$, and $\chi_d=\mu_t$, $t=\rem( d, J)$, $\forall d \in \{3D+1,\ldots,4D\}$.
Furthermore, $\mathcal{V}=\mathcal{G}\cap\mathcal{H}$ is the feasible set where $\mathcal{G}$ is given by \vspace*{-2mm}
\begin{eqnarray}
\hspace*{-3mm}\mathcal{G}\hspace*{-3mm}&=&\hspace*{-3mm}\Big\{ (\mathbf{z},\mathbf{t}) \mid 1\le z_d \le \frac{f_d(\tilde{\mathbf{W}},\hspace*{-0.5mm}\tilde{\mathbf{p}},\hspace*{-0.5mm}\mathbf{Z})} {g_d(\tilde{\mathbf{W}},\hspace*{-0.5mm}\tilde{\mathbf{p}},\hspace*{-0.5mm}\mathbf{Z})},
(\mathbf{z},\mathbf{t}) \in \mathcal{Q}, (\tilde{\mathbf{W}},\tilde{\mathbf{p}},\mathbf{Z})\in\mathcal{P}, \forall d\Big\},
\end{eqnarray}
with $\mathcal{P}$ and $\mathcal{Q}$ being the feasible sets spanned by constraints $\mbox{C2, C5--C8, C11--C13}$ and $\overline{\mbox{C1}}\mbox{a}$, respectively. Feasible set $\mathcal{H}$ is spanned by constraints $\mbox{C1b}$, C3, and C4.

\subsubsection{Optimal Algorithm}
Now, we note that the objective function and all functions appearing in the constraints in \eqref{MO-pro} are monotonically increasing functions. Therefore, problem \eqref{MO-pro} is in the canonical form of a monotonic optimization problem. Using a similar approach as in our previous work on FD SISO MC-NOMA \cite{sun2016optimalJournal}, we employ the polyblock outer approximation approach for solving the monotonic optimization problem in \eqref{MO-pro}.
However, the design of the corresponding iterative resource allocation algorithm is much more challenging for the considered FD MISO MC-NOMA system compared to the SISO MC-NOMA case in \cite{sun2016optimal,sun2016optimalJournal}. In particular, constraints C5 and C6 involve an infinite number of inequality constraints and constraint C13 restricts the rank of the optimization variables $\tilde{\mathbf{W}}$.
In the following, we first present the polyblock outer approximation-based iterative resource allocation algorithm framework employed for solving the monotonic optimization problem in \eqref{MO-pro}.
Then, we handle non-convex constraints C5, C6, and C13 appearing in each iteration of the proposed resource allocation algorithm.

\begin{table}\vspace*{-8mm}
\begin{algorithm} [H]                    
\caption{Polyblock Outer Approximation Algorithm}          
\label{alg1}                           
\begin{algorithmic} [1]
\small          
\STATE Initialize polyblock $\mathcal{D}^{(1)}$. The vertex $\bm{\upsilon}^{(1)}=(\mathbf{z}^{(1)},\mathbf{t}^{(1)}) $ is initialized by setting its elements:
$u_{m,n,r,t}^{[i]}  =  1  +  \Tr(\mathbf{H}_{m}^{[i]}) P_{\mathrm{max}}^{\mathrm{DL}}/\sigma_{m}^2$,
$v_{m,n,r,t}^{[i]}  =  1  +  \Tr(\mathbf{H}_{n}^{[i]}) P_{\mathrm{max}}^{\mathrm{DL}}/\sigma_{n}^2$,
$\zeta_{m,n,r,t}^{[i]}  =  1  +  \Tr(\mathbf{G}_{r}^{[i]} \mathbf{V}_{r}^{[i]}) P_{\mathrm{max}_r}^{\mathrm{UL}}/(\sigma_{\mathrm{BS}}^2 \hspace*{-0.5mm} \Tr( \hspace*{-0.5mm} \mathbf{V}_r^{[i]}) \hspace*{-0.5mm} )$,
$\xi_{m,n,r,t}^{[i]}  =  1  + \Tr(\mathbf{G}_{t}^{[i]} \mathbf{V}_{t}^{[i]}) P_{\mathrm{max}_t}^{\mathrm{UL}}/(\sigma_{\mathrm{BS}}^2 \hspace*{-0.5mm} \Tr( \hspace*{-0.5mm} \mathbf{V}_t^{[i]} \hspace*{-0.5mm} ))$,
and $\varsigma_{m,n,r,t}^{[i]}  =  \log_2   ( 1  +  P_{\mathrm{max}}^{\mathrm{DL}}/\sigma_{m}^2  )$, $\forall i,m,n,r,t$

\STATE Set error tolerance $\epsilon \ll 1$ and iteration index $k=1$

\REPEAT [Main Loop]
\STATE Construct a smaller polyblock $\mathcal{D}^{(k+1)}$ with vertex set $\bm{\Upsilon}^{{(k+1)}}$ by replacing $\bm{\upsilon}^{(k)}$ with $5D$ new vertices $\big\{\tilde{\bm{\upsilon}}^{(k)}_{1}, \ldots,\tilde{\bm{\upsilon}}^{(k)}_{5D}\big\}$. The new vertex $\tilde{\bm{\upsilon}}^{(k)}_{j}$ is generated as $\tilde{\bm{\upsilon}}^{(k)}_{j} =\bm{\upsilon}^{(k)}-\big(\upsilon^{(k)}_j -\phi_j(\bm{\upsilon}^{(k)})\big)\mathbf{u}_j$, where $\upsilon^{(k)}_j$ and $\phi_j(\bm{\upsilon}^{(k)})$ are the $j$-th elements of $\bm{\upsilon}^{(k)}$ and $\mathbf{\Phi}(\bm{\upsilon}^{(k)})$, respectively. $\bm{\Phi}(\bm{\upsilon}^{(k)})$ is obtained by $\textbf{Algorithm 2}$

\STATE  Find $\bm{\upsilon}^{(k+1)}$ as that vertex of $\bm{\Upsilon}^{{(k+1)}} \cap \mathcal{H}$ whose projection maximizes the objective function of the problem, i.e., $\bm{\upsilon}^{(k+1)}=\underset{\bm{\upsilon} \in\bm{\Upsilon}^{(k+1)}\cap \mathcal{H}}{\arg \max} \Big\{ \sum_{j=1}^{4D} \log_2\big(\phi_j(\bm{\upsilon})\big)^{\chi_j}\Big\}$, and set $k=k+1$

\UNTIL $\frac{\norm{\bm{\upsilon}^{(k)} -\mathbf{\Phi}(\bm{\upsilon}^{(k)})}} {\norm{\bm{\upsilon}^{(k)}}} \le \epsilon$

\STATE $\bm{\upsilon}^{*} =\mathbf{\Phi}(\bm{\upsilon}^{(k)})$ and $(\tilde{\mathbf{W}}^*, \tilde{\mathbf{p}}^*, \mathbf{Z}^*)$ are obtained when calculating $\mathbf{\Phi}(\bm{\upsilon}^{(k)})$
\end{algorithmic}
\end{algorithm}\vspace*{-10mm}
\end{table}

\begin{figure}
\centering \vspace*{-0mm}
  \subfigure[]{
    \label{fig:polyblock1} 
    \begin{minipage}[b]{0.2\textwidth}
      \centering \vspace*{-3mm}
      \includegraphics[width=1.25in]{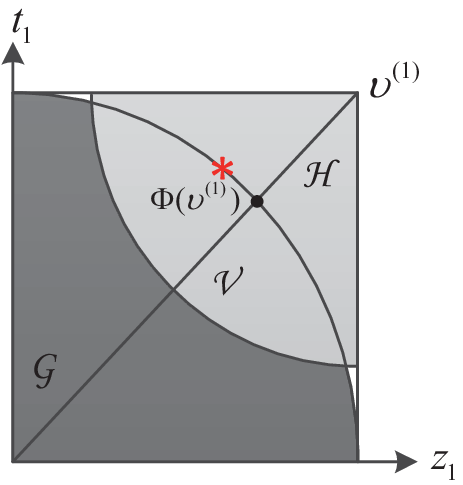}\vspace*{-3mm}
    \end{minipage}}%
  \subfigure[]{
    \label{fig:polyblock2}
    \begin{minipage}[b]{0.2\textwidth}
      \centering \vspace*{-3mm}
      \includegraphics[width=1.25in]{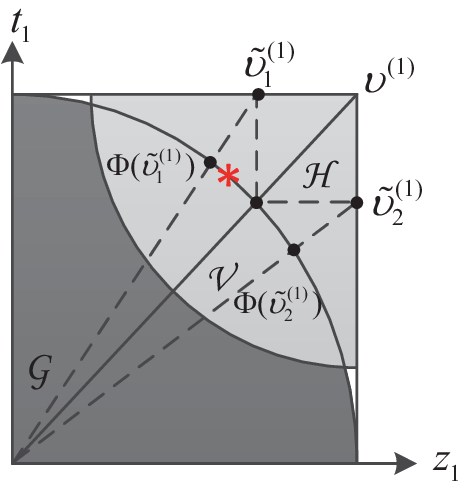}\vspace*{-3mm}
    \end{minipage}}
    \subfigure[]{
    \label{fig:polyblock3}
    \begin{minipage}[b]{0.2\textwidth}
      \centering \vspace*{-3mm}
      \includegraphics[width=1.25in]{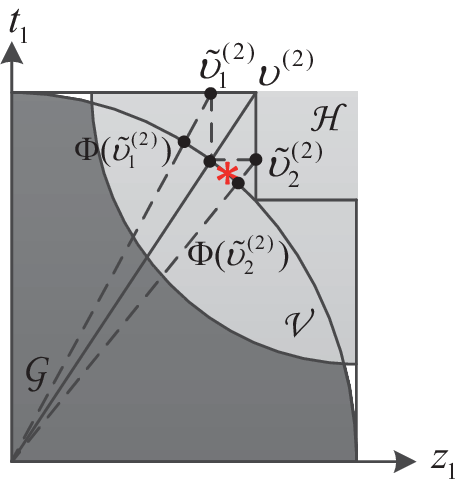}\vspace*{-3mm}
    \end{minipage}}%
    \subfigure[]{
    \label{fig:polyblock4}
    \begin{minipage}[b]{0.2\textwidth}
      \centering \vspace*{-3mm}
      \includegraphics[width=1.25in]{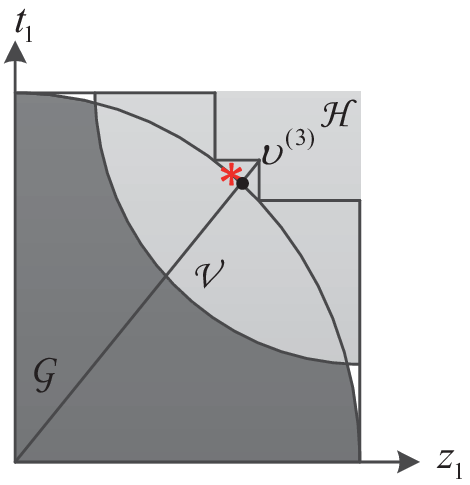}\vspace*{-3mm}
    \end{minipage}} \vspace*{-3mm}
  \caption{Illustration of the polyblock outer approximation algorithm. The red star is an optimal point on the boundary of the feasible set $\mathcal{V}$.}
  \label{fig:polyblock}  \vspace*{-7mm}
\end{figure}

According to monotonic optimization theory \cite{tuy2000monotonic,zhang2013monotonic},
the optimal solution of problem \eqref{MO-pro} is at the boundary of the feasible set $\mathcal{V}\hspace*{-1mm}=\hspace*{-1mm}\mathcal{G} \hspace*{-0.5mm} \cap \hspace*{-0.5mm} \mathcal{H}$. However, due to the non-convexity of the constraints in \eqref{MO-pro}, we cannot present the boundary of $\mathcal{V}$ analytically.
Therefore, we construct a sequence of polyblocks for approaching the boundary of $\mathcal{V}$ iteratively. First, we construct a polyblock $\mathcal{D}^{(1)}$ enclosing the feasible set $\mathcal{V}\hspace*{-1mm}=\hspace*{-1mm}\mathcal{G} \hspace*{-1mm}\cap \hspace*{-1mm} \mathcal{H}$ and the vertex set of $\mathcal{D}^{(1)}$, denoted as $\bm{\Upsilon}^{(1)}$, contains only one vertex $\bm{\upsilon}^{(1)}$. Here, vertex $\bm{\upsilon}^{(1)}$ is defined as $\bm{\upsilon}^{(1)}\triangleq (\mathbf{z}^{(1)}, \hspace*{-0.6mm} \mathbf{t}^{(1)})$ representing the optimization variables in \eqref{MO-pro}.
Then, we shrink polyblock $\mathcal{D}^{(1)}$ by replacing $\bm{\upsilon}^{(1)}$ with $5D$ new vertices $\tilde{\bm{\Upsilon}}^{(1)} =\big\{\tilde{\bm{\upsilon}}^{(1)}_{1},\hspace*{-0.6mm} \ldots,\hspace*{-0.6mm}\tilde{\bm{\upsilon}}^{(1)}_{5D}\big\}$.
The $5D$ new vertices, i.e., $\tilde{\bm{\upsilon}}^{(1)}_{j}$, $j\in\{1,\hspace*{-0.5mm}\ldots, \hspace*{-0.5mm}5D\}$, are generated based on vertex $\bm{\upsilon}^{(1)}$, as $\tilde{\bm{\upsilon}}^{(1)}_{j}\hspace*{-0.6mm} =\hspace*{-0.6mm}\bm{\upsilon}^{(1)} \hspace*{-0.6mm}-\hspace*{-0.6mm}\big(\upsilon^{(1)}_j \hspace*{-0.6mm} - \hspace*{-0.6mm} \phi_j(\bm{\upsilon}^{(1)})\big)\mathbf{u}_j$, where $\upsilon^{(1)}_j$ and $\phi_j(\bm{\upsilon}^{(1)})$ are the $j$-th elements of $\bm{\upsilon}^{(1)}$ and $\mathbf{\Phi}(\bm{\upsilon}^{(1)})$, respectively. Here, $\mathbf{\Phi}(\bm{\upsilon}^{(1)})\in \mathbb{C}^{5D \times 1}$ is the projection of $\bm{\upsilon}^{(1)}$ onto set $\mathcal{G}$,  and $\mathbf{u}_j$ is a unit vector containing only one non-zero element at position $j$.
Thus, the new vertex set $\bm{\Upsilon}^{(2)}\hspace*{-0.6mm} =\hspace*{-0.6mm}\big(\bm{\Upsilon}^{(1)} \hspace*{-0.6mm}-\hspace*{-0.6mm}\bm{\upsilon}^{(1)}\big) \hspace*{-0.6mm} \cup \hspace*{-0.6mm} \tilde{\bm{\Upsilon}}^{(1)}$ constitutes a new polyblock $\mathcal{D}^{(2)}$ which is smaller than $\mathcal{D}^{(1)}$, yet still encloses the feasible set $\mathcal{V}$.
Then, we choose $\bm{\upsilon}^{(2)}$ as the optimal vertex of $\bm{\Upsilon}^{(2)} \hspace*{-1mm} \cap \hspace*{-1mm} \mathcal{H}$ whose projection maximizes the objective function of the problem in \eqref{MO-pro}, i.e., $\bm{\upsilon}^{(2)}\hspace*{-0.5mm} =\hspace*{-0.5mm}\underset{\bm{\upsilon} \in\bm{\Upsilon}^{(2)}\cap \mathcal{H}}{\arg \max} \Big\{ \hspace*{-0.5mm} \sum_{j=1}^{4D} \hspace*{-0.5mm} \log_2\big(\phi_j(\bm{\upsilon})\big)^{\chi_j}\hspace*{-0.5mm}  \Big\}$. Similarly, we repeat the above procedure to shrink $\mathcal{D}^{(2)}$ based on $\bm{\upsilon}^{(2)}$, constructing a smaller polyblock and so on, i.e., $\mathcal{D}^{(1)} \hspace*{-1mm} \supset \hspace*{-1mm} \mathcal{D}^{{(2)}} \hspace*{-1mm} \supset \hspace*{-1mm} \dots \hspace*{-1mm} \supset \hspace*{-1mm}  \mathcal{V}$. The algorithm terminates if  $\frac{\norm{\bm{\upsilon}^{(k)} -\mathbf{\Phi}(\bm{\upsilon}^{(k)})}} {\norm{\bm{\upsilon}^{(k)}}} \hspace*{-0.6mm} \le \hspace*{-0.6mm} \epsilon$, where the error tolerance constant $\epsilon \hspace*{-0.6mm} > \hspace*{-0.6mm} 0$ specifies the accuracy of the approximation.
The algorithm is illustrated for a simple case in Figure \ref{fig:polyblock}, where, for simplicity of presentation, $\mathbf{z}$ and $\mathbf{t}$ contain only one element, respectively, i.e., $z_1$ and $t_1$.
We summarize the proposed polyblock outer approximation algorithm in \textbf{Algorithm 1}.

We can  acquire the optimal subcarrier allocation policy from the optimal vertex $\bm{\upsilon}^*$ obtained in line 7 of \textbf{Algorithm 1}. In particular, we can restore the values of $u_{m,n,r,t}^{[i]}$, $v_{m,n,r,t}^{[i]}$, $\zeta_{m,n,r,t}^{[i]}$, and $\xi_{m,n,r,t}^{[i]}$ from $\mathbf{z}^*$.
Besides, we note that $u_{m,n,r,t}^{[i]}$, $v_{m,n,r,t}^{[i]}$, $\zeta_{m,n,r,t}^{[i]}$, and $\xi_{m,n,r,t}^{[i]}$ are larger than one only if DL users $m$, $n$ and UL users $r$, $t$ are multiplexed on subcarrier $i$. Thus,
we can obtain the optimal subcarrier allocation policy $\mathbf{s}^*$ as: $s_{m,n,r,t}^{[i]}=1$ if $u_{m,n,r,t}^{[i]} > 1$, $v_{m,n,r,t}^{[i]} >1$, $\zeta_{m,n,r,t}^{[i]} >1$,  and $\xi_{m,n,r,t}^{[i]} >1$; otherwise, $s_{m,n,r,t}^{[i]}=0$.

In the following, we explain in detail how the projection of $\bm{\upsilon}^{(k)}$ required in each iteration of \textbf{Algorithm 1} is computed.
\subsubsection{Computation of Projection}
In each iteration of \textbf{Algorithm 1}, we obtain the projection of $\bm{\upsilon}^{(k)} $, i.e., $\mathbf{\Phi}(\hspace*{-0.5mm}\bm{\upsilon}^{(k)}\hspace*{-0.5mm}) \hspace*{-1mm} = \hspace*{-1mm} \lambda \bm{\upsilon}^{(k)} \hspace*{-1mm} = \hspace*{-1mm} \lambda (\hspace*{-0.5mm}\mathbf{\mathbf{z}}^{(k)},\hspace*{-1mm}\mathbf{\mathbf{t}}^{(k)}\hspace*{-0.5mm})$,  by solving \vspace*{-2mm}
\begin{eqnarray} \label{lambda}
\hspace*{-5mm}\lambda\hspace*{-3mm}&=&\hspace*{-3mm}\max\{ \beta  \mid \beta \bm{\upsilon}^{(k)}  \in \mathcal{G}\}  = \max\{ \beta \mid \beta  (\mathbf{z}^{(k)},\mathbf{t}^{(k)})  \in  \mathcal{G}\} \notag \\[-1mm]
\hspace*{-8mm}&=&\hspace*{-3mm} \max \hspace*{-0.5mm} \Big\{ \hspace*{-0.5mm} \beta \hspace*{-0.5mm} \mid  \hspace*{-0.5mm} \beta \hspace*{-0.5mm}  \le \hspace*{-2mm} \underset{(\tilde{\mathbf{W}},\tilde{\mathbf{p}},\mathbf{Z})\in\mathcal{P}}{\underset{1\le d \le 4D}{\min}} \hspace*{-0.5mm} \frac{f_d(\tilde{\mathbf{W}},\hspace*{-0.5mm}\tilde{\mathbf{p}},\hspace*{-0.5mm}\mathbf{Z})}{z_d^{(k)} g_d(\tilde{\mathbf{W}},\hspace*{-0.5mm}\tilde{\mathbf{p}},\hspace*{-0.5mm}\mathbf{Z})}, \beta (\mathbf{z}^{(k)},\mathbf{t}^{(k)}) \in \mathcal{Q} \Big\} \notag\\[-4mm]
\hspace*{-8mm}&=&\hspace*{-3mm} \max \hspace*{-0.5mm} \Big\{ \hspace*{-0.5mm} \beta \hspace*{-0.5mm} \mid  \hspace*{-0.5mm} \beta \hspace*{-0.5mm}  \le \underset{(\tilde{\mathbf{W}},\tilde{\mathbf{p}},\mathbf{Z})\in\mathcal{P}}{\underset{1\le d \le 4D}{\min}} \frac{f_d(\tilde{\mathbf{W}},\hspace*{-0.5mm}\tilde{\mathbf{p}},\hspace*{-0.5mm}\mathbf{Z})}{z_d^{(k)} g_d(\tilde{\mathbf{W}},\hspace*{-0.5mm}\tilde{\mathbf{p}},\hspace*{-0.5mm}\mathbf{Z})}, \log_2 \hspace*{-0mm} ( \beta z_{\Delta}^{(k)} ) \hspace*{-0mm} + \hspace*{-0mm} \beta t_{\Delta}^{(k)}  \hspace*{-0mm} \le \hspace*{-0mm} \log_2 \hspace*{-0mm} (\hspace*{-0mm} 1 \hspace*{-0mm} + \hspace*{-0mm} P_{\mathrm{max}}^{\mathrm{DL}}/\sigma_{m}^2 \hspace*{-0mm}), \forall \Delta \Big\} \notag\\[-4mm]
\hspace*{-8mm}&=&\hspace*{-3mm} \min \Big\{\underset{(\tilde{\mathbf{W}},\tilde{\mathbf{p}},\mathbf{Z})\in\mathcal{P}}{\max}\underset{1\le d \le 4D}{\min}  \frac{f_d(\tilde{\mathbf{W}},\hspace*{-0.5mm}\tilde{\mathbf{p}},\hspace*{-0.5mm}\mathbf{Z})}{z_d^{(k)} \hspace*{-0.5mm} g_d(\tilde{\mathbf{W}},\hspace*{-0.5mm}\tilde{\mathbf{p}},\hspace*{-0.5mm}\mathbf{Z})},
\beta^*   \Big\},
\end{eqnarray}
where $\beta^*$ is obtained by solving $\underset{0 \le \beta \le 1,\forall \Delta}{\max} \beta  \,\,\,\, \mbox{s.t.}  \log_2 \hspace*{-0.5mm} ( \beta z_{\Delta}^{(k)} ) \hspace*{-0.5mm} + \hspace*{-0.5mm} \beta t_{\Delta}^{(k)}  \hspace*{-1mm} \le \hspace*{-0.5mm} \log_2 \hspace*{-0.5mm} (\hspace*{-0.5mm} 1 \hspace*{-0.5mm} + \hspace*{-0.5mm} P_{\mathrm{max}}^{\mathrm{DL}}/\sigma_{m}^2 \hspace*{-0.5mm})$ via the bisection search method \cite{book:convex}.
In \eqref{lambda}, $\underset{(\tilde{\mathbf{W}},\tilde{\mathbf{p}},\mathbf{Z})\in\mathcal{P}}{\max}\underset{1\le d \le 4D}{\min} \,\, \frac{f_d(\tilde{\mathbf{W}},\tilde{\mathbf{p}},\mathbf{Z})}{z_d^{(k)} g_d(\tilde{\mathbf{W}},\tilde{\mathbf{p}},\mathbf{Z})}$ is a fractional programming problem. Therefore, we propose a Dinkelbach-type algorithm \cite{JR:fractional} for solving problem \eqref{lambda}. In particular, the proposed algorithm is summarized in \textbf{Algorithm 2}.
Specifically, $\tilde{\mathbf{W}}_x^*$, $\tilde{\mathbf{p}}_x^*$, and $\mathbf{Z}_x^*$ in line 3 are obtained by solving the following non-convex problem: \vspace*{-2mm}
\begin{eqnarray}\label{max-min-pro}
&&\hspace*{30mm}(\tilde{\mathbf{W}}_x^*,\tilde{\mathbf{p}}_x^*,\mathbf{Z}_x^*)=
\underset{\tilde{\mathbf{W}},\tilde{\mathbf{p}},\mathbf{Z},\tau}{\arg \max} \,\,\,\, \tau  \\[-2mm]
\mathrm{s.t.}
&&\hspace*{-6mm}\mbox{C2, C5--C8, C11--C13,} \mbox{ C14:}f_d(\tilde{\mathbf{W}},\tilde{\mathbf{p}},\mathbf{Z}) \hspace*{-1mm} - \hspace*{-1mm} \alpha_x z_d^{(k)} \hspace*{-0.5mm} g_d ( \tilde{\mathbf{W}},\tilde{\mathbf{p}},\mathbf{Z}) \hspace*{-1mm} \ge \hspace*{-1mm} \tau,  \forall d \hspace*{-0.7mm} \in \hspace*{-0.7mm} \{1,\hspace*{-0.5mm}\ldots\hspace*{-0.5mm},4D\},\notag
\end{eqnarray}
where $\tau$ is an auxiliary variable. We note that the problem in \eqref{max-min-pro} needs to be solved in each iteration of \textbf{Algorithm 2}. Besides, constraints C5 and C6 are non-convex constraints due to the inverse matrix $(\mathbf{X}^{[i]})^{-1}$ in the logarithmic determinant. Also, constraints C5 and C6 involve an infinite number of inequality constraints which are difficult to tackle. Hence, we first establish the following proposition for transforming constraints C5 and C6 into linear matrix inequality (LMI) constraints.
\begin{Prop} \label{prop_robust_variable}
For any $R_{\mathrm{tol}_{m}}^{[i]\mathrm{DL}} > 0$ and $R_{\mathrm{tol}_{r}}^{[i]\mathrm{UL}} > 0$, if $\Rank(\tilde{\mathbf{W}}_{m,n,r,t}^{[i]\mathrm{DL}_m}) \le 1$, constraints ${\mbox{C5}}$ and ${\mbox{C6}}$ of problem \eqref{eqv-pro} can be equivalently expressed in form of the following matrix inequalities: \vspace*{-2mm}
\begin{eqnarray}
\label{eqn:DL-tol-rate-QC}{\mbox{C5}}\hspace*{-2mm}&\Leftrightarrow&\hspace*{-2mm} {\widetilde{\mbox{C5}}}\mbox{: } \mathbf{L}^{[i]H}\tilde{\mathbf{W}}_{m,n,r,t}^{[i]\mathrm{DL}_m} \mathbf{L}^{[i]} \hspace*{1mm}\preceq\hspace*{1mm} \vartheta_{\mathrm{DL}_m}^{[i]}\mathbf{X}^{[i]}, \,\, \forall \mathbf{L}^{[i]}\in \mathbf{\Omega}_{\mathrm{DL}}^{[i]},\, \forall i,m,n,r,t, \,\,\, \text{and}\\[-1mm]
\label{eqn:UL-tol-rate-QC}{\mbox{C6}}\hspace*{-2mm}&\Leftrightarrow&\hspace*{-2mm} {\widetilde{\mbox{C6}}}\mbox{: } \tilde{P}_{m,n,r,t}^{[i]\mathrm{UL}_r} \mathbf{e}_{r}^{[i]} \mathbf{e}_{r}^{[i]H} \hspace*{1mm}\preceq\hspace*{1mm} \vartheta_{\mathrm{UL}_r}^{[i]} \mathbf{X}^{[i]}, \, \forall \mathbf{e}_r^{[i]}\in \mathbf{\Omega}_{\mathrm{UL}_r}^{[i]},\, \forall \mathbf{L}^{[i]}\in \mathbf{\Omega}_{\mathrm{DL}}^{[i]},\, \forall i,m,n,r,t,
\end{eqnarray}
where $\vartheta_{\mathrm{DL}_m}^{[i]}=2^{R_{\mathrm{tol}_{m}}^{[i]\mathrm{DL}}}-1$ and $\vartheta_{\mathrm{UL}_r}^{[i]}=2^{R_{\mathrm{tol}_{r}}^{[i]\mathrm{UL}}}-1$.
\end{Prop}
\emph{\quad Proof: }  Please refer to Appendix A. \hfill\qed

\begin{table}[t]\vspace*{-10mm}
\begin{algorithm} [H]                    
\caption{Projection Algorithm}          
\label{alg2}                           
\begin{algorithmic} [1]
\small          
\STATE Initialize $\alpha_1=0$ and set the iteration index to $x=1$ and the error tolerance $\delta \ll 1$ \vspace*{-2mm}

\REPEAT\vspace*{-0mm}
\STATE $(\tilde{\mathbf{W}}_x^*,\tilde{\mathbf{p}}_x^*,\mathbf{Z}_x^*)=\underset{(\tilde{\mathbf{W}},\tilde{\mathbf{p}},\mathbf{Z})\in\mathcal{P}}{\arg \max} \Big\{\underset{1\le d\le 4D}{\min}\big\{f_d(\tilde{\mathbf{W}},\tilde{\mathbf{p}},\mathbf{Z})-\alpha_x z_d^{(k)}g_d (\tilde{\mathbf{W}},\tilde{\mathbf{p}},\mathbf{Z})\big\}\Big\}$ \vspace*{-0mm}

\STATE $\alpha_{x+1}=\underset{1\le d \le 4D}{\min}\frac{f_d (\tilde{\mathbf{W}}_x^*,\tilde{\mathbf{p}}_x^*,\mathbf{Z}_x^*)}{z_d^{(k)}g_d (\tilde{\mathbf{W}}_x^*,\tilde{\mathbf{p}}_x^*,\mathbf{Z}_x^*)}$ $\,\,$ and set $x=x+1$

\UNTIL $\underset{1\le d\le 4D}{\min}\big\{f_d(\tilde{\mathbf{W}}_{x-1}^*,\tilde{\mathbf{p}}_{x-1}^*,\mathbf{Z}_{x-1}^*)-\alpha_{x} z_d^{(k)}g_d (\tilde{\mathbf{W}}_{x-1}^*,\tilde{\mathbf{p}}_{x-1}^*,\mathbf{Z}_{x-1}^*)\big\} \le \delta$ \vspace*{-1mm}

\STATE Obtain $\beta^*=\underset{0 \le \beta \le 1,\forall \Delta}{\max} \beta  \,\,\,\, \mbox{s.t.}  \log_2 \hspace*{-0.5mm} ( \beta z_{\Delta}^{(k)} ) \hspace*{-0.5mm} + \hspace*{-0.5mm} \beta t_{\Delta}^{(k)}  \hspace*{-1mm} \le \hspace*{-0.5mm} \log_2 \hspace*{-0.5mm} (\hspace*{-0.5mm} 1 \hspace*{-0.5mm} + \hspace*{-0.5mm} P_{\mathrm{max}}^{\mathrm{DL}}/\sigma_{m}^2 \hspace*{-0.5mm})$ by the  bisection search method

\STATE $\lambda=\min\{\alpha_{x},\beta^*\}$ and the projection is $\mathbf{\Phi}(\bm{\upsilon}^{(k)}) =\lambda \bm{\upsilon}^{(k)}$ and line $3$ provides the corresponding resource allocation policy $\tilde{\mathbf{W}}^*$

\end{algorithmic}
\end{algorithm}\vspace*{-13mm}
\end{table}

The resulting LMI constraints $\widetilde{\mbox{C5}}$ and $\widetilde{\mbox{C6}}$ still involve an infinite number of inequality constraints. Hence, we introduce the following Lemma for further simplifying $\widetilde{\mbox{C5}}$ and $\widetilde{\mbox{C6}}$.

\begin{Lem}[Generalized S-Procedure \cite{luo2004multivariate}] \label{lem:gernal-s-proc}Let $h(\mathbf{\Theta})=\mathbf{\Theta}^H\mathbf{A}\mathbf{\Theta}+\mathbf{\Theta}^H\mathbf{B}+\mathbf{B}^H\mathbf{\Theta}+\mathbf{C}$, where $\mathbf{\Theta}, \mathbf{B} \in \mathbb{C}^{N\times M}$, $\mathbf{A}\in \mathbb{H}^{N}$, $\mathbf{C}\in \mathbb{C}^{M \times M}$, and $\mathbf{D}\in \mathbb{C}^{N \times N}$ and $\mathbf{D}\succeq \mathbf{0}$. There exists a $\varrho \ge 0$ such that $h(\mathbf{\Theta})\succeq \mathbf{0}, \forall \mathbf{\Theta} \in \Big\{\mathbf{\Theta} |\Tr(\mathbf{D\Theta \Theta}^H) \le 1 \Big\}$, is equivalent to \vspace*{-2mm}
\begin{eqnarray}
\label{eqn:General-S-proc-LMI}
&&\begin{bmatrix}
       \mathbf{C} & \mathbf{B}^H          \\
       \mathbf{B} & \mathbf{A}           \\
           \end{bmatrix} -\varrho\begin{bmatrix}
       \mathbf{I}_{M} & \mathbf{0}          \\
       \mathbf{0} & -\mathbf{D}           \\
           \end{bmatrix}          \succeq \mathbf{0}.
\end{eqnarray}
\end{Lem}

By substituting $\mathbf{L}^{[i]}\hspace*{-0mm}=\hspace*{-0mm}\mathbf{\hat L}^{[i]} \hspace*{-0mm}+ \hspace*{-0mm}\Delta\mathbf{L}^{[i]}$ into (\ref{eqn:DL-tol-rate-QC}), we can rewrite constraint $\widetilde{\mbox{C5}}$ as \vspace*{-2mm}
\begin{eqnarray}
\label{eqn:BS-EVE-err-QC}\mathbf{0} \hspace*{-2mm} &\preceq& \hspace*{-2mm} \Delta\mathbf{L}^{[i]H}(\vartheta_{\mathrm{DL}_m}^{[i]}\mathbf{Z}^{[i]}-\tilde{\mathbf{W}}_{m,n,r,t}^{[i]\mathrm{DL}_m})\Delta\mathbf{L}^{[i]} +\Delta\mathbf{L}^{[i]H}(\vartheta_{\mathrm{DL}_m}^{[i]}\mathbf{Z}^{[i]}-\tilde{\mathbf{W}}_{m,n,r,t}^{[i]\mathrm{DL}_m})\mathbf{\hat L}^{[i]} \notag\\[-2mm]
\hspace*{-2mm}&+&\hspace*{-2mm}\mathbf{\hat L}^{[i]H}(\vartheta_{\mathrm{DL}_m}^{[i]}\mathbf{Z}^{[i]}-\tilde{\mathbf{W}}_{m,n,r,t}^{[i]\mathrm{DL}_m})\Delta\mathbf{L}^{[i]} +\mathbf{\hat L}^{[i]H}(\vartheta_{\mathrm{DL}_m}^{[i]}\mathbf{Z}^{[i]}-\tilde{\mathbf{W}}_{m,n,r,t}^{[i]\mathrm{DL}_m})\mathbf{\hat L}^{[i]} +\vartheta_{\mathrm{DL}_m}^{[i]}\sigma_{\mathrm{E}}^2\mathbf{I}_{M},
\end{eqnarray}
for $\Delta\mathbf{L}^{[i]} \in \Big\{\Delta\mathbf{L}^{[i]} |\Tr((\varepsilon_{\mathrm{DL}}^{[i]})^{-2}\Delta\mathbf{L}^{[i]}\Delta\mathbf{L}^{[i]H}) \le 1 \Big\}$, $\forall i,m,n,r,t$.
Then, by applying Lemma \ref{lem:gernal-s-proc}, constraint $\widetilde{\mbox{C5}}$ is equivalently represented by \vspace*{-2mm}
\begin{eqnarray}\label{eqn:LMI-C3}
\hspace*{-11mm}&&\overline{\text{C5}}\mbox{: } \mathbf{R}_{{\overline{\mathrm{C5}}}_{m,n,r,t}^{[i]\mathrm{DL}_m}} \big(\tilde{\mathbf{W}}_{m,n,r,t}^{[i]\mathrm{DL}_m},\mathbf{Z}^{[i]},\varrho_{m,n,r,t}^{[i]\mathrm{DL}_m}\big) \notag\\[-1mm]
\hspace*{-11mm}&=&\hspace*{-3mm}
          \begin{bmatrix}
       \vartheta_{\mathrm{DL}_m}^{[i]}\mathbf{\hat L}^{[i]H}\mathbf{Z}\mathbf{\hat L}^{[i]}\hspace*{-1mm}+\hspace*{-1mm}(\vartheta_{\mathrm{DL}_m}^{[i]}\sigma_{\mathrm{E}}^2 \hspace*{-1mm} - \hspace*{-1mm}\varrho_{m,n,r,t}^{[i]\mathrm{DL}_m})\mathbf{I}_{M} \hspace*{-1mm}&\hspace*{-1mm} \vartheta_{\mathrm{DL}_m}^{[i]}\mathbf{\hat L}^{[i]H}\mathbf{Z}^{[i]}          \\
       \vartheta_{\mathrm{DL}_m}^{[i]}\mathbf{Z}^{[i]}\mathbf{\hat L}^{[i]}    \hspace*{-1mm}&\hspace*{-11mm}   \vartheta_{\mathrm{DL}_m}^{[i]}\mathbf{Z}^{[i]}\hspace*{-1mm}+\hspace*{-1mm}  \varrho_{m,n,r,t}^{[i]\mathrm{DL}_m}(\varepsilon_{\mathrm{DL}}^{[i]})^{-2}\mathbf{I}_{N_\mathrm{T}}     \\
           \end{bmatrix}
           \hspace*{-1mm}  - \hspace*{-1mm} \mathbf{B}_{\mathbf{L}^{[i]}}^H\tilde{\mathbf{W}}_{m,n,r,t}^{[i]\mathrm{DL}_m}\mathbf{B}_{\mathbf{L}^{[i]}} \hspace*{-1mm} \succeq \hspace*{-1mm} \mathbf{0},
\end{eqnarray}
for $\varrho_{m,n,r,t}^{[i]\mathrm{DL}_m} \ge 0$, $\forall i, m, n,r,t$, where $\mathbf{B}_{\mathbf{L}^{[i]}}=\big[\mathbf{\hat L}^{[i]}\quad\mathbf{I}_{N_\mathrm{T}}\big]$.
On the other hand, we can also apply Lemma \ref{lem:gernal-s-proc}  to handle the CSI estimation error variables in constraint ${\widetilde{\mbox{C6}}}$, i.e., $\Delta\mathbf{e}_{r}^{[i]}$ and $\Delta\mathbf{L}^{[i]}$.
To facilitate the application of Lemma \ref{lem:gernal-s-proc}, we represent constraint ${\widetilde{\mbox{C6}}}$ in the equivalent form: \vspace*{-2mm}
\begin{eqnarray}
{\widetilde{\mbox{C6a}}}\mbox{: }\tilde{P}_{m,n,r,t}^{[i]\mathrm{UL}_r} \mathbf{e}_{r}^{[i]}\mathbf{e}_{r}^{[i]H} \hspace*{-0.7mm} \preceq \hspace*{-0.7mm} \mathbf{M}_{m,n,r,t}^{[i]\mathrm{UL}_r},\, \forall \mathbf{e}_{r}^{[i]} \in \mathbf{\Omega}_{\mathrm{UL}_{r}}^{[i]},\,  {\widetilde{\mbox{C6b}}}\mbox{: }\mathbf{M}_{m,n,r,t}^{[i]\mathrm{UL}_r} \hspace*{-0.7mm} \preceq \hspace*{-0.7mm} (\vartheta_{\mathrm{UL}_r}^{[i]}-1) \mathbf{X}^{[i]},\, \forall \mathbf{L}^{[i]}\in \mathbf{\Omega}_{\mathrm{DL}}^{[i]},\,
\end{eqnarray}
where $\mathbf{M}_{m,n,r,t}^{[i]\mathrm{UL}_r} \in \mathbb{H}^{M}$ is a slack matrix variable. We note that constraints ${\widetilde{\mbox{C6a}}}$ and ${\widetilde{\mbox{C6b}}}$ involve only $\Delta\mathbf{e}_{r}^{[i]}$ and $\Delta\mathbf{L}^{[i]}$, respectively. Then, by applying Lemma \ref{lem:gernal-s-proc} to constraints ${\widetilde{\mbox{C6a}}}$ and ${\widetilde{\mbox{C6b}}}$, respectively, we obtain the equivalent LMIs: \vspace*{-2mm}
\begin{eqnarray}\label{eqn:LMI-C6}
\hspace*{-35mm}&&\hspace*{-1mm} \overline{\text{C6}}\mbox{a: }\mathbf{R}_{{\overline{\mathrm{C6}}\mathrm{a}}_{m,n,r,t}^{[i]\mathrm{UL}_r}}\hspace*{-0.5mm}\big(\hspace*{-0.5mm} \mathbf{M}_{m,n,r,t}^{[i]\mathrm{UL}_r},\tilde{P}_{m,n,r,t}^{[i]\mathrm{UL}_r},\alpha_{m,n,r,t}^{[i]\mathrm{UL}_r}\hspace*{-0.5mm} \big) \notag \\[-2mm]
\hspace*{-35mm} &&\hspace*{-3mm}=
          \begin{bmatrix}
       -\tilde{P}_{m,n,r,t}^{[i]\mathrm{UL}_r}\mathbf{\hat e}_{r}^{[i]}\mathbf{\hat e}_{r}^{[i]H} \hspace*{-1.5mm}+\hspace*{-1mm} \mathbf{M}_{m,n,r,t}^{[i]\mathrm{UL}_r}\hspace*{-1.5mm} -\hspace*{-0.5mm} \alpha_{m,n,r,t}^{[i]\mathrm{UL}_r}\mathbf{I}_{M}  \hspace*{-1mm}  & \hspace*{-1mm} -\tilde{P}_{m,n,r,t}^{[i]\mathrm{UL}_r}\mathbf{\hat e}_{r}^{[i]}          \\
       \hspace*{-1.5mm} -\tilde{P}_{m,n,r,t}^{[i]\mathrm{UL}_r}\mathbf{\hat e}_{r}^{[i]H}  \hspace*{-1mm}  & \hspace*{-3.5mm} -\tilde{P}_{m,n,r,t}^{[i]\mathrm{UL}_r}\hspace*{-0.5mm}+\hspace*{-0.5mm} \alpha_{m,n,r,t}^{[i]\mathrm{UL}_r}(\varepsilon_{\mathrm{UL}_{r}}^{[i]})^{-2}    \\
           \end{bmatrix}\hspace*{-0.5mm}\succeq\hspace*{-0.5mm} \mathbf{0},
\end{eqnarray}
\begin{eqnarray}\label{eqn:LMI-C7}
\hspace*{-10mm}&&\hspace*{-5mm} \overline{\text{C6}}\mbox{b: }\mathbf{R}_{{\overline{\mathrm{C6}}\mathrm{b}}_{m,n,r,t}^{[i]\mathrm{UL}_r}} \hspace*{-0.5mm}\big(\mathbf{Z}^{[i]},\mathbf{M}_{m,n,r,t}^{[i]\mathrm{UL}_r},\beta_{m,n,r,t}^{[i]\mathrm{UL}_r}\big)\notag\\[-2mm]
\hspace*{-10mm} &=&\hspace*{-3mm}
          \begin{bmatrix}
       \vartheta_{\mathrm{UL}_r}^{[i]}\mathbf{\hat L}^{[i]H}\mathbf{Z}^{[i]}\mathbf{\hat L}^{[i]} \hspace*{-0.5mm}+\hspace*{-0.5mm}(\vartheta_{\mathrm{UL}_r}^{[i]}\sigma_{\mathrm{E}}^2 \hspace*{-0.5mm}- \hspace*{-0.5mm} \beta_{m,n,r,t}^{[i]\mathrm{UL}_r})\mathbf{I}_{M}\hspace*{-0.5mm} -\hspace*{-0.5mm} \mathbf{M}_{m,n,r,t}^{[i]\mathrm{UL}_r} \hspace*{-1mm} & \hspace*{-5mm} \vartheta_{\mathrm{UL}_r}^{[i]}\mathbf{\hat L}^{[i]H}\mathbf{Z}^{[i]}          \\
       \vartheta_{\mathrm{UL}_r}^{[i]}\mathbf{Z}^{[i]}\mathbf{\hat L}^{[i]}   \hspace*{-1mm} & \hspace*{-5mm}  \vartheta_{\mathrm{UL}_r}^{[i]}\mathbf{Z}^{[i]} \hspace*{-0.5mm}+\hspace*{-0.5mm}  \beta_{m,n,r,t}^{[i]\mathrm{UL}_r}(\varepsilon_{\mathrm{DL}}^{[i]})^{-2}\mathbf{I}_{N_\mathrm{T}}     \\
           \end{bmatrix}  \succeq \mathbf{0},
\end{eqnarray}
where $\alpha_{m,n,r,t}^{[i]\mathrm{UL}_r} \ge 0$ and $\beta_{m,n,r,t}^{[i]\mathrm{UL}_r} \ge 0$, $\forall i,m,n,r,t$.

Now, the remaining non-convexity of problem \eqref{max-min-pro} is due to rank-one constraint C13. Hence, we adopt SDP relaxation by removing constraint C13 from the problem formulation, such that problem \eqref{max-min-pro} becomes a convex SDP given by \vspace*{-2mm}
\begin{eqnarray}\label{sdp-max-min-pro}
&&\hspace*{30mm}(\tilde{\mathbf{W}}_x^*,\tilde{\mathbf{p}}_x^*,\mathbf{Z}_x^*)=
\underset{\tilde{\mathbf{W}},\tilde{\mathbf{p}},\mathbf{Z},\tau, \mathbf{M},\bm{\varrho}, \bm{\alpha}, \bm{\beta}}{\arg \max} \,\,\,\, \tau  \\[-2mm]
\mbox{s.t.}
&&\hspace*{-6mm}\mbox{C2}, \overline{\text{C5}}, \overline{\text{C6}}\mbox{a}, \overline{\text{C6}}\mbox{b}, \mbox{C7, C8, C11, C12,} \notag \mbox{ C14:}f_d(\tilde{\mathbf{W}},\tilde{\mathbf{p}},\mathbf{Z}) \hspace*{-1mm} - \hspace*{-1mm} \alpha_x z_d^{(k)} \hspace*{-0.5mm} g_d ( \tilde{\mathbf{W}},\tilde{\mathbf{p}},\mathbf{Z}) \hspace*{-1mm} \ge \hspace*{-1mm} \tau,  \forall d, \notag
\end{eqnarray}
where $\mathbf{M}$, $\bm{\varrho}$, $\bm{\alpha}$, and $\bm{\beta}$ are the collections of all $M_{m,n,r,t}^{[i]\mathrm{UL}_r}$, $\varrho_{m,n,r,t}^{[i]\mathrm{UL}_r}$, $\alpha_{m,n,r,t}^{[i]\mathrm{UL}_r}$, and $\beta_{m,n,r,t}^{[i]\mathrm{UL}_r}$, respectively. We note that \eqref{sdp-max-min-pro} can be solved by standard convex program solvers such as CVX \cite{website:CVX} and the tightness of the SDP relaxation is verified by the following theorem.

\begin{Thm}\label{thm:rankone_condition}
If $P_{\max}^{\mathrm{DL}}>0$, the optimal beamforming matrix
$\tilde{\mathbf{W}}_{x}^*$ in \eqref{sdp-max-min-pro} is a rank-one matrix.
\end{Thm}
\emph{\quad Proof: } Please refer to Appendix B. \hfill\qed



\subsubsection{Summary}
From the above discussion, we note that the globally optimal precoding and subcarrier allocation policy can be obtained via the proposed monotonic optimization based resource allocation algorithm. However, the computational complexity of the algorithm grows exponentially with the number of vertices, $5D$, used in each iteration.
In order to strike a balance between complexity and optimality,  in the next section we develop a suboptimal scheme which has only polynomial time computational complexity.
Nevertheless, the optimal algorithm is useful as the corresponding performance can serve as a performance benchmark for any suboptimal algorithm.

\vspace*{-2mm}
\subsection{Suboptimal Resource Allocation Scheme}
In this section, we propose a suboptimal algorithm requiring a low computational complexity to obtain a locally optimal solution for the optimization problem in \eqref{pro}.
We focus on solving problem \eqref{eqv-pro} since this is equivalent to solving problem \eqref{pro}.
First, we note that the auxiliary variables $\tilde{\mathbf{W}}_{m,n,r,t}^{[i]\mathrm{DL}_m} = s_{m,n,r,t}^{[i]} \mathbf{W}_{m}^{[i]}$ and $\tilde{P}_{m,n,r,t}^{[i]\mathrm{UL}_r} = s_{m,n,r,t}^{[i]} P_r^{[i]}$ in \eqref{eqv-pro} are products of integer variables and continuous variables, which is an obstacle for solving problem \eqref{eqv-pro} efficiently. Hence, we adopt the big-M method to overcome this difficulty \cite{lee2011mixed}. In particular, we decompose the product terms by imposing the following additional constraints:  \vspace*{-2mm}
\begin{eqnarray}
&& \hspace*{-15mm}
\mbox{C15: } \hspace*{-1mm} \tilde{\mathbf{W}}_{m,n,r,t}^{[i]\mathrm{DL}_m} \hspace*{-1mm} \preceq \hspace*{-1mm}  P_{\mathrm{max}}^{\mathrm{DL}} \mathbf{I}_{N_{\mathrm{T}}} s_{m,n,r,t}^{[i]}, \quad
\mbox{C16: } \hspace*{-1mm} \tilde{\mathbf{W}}_{m,n,r,t}^{[i]\mathrm{DL}_m} \hspace*{-1mm} \preceq \hspace*{-1mm} \mathbf{W}_m^{[i]}, \quad
\mbox{C17: }  \hspace*{-1mm} \tilde{\mathbf{W}}_{m,n,r,t}^{[i]\mathrm{DL}_m} \hspace*{-1mm} \succeq \hspace*{-1mm} \mathbf{0}, \\ [-2mm]
&& \hspace*{-15mm}
\mbox{C18: } \hspace*{-1mm} \tilde{\mathbf{W}}_{m,n,r,t}^{[i]\mathrm{DL}_m} \hspace*{-1mm} \succeq \hspace*{-1mm} \mathbf{W}_m^{[i]} \hspace*{-1mm} - \hspace*{-1mm} (1 \hspace*{-1mm} - \hspace*{-1mm}s_{m,n,r,t}^{[i]})P_{\mathrm{max}}^{\mathrm{DL}} \mathbf{I}_{N_{\mathrm{T}}},
\quad
\mbox{C19: } \hspace*{-1mm} \tilde{P}_{m,n,r,t}^{[i]\mathrm{UL}_r} \hspace*{-1mm} \ge \hspace*{-1mm} P_r^{[i]} \hspace*{-1mm} - \hspace*{-1mm} (1 \hspace*{-1mm} - \hspace*{-1mm}s_{m,n,r,t}^{[i]})P_{\mathrm{max}_r}^{\mathrm{UL}}, \\ [-2mm]
&& \hspace*{-15mm}
\mbox{C20: } \hspace*{-1mm} \tilde{P}_{m,n,r,t}^{[i]\mathrm{UL}_r} \hspace*{-1mm} \le \hspace*{-1mm}  P_{\mathrm{max}_r}^{\mathrm{UL}} s_{m,n,r,t}^{[i]}, \quad
\mbox{C21: } \hspace*{-1mm} \tilde{P}_{m,n,r,t}^{[i]\mathrm{UL}_r} \hspace*{-1mm} \le \hspace*{-1mm} P_r^{[i]}, \quad
\mbox{C22: } \hspace*{-1mm} \tilde{P}_{m,n,r,t}^{[i]\mathrm{UL}_r} \hspace*{-1mm} \ge \hspace*{-1mm} 0.
\end{eqnarray}
Besides, in order to handle the non-convex integer constraint C9 in problem \eqref{eqv-pro}, we rewrite constraint C9 in equivalent form as: \vspace*{-2mm}
\begin{eqnarray}
&&\hspace*{-16mm}\text{C9}\mbox{a: } \overset{{N_{\mathrm{F}}}}{\underset{i=1}{\sum}} \overset{K}{\underset{m=1}{\sum}} \overset{K}{\underset{n=1}{\sum}} \overset{J}{\underset{r=1}{\sum}} \overset{J}{\underset{t=1}{\sum}} s_{m,n,r,t}^{[i]} - (s_{m,n,r,t}^{[i]})^2 \le 0 \,\,\,  \text{and} \,\,\, \text{C9}\mbox{b: } 0 \le s_{m,n,r,t}^{[i]} \le 1,\,\, \forall i,m,n,r,t,
\end{eqnarray}
i.e., the optimization variables $s_{m,n,r,t}^{[i]}$ are relaxed to a continuous interval between zero and one. However, constraint $\mbox{C9a}$ is a reverse convex function \cite{ng2015power,dinh2010local} which is non-convex.
To resolve this issue, we reformulate the problem in \eqref{eqv-pro} as \vspace*{-1mm}
\begin{eqnarray} \label{penalty-pro}
\hspace*{-5mm}&&\hspace*{-8mm}\underset{\tilde{\mathbf{W}},\tilde{\mathbf{p}},\mathbf{Z},\mathbf{s},\mathbf{M},\bm{\varrho}, \bm{\alpha}, \bm{\beta}}{\mino}  \hspace*{-0.5mm} \sum_{i=1}^{{N_{\mathrm{F}}}} \hspace*{-0mm} \sum_{m=1}^{K} \hspace*{-0.8mm} \sum_{n=1}^{K} \hspace*{-0.8mm} \sum_{r=1}^{J} \sum_{t=1}^{J} -U_{m,n,r,t}^{[i]}(\tilde{\mathbf{W}},\tilde{\mathbf{p}},\mathbf{Z}) \hspace*{-0.5mm} + \hspace*{-0.5mm} \eta \big(s_{m,n,r,t}^{[i]} \hspace*{-0.5mm} - \hspace*{-0.5mm} (s_{m,n,r,t}^{[i]})^2\big) \notag \\
&&\hspace*{-8mm} \mathrm{s.t.}\,\,\mbox{C1--C4},\overline{\text{C5}}, \overline{\text{C6}}\mbox{a}, \overline{\text{C6}}\mbox{b}, \mbox{C7},\mbox{C8},\text{C9}\mbox{b},\mbox{C10--C22},
\end{eqnarray}
where $\eta \gg 1$ acts as a penalty factor for penalizing the objective function for any $s_{m,n,r,t}^{[i]}$ that is not equal to $0$ or $1$. According to \cite{ng2015power,sun2016optimalJournal}, \eqref{penalty-pro} and \eqref{eqv-pro} are equivalent for $\eta \gg 1$. Here, we note that Proposition \ref{prop_robust_variable} and Lemma \ref{lem:gernal-s-proc} are employed for obtaining $\overline{\text{C5}}$, $\overline{\text{C6}}\mbox{a}$, and $\overline{\text{C6}}\mbox{b}$. The non-convexity of problem \eqref{penalty-pro} is also due to constraints $\mbox{C1}$, $\mbox{C3--C4}$, $\mbox{C13}$, and the objective function. However, \eqref{penalty-pro} can be rewritten in form of a standard difference of convex  programming problem \cite{ng2015power} as: \vspace*{-2mm}
\begin{eqnarray}\label{dc-penalty-pro}
\hspace*{-5mm}&&\hspace*{-0mm} \underset{\tilde{\mathbf{W}},\tilde{\mathbf{p}},\mathbf{Z},\mathbf{s}}{\mino}\,\, \,\, T(\tilde{\mathbf{W}},\tilde{\mathbf{p}},\mathbf{Z}) - Q(\tilde{\mathbf{W}},\tilde{\mathbf{p}},\mathbf{Z})+ \eta (R(\mathbf{s})-S(\mathbf{s}))   \\ [-1mm]
\hspace*{-5mm}\mathrm{s.t.}\hspace*{-8mm} &&\hspace*{4mm}
\mbox{C1: }   F_{m,n,r,t}^{[i]}(\tilde{\mathbf{W}},\tilde{\mathbf{p}},\mathbf{Z}) - G_{m,n,r,t}^{[i]}(\tilde{\mathbf{W}},\tilde{\mathbf{p}},\mathbf{Z}) \le 0,  \notag \quad \mbox{C3: }    H_l(\tilde{\mathbf{W}},\tilde{\mathbf{p}},\mathbf{Z}) - M_l(\tilde{\mathbf{W}},\tilde{\mathbf{p}},\mathbf{Z}) \le 0, \notag\\[-1mm]
\hspace*{-5mm}&&\hspace*{4mm} \mbox{C4: }    B_h(\tilde{\mathbf{W}},\tilde{\mathbf{p}},\mathbf{Z}) - D_h(\tilde{\mathbf{W}},\tilde{\mathbf{p}},\mathbf{Z}) \le 0, \notag \quad  \mbox{C2},\overline{\text{C5}}, \overline{\text{C6}}\mbox{a}, \overline{\text{C6}}\mbox{b}, \mbox{C7},\mbox{C8},\text{C9}\mbox{b},\mbox{C10--C22},
\end{eqnarray}
where \vspace*{-2mm}
\begin{eqnarray}
\hspace*{-5mm}T(\tilde{\mathbf{W}},\tilde{\mathbf{p}},\mathbf{Z}) \hspace*{-3mm}&=&\hspace*{-3mm} \sum_{d=1}^{4D} \hspace*{-0.5mm} - \log_2(f_{d}(\tilde{\mathbf{W}},\tilde{\mathbf{p}},\mathbf{Z}))^{\chi_d},
\quad
Q(\tilde{\mathbf{W}},\tilde{\mathbf{p}},\mathbf{Z}) \hspace*{-1mm}=\hspace*{-1mm} \sum_{d=1}^{4D} \hspace*{-0.5mm} - \log_2(g_{d}(\tilde{\mathbf{W}},\tilde{\mathbf{p}},\mathbf{Z}))^{\chi_d},
\end{eqnarray}
\begin{eqnarray}
\hspace*{-6mm}R(\mathbf{s}) \hspace*{-3mm}&=&\hspace*{-3mm} \overset{{N_{\mathrm{F}}}}{\underset{i=1}{\sum}} \overset{K}{\underset{m=1}{\sum}} \overset{K}{\underset{n=1}{\sum}} \overset{J}{\underset{r=1}{\sum}} \overset{J}{\underset{t=1}{\sum}} s_{m,n,r,t}^{[i]},  \quad
S(\mathbf{s}) \hspace*{-1mm}=\hspace*{-1mm} \overset{{N_{\mathrm{F}}}}{\underset{i=1}{\sum}} \overset{K}{\underset{m=1}{\sum}} \overset{K}{\underset{n=1}{\sum}}
\overset{J}{\underset{r=1}{\sum}} \overset{J}{\underset{t=1}{\sum}} (s_{m,n,r,t}^{[i]})^2, \\
\hspace*{-6mm}F_{m,n,r,t}^{[i]}(\tilde{\mathbf{W}},\tilde{\mathbf{p}},\mathbf{Z}) \hspace*{-3mm}&=&\hspace*{-3mm} \log_2 \hspace*{-0.5mm} \Big(\hspace*{-0.5mm} \Tr \hspace*{-0.5mm} \big(\mathbf{H}_{m}^{[i]} (\tilde{\mathbf{W}}_{m,n,r,t}^{[i]\mathrm{DL}_m} \hspace*{-0.5mm} + \hspace*{-0.5mm} \tilde{\mathbf{W}}_{m,n,r,t}^{[i]\mathrm{DL}_n} \hspace*{-0.5mm} + \hspace*{-0.5mm} \mathbf{Z}^{[i]}) \big) \hspace*{-0.5mm} + \hspace*{-0.5mm} \tilde{\mathbf{p}}_{m,n,r,t}^{[i]} \mathbf{f}_{r,t,m}^{[i]}   \hspace*{-0.5mm} +  \hspace*{-0.5mm} \sigma_{m}^2 \Big) \notag \\
\hspace*{-3mm}&+&\hspace*{-3mm} \log_2 \hspace*{-0.5mm} \Big( \hspace*{-0.5mm} \Tr \big(\mathbf{H}_{n}^{[i]} (\tilde{\mathbf{W}}_{m,n,r,t}^{[i]\mathrm{DL}_n} + \mathbf{Z}^{[i]}) \big) \hspace*{-1mm} + \hspace*{-1mm} \tilde{\mathbf{p}}_{m,n,r,t}^{[i]} \mathbf{f}_{r,t,n}^{[i]}   \hspace*{-1mm} + \hspace*{-0.5mm} \sigma_{n}^2 \Big), \\
\hspace*{-4mm}G_{m,n,r,t}^{[i]}(\tilde{\mathbf{W}},\tilde{\mathbf{p}},\mathbf{Z}) \hspace*{-3mm}&=&\hspace*{-3mm}  \log_2 \hspace*{-0.5mm}\Big( \hspace*{-0.5mm} \Tr \hspace*{-0.5mm} \big(\mathbf{H}_{m}^{[i]}  (\tilde{\mathbf{W}}_{m,n,r,t}^{[i]\mathrm{DL}_n} \hspace*{-0.5mm} + \hspace*{-0.5mm} \mathbf{Z}^{[i]})\big) \hspace*{-0.5mm} + \hspace*{-0.5mm} \tilde{\mathbf{p}}_{m,n,r,t}^{[i]} \mathbf{f}_{r,t,m}^{[i]}   \hspace*{-0.5mm} +  \hspace*{-0.5mm} \sigma_{m}^2 \Big) \notag\\
\hspace*{-3mm}&+&\hspace*{-3mm}  \log_2 \hspace*{-0.5mm}\Big( \hspace*{-0.5mm} \Tr \hspace*{-0.5mm} \big(\mathbf{H}_{n}^{[i]} (\tilde{\mathbf{W}}_{m,n,r,t}^{[i]\mathrm{DL}_m} \hspace*{-0.5mm} + \hspace*{-0.5mm} \tilde{\mathbf{W}}_{m,n,r,t}^{[i]\mathrm{DL}_n} \hspace*{-0.5mm} + \hspace*{-0.5mm} \mathbf{Z}^{[i]}) \big) \hspace*{-0.5mm} + \hspace*{-0.5mm} \tilde{\mathbf{p}}_{m,n,r,t}^{[i]} \mathbf{f}_{r,t,n}^{[i]}   \hspace*{-0.5mm} +  \hspace*{-0.5mm} \sigma_{n}^2\Big).
\end{eqnarray}
The definitions of $H_l(\tilde{\mathbf{W}}\hspace*{-0.5mm}, \hspace*{-0.5mm}\tilde{\mathbf{p}}, \hspace*{-0.5mm}\mathbf{Z}) $ and $M_l(\tilde{\mathbf{W}}\hspace*{-0.5mm}, \hspace*{-0.5mm}\tilde{\mathbf{p}}, \hspace*{-0.5mm}\mathbf{Z})$ in C3, and the definitions of $B_{h}(\tilde{\mathbf{W}}\hspace*{-0.5mm}, \hspace*{-0.5mm}\tilde{\mathbf{p}}, \hspace*{-0.5mm}\mathbf{Z})$ and $D_{h}(\tilde{\mathbf{W}}\hspace*{-0.5mm}, \hspace*{-0.5mm}\tilde{\mathbf{p}}, \hspace*{-0.5mm}\mathbf{Z})$ in C4 are similar to those of $T(\tilde{\mathbf{W}},\tilde{\mathbf{p}},\mathbf{Z})$ and $Q(\tilde{\mathbf{W}},\tilde{\mathbf{p}},\mathbf{Z})$. In particular, constraints C3 and C4 are written as differences of logarithmic functions.  We note that the problems in \eqref{eqv-pro} and \eqref{dc-penalty-pro} are equivalent in the sense that they have the same optimal solution.
We can obtain a locally optimal solution of \eqref{dc-penalty-pro} by applying sequential convex approximation \cite{dinh2010local}.
However, the traditional sequential convex approximation approach in \cite{dinh2010local} needs a favorable initialization for achieving good performance, which results in high computational complexity. Thus, we propose an algorithm based on the penalized sequential convex approximation \cite{Yang_penalizedDC16} which can start from an arbitrary initial point.
In particular, the following inequality holds for any point $\tilde{\mathbf{W}}^{(k)}, \vspace*{-1mm} \tilde{\mathbf{p}}^{(k)}, \vspace*{-1mm} \mathbf{Z}^{(k)}$, and $\mathbf{s}^{(k)}$: \vspace*{-2mm}
\begin{eqnarray}\label{ineq1}
\hspace*{-1mm}Q(\tilde{\mathbf{W}},\tilde{\mathbf{p}},\mathbf{Z})
\hspace*{-3mm} &\ge& \hspace*{-3mm} Q(\tilde{\mathbf{W}}^{(k)}, \tilde{\mathbf{p}}^{(k)}, \mathbf{Z}^{(k)}) \hspace*{-1mm} + \hspace*{-1mm} \Tr(\nabla_{\tilde{\mathbf{W}}} Q(\tilde{\mathbf{W}}^{(k)}, \tilde{\mathbf{p}}^{(k)}, \mathbf{Z}^{(k)})^{T} (\tilde{\mathbf{W}} \hspace*{-1mm} - \hspace*{-1mm} \tilde{\mathbf{W}}^{(k)}) ) \notag \\[-2mm]
\hspace*{-2mm}&+& \hspace*{-3mm}  \Tr(\nabla_{\mathbf{Z}} Q (\tilde{\mathbf{W}}^{(k)}, \hspace*{-0.5mm} \tilde{\mathbf{p}}^{(k)}, \hspace*{-0.5mm} \mathbf{Z}^{(k)})^{T}(\mathbf{Z} \hspace*{-1mm} - \hspace*{-1mm}\mathbf{Z}^{(k)}) ) \hspace*{-1mm} +  \hspace*{-1mm} \Tr(\nabla_{\tilde{\mathbf{p}}} Q(\tilde{\mathbf{W}}^{(k)}, \hspace*{-0.5mm} \tilde{\mathbf{p}}^{(k)}, \hspace*{-0.5mm} \mathbf{Z}^{(k)})^{T} (\tilde{\mathbf{p}} \hspace*{-1mm} - \hspace*{-1mm} \tilde{\mathbf{p}}^{(k)}) ) \notag \\[-2mm]
\hspace*{-2mm} &\triangleq& \hspace*{-3mm} \overline{Q} (\bm{\Psi}, \bm{\Psi}^{(k)}),
\end{eqnarray}
where $\bm{\Psi}$ and $\bm{\Psi}^{(k)}$ are defined as the collection of variables $\{\tilde{\mathbf{W}},\tilde{\mathbf{p}},\mathbf{Z}\}$ and $\{\tilde{\mathbf{W}}^{(k)}, \tilde{\mathbf{p}}^{(k)}, \mathbf{Z}^{(k)}\}$, respectively, and the right hand side of \eqref{ineq1} is an affine function and constitutes a global underestimation of $Q(\bm{\Psi})$. Similarly, we denote $\overline{S}(\mathbf{s},\mathbf{s}^{(k)})$, $\overline{G}_{m,n,r,t}^{[i]} (\bm{\Psi}, \bm{\Psi}^{(k)})$, $\overline{M}_{l} ( \bm{\Psi}, \bm{\Psi}^{(k)} )$, and $\overline{D}_{h} ( \bm{\Psi}, \bm{\Psi}^{(k)} )$ as the global underestimations of $S(\mathbf{s})$, $G_{m,n,r,t}^{[i]}(\bm{\Psi})$, $M_{l}(\bm{\Psi})$, and $D_{h}(\bm{\Psi})$, respectively, which can be obtained in a similar manner as $\overline{Q} (\bm{\Psi}, \bm{\Psi}^{(k)})$.
Therefore, for any given $\bm{\Psi}^{(k)}$ and $\mathbf{s}^{(k)}$, we can find a lower bound of \eqref{dc-penalty-pro} by solving the following optimization problem: \vspace*{-1mm}
\begin{eqnarray}\label{dc}
\hspace*{-1mm}&&\hspace*{-8mm}\underset{\bm{\Psi},\mathbf{s}}{\mino}\,\, T(\hspace*{-0.3mm}\bm{\Psi}\hspace*{-0.3mm}) \hspace*{-0.6mm} - \hspace*{-0.6mm} \overline{Q} (\hspace*{-0.3mm}\bm{\Psi},\hspace*{-0.3mm} \bm{\Psi}^{(k)}\hspace*{-0.3mm})\hspace*{-0.6mm}  + \hspace*{-0.6mm}\eta\big(\hspace*{-0.3mm} R( \hspace*{-0.3mm} \mathbf{s} \hspace*{-0.3mm}) \hspace*{-0.6mm} - \hspace*{-0.6mm} \overline{S} (\hspace*{-0.3mm} \mathbf{s},\mathbf{s}^{(k)}\hspace*{-0.3mm} )\big) \hspace*{-0.5mm}   + \hspace*{-0.5mm} \varpi^{(k)}\hspace*{-0.3mm}\Big(\hspace*{-0.3mm}\sum_{i=1}^{{N_{\mathrm{F}}}} \hspace*{-1mm} \sum_{m=1}^{K}\hspace*{-1mm} \sum_{n=1}^{K}\hspace*{-1mm} \sum_{r=1}^{J} \hspace*{-1mm}\sum_{t=1}^{J} \hspace*{-1mm} a_{m,n,r,t}^{[i]} \hspace*{-1mm} + \hspace*{-1mm} \sum_{l=1}^{K} b_l \hspace*{-1mm} + \hspace*{-1mm} \sum_{h=1}^{J} c_h \Big) \notag \\[-1mm]
\hspace*{-1mm}\mathrm{s.t.}\hspace*{-3mm}
&& \hspace*{-3mm} \mbox{C1: } \hspace*{-1mm}  F_{m,n,r,t}^{[i]} \hspace*{-0.5mm} (\bm{\Psi}) \hspace*{-1mm} - \overline{G}_{m,n,r,t}^{[i]} (\bm{\Psi}, \bm{\Psi}^{(k)}) \le a_{m,n,r,t}^{[i]}, \quad \mbox{C3: } H_l(\bm{\Psi}) - \overline{M}_{l} (\bm{\Psi}, \bm{\Psi}^{(k)}) \le b_l, \,\, \forall l, \notag \\[-2mm]
\hspace*{-1mm}&&\hspace*{-3mm} \mbox{C4: } B_h(\bm{\Psi}) - \overline{D}_{h} (\bm{\Psi}, \bm{\Psi}^{(k)}) \le c_h, \,\, \forall h, \hspace*{16mm} \mbox{C2},\overline{\text{C5}}, \overline{\text{C6}}\mbox{a}, \overline{\text{C6}}\mbox{b}, \mbox{C7},\mbox{C8},\text{C9}\mbox{b},\mbox{C10--C22},
\end{eqnarray}
where $a_{m,n,r,t}^{[i]}$, $b_l$, and $c_h$ are auxiliary variables. Besides, $\varpi^{(k)}$ is a penalty term for the $k$-th iteration. For any non-zero $a_{m,n,r,t}^{[i]}$, $b_l$, and $c_h$, the penalty term $\varpi^{(k)}$ penalizes the violation of constraints C1, C3, and C4, respectively.
In problem \eqref{dc}, the only remaining non-convex constraint is the rank-one constraint C13. Similar to the optimal algorithm, we apply SDP relaxation to \eqref{dc} by removing constraint C13. Then, we employ an iterative algorithm to tighten the obtained lower bound as summarized in \textbf{Algorithm 3}.
In each iteration, the convex problem in \eqref{dc} can be solved efficiently by standard convex program solvers such as CVX \cite{website:CVX}.
By solving the convex lower bound problem in \eqref{dc}, the proposed iterative scheme generates a sequence of solutions $\bm{\Psi}^{(k+1)}$ and $\mathbf{s}^{(k+1)}$ successively.
The proposed suboptimal iterative algorithm converges to a locally optimal solution of \eqref{dc-penalty-pro} with a polynomial time computational complexity \cite{dinh2010local,Yang_penalizedDC16}. Besides, we note that for $P_{\mathrm{max}}^{\mathrm{DL}} > 0$, the obtained beamforming matrix in \textbf{Algorithm 3} is a rank-one matrix, which can be proved in a similar manner as was done for the proposed optimal solution in Appendix B. The proof is omitted here because of space constraints.

\begin{table} \vspace*{-8mm}
\begin{algorithm} [H]                    
\caption{Penalized Sequential Convex Approximation}          
\label{alg3}                           
\begin{algorithmic} [1]
\small          
\STATE Initialize the  maximum number of iterations $N_{\mathrm{max}}$, penalty factor $\eta \hspace*{-0.8mm} \gg \hspace*{-0.8mm} 1$, $\epsilon_1 > 0$,  $\epsilon_2 > 0$, iteration index $k=1$, and initial point $\bm{\Psi}^{(1)}=\{\tilde{\mathbf{W}}^{(1)},\tilde{\mathbf{p}}^{(1)},\mathbf{Z}^{(1)}\}$, $\mathbf{s}^{(1)}$, and $\varpi^{(1)}$

\REPEAT

\STATE For a given $\bm{\Psi}^{(k)}$, $\mathbf{s}^{(k)}$, and $\varpi^{(k)}$, solve \eqref{dc} and save the obtained solutions $\bm{\Psi}$ and $\mathbf{s}$ as the intermediate resource allocation policy

\STATE Update $\varpi$: Define $\varphi_{\mathrm{1}}^{(k)}$, $\varphi_{\mathrm{2}}^{(k)}$, and $\varphi_{\mathrm{3}}^{(k)}$ as the Lagrange multipliers for constraints C1, C3, and C4 in \eqref{dc} in  iteration $k$ and let $e^{(k)} \triangleq \min \Big\{ \norm{\bm{\Psi}- \bm{\Psi}^{(k)}}^{-1}, \sum_{j=1}^{3}\varphi_{j}^{(k)} + \epsilon_1 \Big\} $. $\varpi$ is updated as follows, if $\varpi^{(k)} \ge e^{(k)}$, $\varpi^{(k+1)}=\varpi^{(k)}$; if $\varpi^{(k)} < e^{(k)}$, $\varpi^{(k+1)}=\varpi^{(k)} + \epsilon_2$

\STATE Set $k=k+1$ and $\bm{\Psi}^{(k)}=\bm{\Psi}$ , $\mathbf{s}^{(k)}=\mathbf{s}$

\UNTIL convergence or $k=N_{\mathrm{max}}$

\STATE $\bm{\Psi}^{*}=\bm{\Psi}^{(k)}$ and $\mathbf{s}^{*}=\mathbf{s}^{(k)}$

\end{algorithmic}
\end{algorithm}\vspace*{-13mm}
\end{table}

\vspace*{-1mm}
\section{Simulation Results}

In this section, the performance of the proposed resource allocation schemes is evaluated via simulations. The simulation parameters are chosen as in Table \ref{tab:parameters}, unless specified otherwise.
We consider a single cell where the FD BS is located at the center of the cell.
The UL, DL, and idle users are distributed randomly and uniformly between the maximum service distance and the reference distance.
To provide fairness in resource allocation, especially for the cell edge users, who suffer from poor channel conditions, the weights of the users are set equal to the normalized distance between the users and the FD BS, i.e., $w_m\hspace*{-1mm}=\hspace*{-1mm}\frac{l_m^{\mathrm{DL}}}{\underset{i\in\{1,\ldots,K\}}{\max}\{l_i^{\mathrm{DL}}\}}$ and $\mu_r\hspace*{-1mm}=\hspace*{-1mm}\frac{l_r^{\mathrm{UL}}}{\underset{i\in\{1,\ldots,J\}}{\max}\{l_i^{\mathrm{UL}}\}}$, where $l_m^{\mathrm{DL}}$ and $l_r^{\mathrm{UL}}$ denote the distance from the FD BS to DL user $m$ and UL user $r$, respectively.
The small-scale fading channels between the FD BS and the users and between the users and the equivalent eavesdropper are modeled as independent and identically  Rayleigh distributed.
We model the multipath fading SI channel on each subcarrier as independent and identically Rician distributed with Rician factor $5$ dB.
Besides, we define the normalized maximum estimation error of the eavesdropping channels between the FD BS and the eavesdropper and between the UL users and the eavesdropper as $\frac{(\varepsilon_{\mathrm{DL}}^{[i]})^2}{\norm{\mathbf{L}^{[i]}}^2_{F}} \hspace*{-1mm}=\hspace*{-1mm}\kappa_{\mathrm{DL}^{[i]}}^2$ and
$\frac{(\hspace*{-0.5mm}\varepsilon_{\mathrm{UL}_{r}}^{[i]}\hspace*{-0.5mm})^2}{\norm{\mathbf{e}_{r}^{[i]}}^2}\hspace*{-1mm}=\hspace*{-1mm}\kappa_{\mathrm{UL}_{r}^{[i]}}^2$, respectively.
The simulation results shown in this section were averaged over $1000$ different path losses and multipath fading realizations.

\begin{table}[t]\vspace*{-2mm}\caption{System parameters used in simulations.}\vspace*{-2mm}\label{tab:parameters} 
\newcommand{\tabincell}[2]{\begin{tabular}{@{}#1@{}}#2\end{tabular}}
\centering
\begin{tabular}{|l|l|}\hline
\hspace*{-1mm}Carrier center frequency and system bandwidth & $2.5$ GHz and  $5$ MHz \\
\hline
\hspace*{-1mm}Number of subcarriers, ${N_{\mathrm{F}}}$, and bandwidth of each subcarrier & $32$ and  $78$ kHz \\
\hline
\hspace*{-1mm}Maximum service distance and reference distance& $400$ meters and  $15$ meters \\
\hline
\hspace*{-1mm}Path loss exponent, SI cancellation constant, $\rho$, and maximum  estimation error $\kappa_{\mathrm{est}}^2$ &  \mbox{$3.6$},  \mbox{$-90$ dB}, and  $6\%$   \\
\hline
\hspace*{-1mm}DL user noise power and UL BS noise power, $\sigma_{\mathrm{z}_{{\mathrm{DL}_m}}}^2$ and $\sigma_{\mathrm{z}_{{\mathrm{BS}}}}^2$ &  \mbox{$-125$ dBm} and \mbox{$-125$ dBm}  \\
\hline
\hspace*{-1mm}Maximum transmit power for UL users and FD BS, i.e., $P_{\mathrm{max}_r}^{\mathrm{UL}}$ and  $P_{\mathrm{max}}^{\mathrm{DL}}$ &  \mbox{$22$ dBm}  and \mbox{$45$ dBm} \\
\hline
\hspace*{-1mm}Number of DL and UL users, $K=J$, and number of potential eavesdroppers, $M$ & $8$ and $2$  \\
\hline
\hspace*{-1mm}Number of antennas at the FD BS, $N_{\mathrm{T}}$, and FD BS antenna gain  &  $6$ and  \mbox{$10$ dBi}  \\
\hline
\hspace*{-1mm}Normalized maximum estimation error for eavesdropping channels, $\kappa_{\mathrm{DL}^{[i]}}^2 =\kappa_{\mathrm{UL}_{r}^{[i]}}^2= \kappa_{\mathrm{est}}^2$ &  $4\%$  \\
\hline
\hspace*{-1mm}Minimum QoS requirements for DL and UL users, $R_{\mathrm{req}_{l}}^{\mathrm{DL}}=R_{\mathrm{req}_{h}}^{\mathrm{UL}}=R_{\mathrm{req}}$  &  $1$ bits/s/Hz  \\
\hline
\hspace*{-1mm}Maximum tolerable data rate at potential eavesdroppers, $R_{\mathrm{tol}_{l}}^{\mathrm{DL}[i]}=R_{\mathrm{tol}_{h}}^{\mathrm{UL}[i]}=R_{\mathrm{tol}}$&  \mbox{$0.001$ bit/s/Hz}  \\
\hline
\hspace*{-1mm}Error tolerances $\epsilon$ and $\delta$ for {\bf{Algorithm 1}} and {\bf{2}}, and $\epsilon_1$ and $\epsilon_2$ for {\bf{Algorithm 3}} &  $0.01$  \\
\hline
\hspace*{-1mm}Penalty term $\eta$ for the proposed suboptimal algorithm &  $10 \log_2(1+\frac{P_{\mathrm{max}}^{\mathrm{DL}}}{\sigma_{\mathrm{z}_m}^2})$  \\
\hline
\end{tabular}
\vspace*{-7mm}
\end{table}

We consider three baseline schemes for comparison.
For baseline scheme 1, we consider an FD MISO MC-NOMA system which employs maximum ratio transmission beamforming (MRT-BF) for DL transmission. i.e., the direction of beamforming vector $\mathbf{w}_m^{[i]}$ is aligned with that of channel vector $\mathbf{h}_m^{[i]}$. Then, we jointly optimize $s_{m,n,r,t}^{[i]}$, $P_r^{[i]}$, $\mathbf{Z}^{[i]}$, and the power allocated to $\mathbf{w}_m^{[i]}$.
For baseline scheme 2, we consider a traditional FD MISO MC-OMA system where the FD BS and the DL users cannot perform SIC for cancelling MUI. For a fair comparison, we assume that at most two UL users and two DL users can be scheduled on each subcarrier. The  resource allocation policy for baseline scheme 2 is obtained by an exhaustive search.
In particular, for all possible subcarrier allocation policies, we determine the joint precoding and power allocation policy with the suboptimal scheme proposed in \cite{sun17WSAsecureFD}.
Then, we choose that subcarrier allocation policy and the corresponding precoding and power allocation policy which maximizes the weighted system throughput.
For baseline scheme 3, we consider an FD MISO MC-NOMA system where the user pair on each subcarrier is randomly selected and we jointly optimize $\mathbf{W}$, $\mathbf{p}$, and $\mathbf{Z}$ subject to constraints C1--C11 as in \eqref{pro}.

\begin{figure}[t]
 \centering\vspace*{-6mm}
 \begin{minipage}[b]{0.45\linewidth} \hspace*{-1cm}
\includegraphics[width=3.55in]{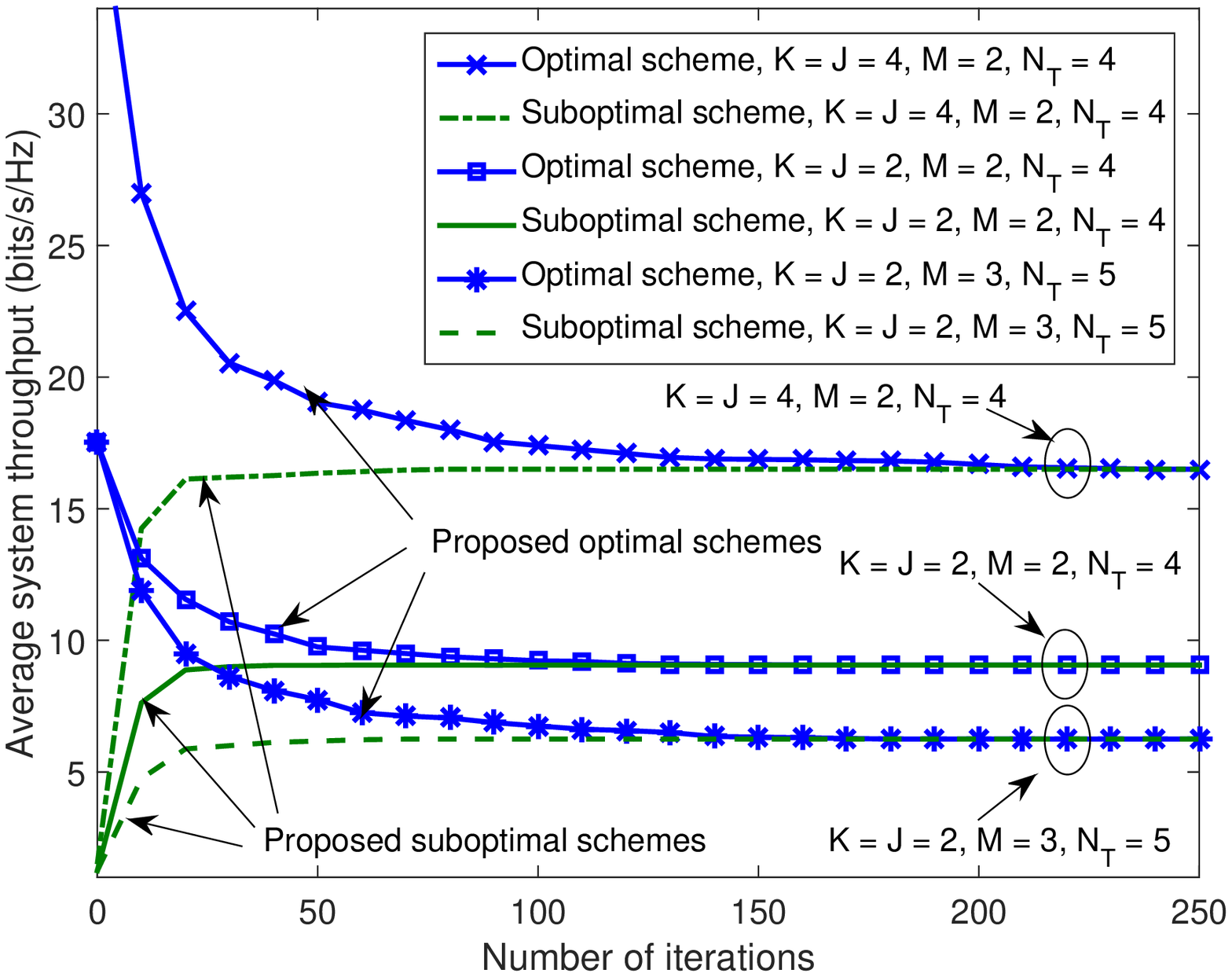}\vspace*{-5mm}
\caption{Convergence of the proposed optimal and suboptimal algorithms for different values of $K$, $J$, $M$, and $N_{\mathrm{T}}$.}
\label{fig:secrecy_vs_iterations}
 \end{minipage}\hspace*{8mm}
 \begin{minipage}[b]{0.45\linewidth} \hspace*{-1cm}
\includegraphics[width=3.55in]{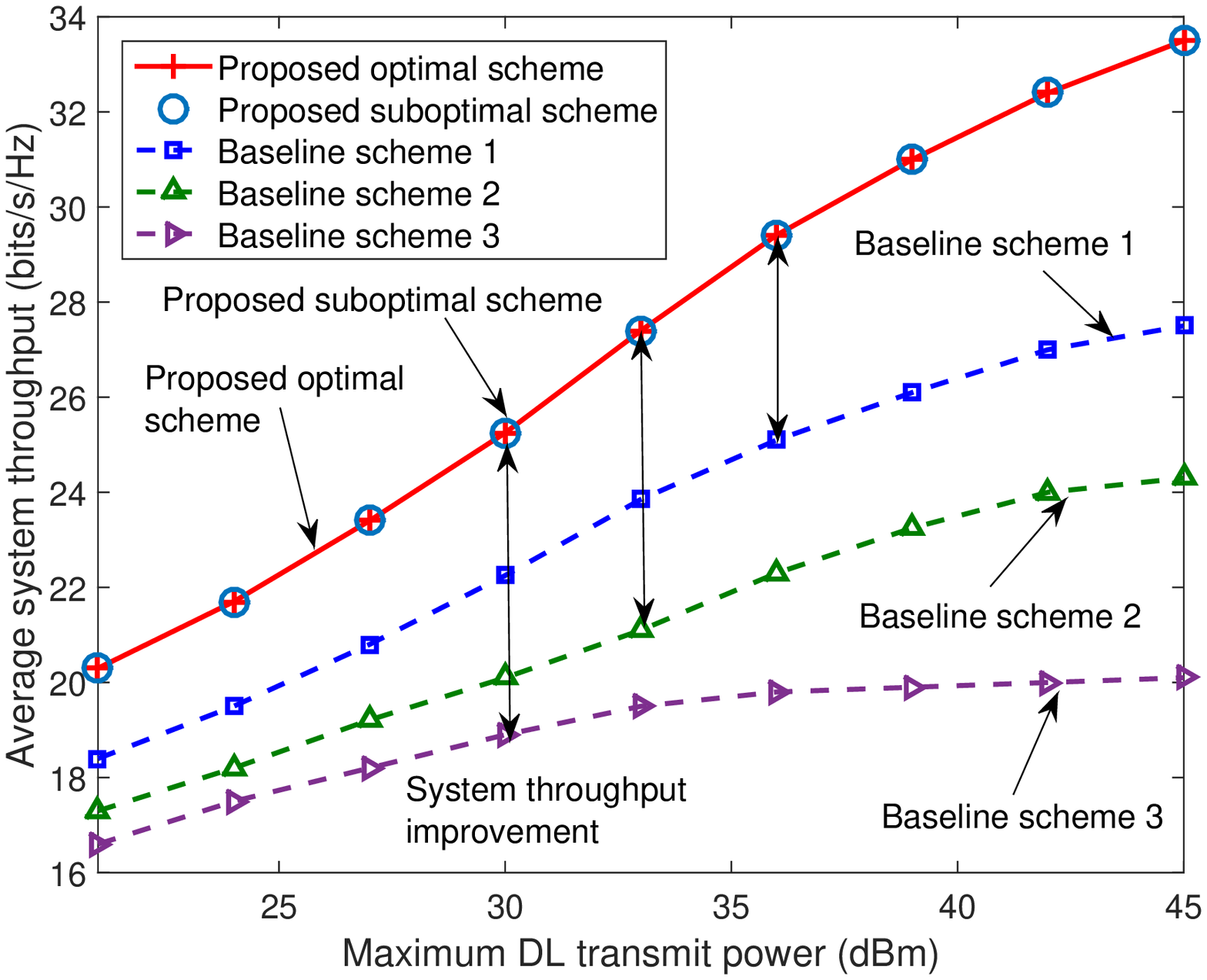}\vspace*{-5mm}
\caption{Average system throughput (bits/s/Hz) versus the maximum DL transmit power at the FD BS, $P_{\mathrm{max}}^{\mathrm{DL}}$ (dBm). }
\label{fig:secrecy_vs_power}
 \end{minipage}\vspace*{-6mm}
\end{figure}

\vspace*{-4mm}
\subsection{Convergence of Proposed Optimal and Suboptimal Schemes}


In Figure \ref{fig:secrecy_vs_iterations}, we investigate the convergence of the proposed optimal and suboptimal algorithms for different numbers of antennas, $N_{\mathrm{T}}$, DL and UL users, $K+J$, and potential eavesdroppers, $M$. As can be observed from Figure \ref{fig:secrecy_vs_iterations}, the proposed optimal and suboptimal schemes converge to the optimal solution for all considered values of $K$, $J$, $M$, and $N_{\mathrm{T}}$. In particular, for $K = J = 2$, $M=2$, and $N_{\mathrm{T}}=4$, the proposed optimal and suboptimal algorithms converge to the optimal solution in less than $130$ and $30$ iterations, respectively. For the case with more potential eavesdroppers and a larger number of antennas, i.e., $K = J = 2$, $M=3$, and $N_{\mathrm{T}}=5$, the proposed optimal algorithm needs on average $20$ iterations more to converge. For the case with more DL and UL users, i.e., $K = J = 4$, $M=2$, and $N_{\mathrm{T}}=4$, the proposed optimal algorithm needs considerably more iterations to converge since the search space for the optimal solution increases exponentially with the number of users. We also note that the number of computations required in each iteration increases with the number of DL and UL users. In contrast, as can be observed from Figure \ref{fig:secrecy_vs_iterations}, the number of iterations required for the proposed suboptimal scheme to converge is less sensitive to the numbers of users, potential eavesdroppers, and transmit antennas at the FD BS.

\vspace*{-3mm}
\subsection{Average System Throughput versus Maximum Transmit Power}
In Figure \ref{fig:secrecy_vs_power}, we investigate the average system throughput versus the maximum DL transmit power at the FD BS, $P_{\mathrm{max}}^{\mathrm{DL}}$. As can be observed from Figure \ref{fig:secrecy_vs_power}, the average system throughput of the optimal and suboptimal resource allocation schemes increases monotonically with the maximum DL transmit power $P_{\mathrm{max}}^{\mathrm{DL}}$ since additional available transmit power at the FD BS is allocated optimally to improve the received SINR at the DL users while limiting the received SINR at the potential eavesdroppers.
Moreover, the rate at which the average system throughput increases diminishes when $P_{\mathrm{max}}^{\mathrm{DL}}$ exceeds $36$ dBm.
The reason behind this is twofold.
First, as the transmit power of the DL information signals increases, the DL transmission also creates more information leakage. Thus, more power has to be allocated to the AN to guarantee the maximum tolerable information leakage requirements, which impedes the improvement of the system throughput.
Second, a higher DL transmit power causes more severe SI, which degrades the quality of the received UL signals, and thus impairs the UL throughput.
Therefore, the reduction in the UL throughput partially neutralizes the improvement in DL throughput which slows down the rate at which the overall system throughput increases.
In addition, from Figure \ref{fig:secrecy_vs_power}, we also observe that the performance of the proposed suboptimal algorithm closely approaches that of the proposed optimal resource allocation scheme.
On the other hand, all considered baseline schemes yield a substantially lower average system throughput compared to the proposed optimal and suboptimal schemes. This is due to the use of suboptimal beamforming, power allocation, and subcarrier allocation policies for baseline schemes 1, 2, and 3, respectively. In particular, baseline scheme 1 employs fixed data beamforming which cannot optimally suppress the SI and MUI. Also, it cannot optimally reduce information leakage, and hence, more power has to be allocated to the AN to degrade the eavesdropping channels. Baseline scheme 2 employs conventional OMA which underutilizes the spectral resources compared to the proposed NOMA-based schemes. Besides, since baseline scheme 2 performs spatially orthogonal beamforming to mitigate MUI, less DoF are available for accommodating the AN to improve communication security.
Baseline scheme 3 employs random subcarrier allocation which cannot exploit multiuser diversity for improving the system throughput.
For the case of $P_{\mathrm{max}}^{\mathrm{DL}} = 45$ dBm, the proposed optimal and suboptimal schemes achieve roughly a $22\%$, $42\%$, and $70\%$ higher average system throughput than baseline schemes 1, 2, and 3, respectively.

\vspace*{-3mm}
\subsection{Average System Secrecy Throughput versus Number of Antennas}
In Figure \ref{fig:secrecy_vs_antennas}, we investigate the average system secrecy throughput versus the number of antennas at the FD BS, $N_{\mathrm{T}}$, for different values of $M$.
As can be observed, the average system secrecy throughput improves as the number of antennas at the FD BS increases. This is due to the fact that the extra DoF offered by additional antennas facilitates a more precise information beamforming and AN injection which leads to higher received SINRs at the DL and UL users and limits the achievable data rate at the equivalent eavesdropper. However, due to the channel hardening effect, the rate at which the system secrecy throughput improves diminishes for large values of $N_{\mathrm{T}}$.
Besides, for a larger value of $M$, the proposed schemes and baseline scheme 1 achieve lower average system secrecy throughputs. This is because the DL and UL transmissions become more vulnerable to eavesdropping when more potential eavesdroppers are in the network. Therefore, more DoF have to be utilized for reducing the achievable data rate at the equivalent eavesdropper which limits the improvement in the system secrecy throughput.
Figure \ref{fig:secrecy_vs_antennas} also shows that the average system secrecy throughput of the proposed schemes grows faster with $N_{\mathrm{T}}$ than those of all baseline schemes. In particular, the average system secrecy throughput of baseline scheme 1 quickly saturates as $N_{\mathrm{T}}$ increase since its fixed beamforming causes severe information leakage.
Besides, since baseline schemes 2 and 3 adopt suboptimal power allocation and subcarrier allocation policies, respectively, they cannot exploit the full benefits of having more antennas available. In addition, all baseline schemes achieve a considerably lower average system secrecy throughput compared to the proposed schemes, even when $N_{\mathrm{T}}$ is relatively large.

\vspace*{-3mm}
\subsection{Average User Secrecy Throughput versus Total Number of Users}
\begin{figure}[t]
 \centering\vspace*{-6mm}
 \begin{minipage}[b]{0.45\linewidth} \hspace*{-1cm}
\includegraphics[width=3.55in]{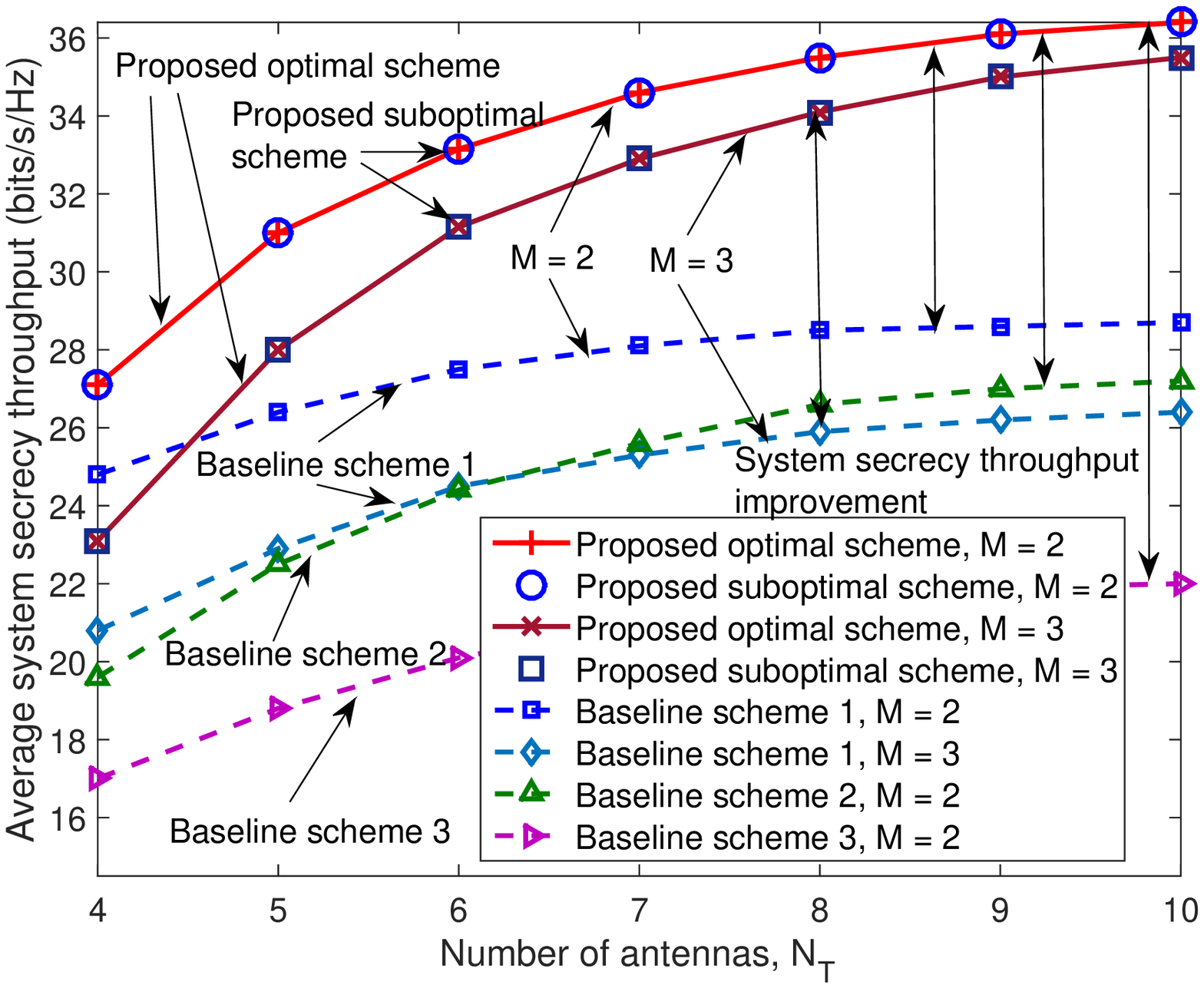}\vspace*{-5mm}
\caption{Average system secrecy throughput (bits/s/Hz) versus the number of antennas at the FD BS, $N_{\mathrm{T}}$, for different values of $M$. }
\label{fig:secrecy_vs_antennas}
 \end{minipage}\hspace*{8mm}
 \begin{minipage}[b]{0.45\linewidth} \hspace*{-1cm}
\includegraphics[width=3.55in]{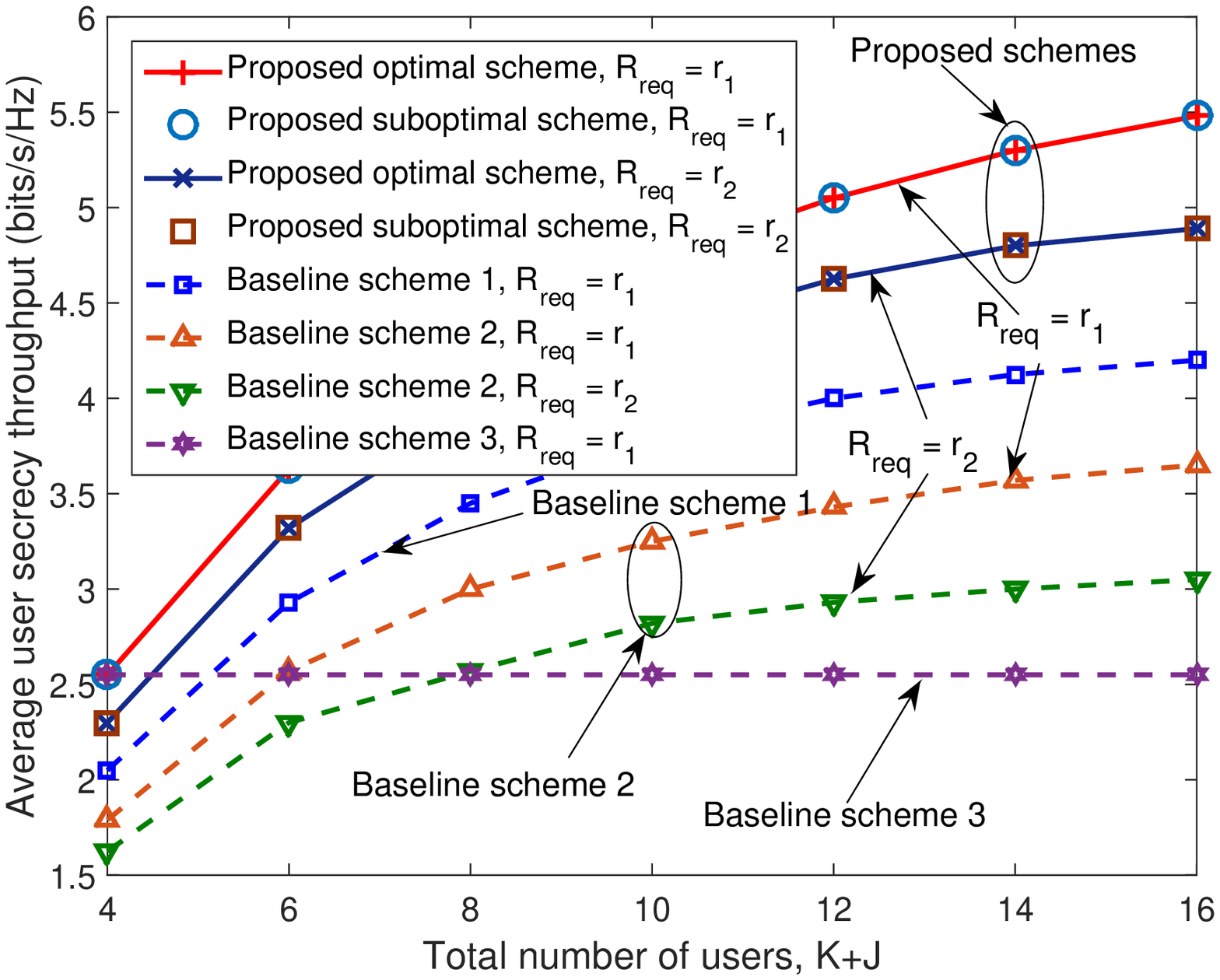}\vspace*{-5mm}
\caption{Average user secrecy throughput (bits/s/Hz) versus the total number of users, $K+J$.}
\label{fig:secrecy_vs_users}
 \end{minipage}\vspace*{-6mm}
\end{figure}

In Figure \ref{fig:secrecy_vs_users}, we investigate the average user secrecy throughput versus the total number of users, i.e., $K+J$, for different minimum QoS requirements, namely $r_1 = 1$ bits/s/Hz and $r_2 = 1.5$ bits/s/Hz. The average user secrecy throughput is calculated as $ \frac{ \sum_{i=1}^{N_{\mathrm{F}}} (\sum_{l=1}^{K} R_{\mathrm{DL}_l}^{[i]\mathrm{Sec}} + \sum_{h=1}^{J} R_{\mathrm{UL}_h}^{[i]\mathrm{Sec}})}{K+J}$.
As can be observed, the average user secrecy throughputs of the proposed optimal/suboptimal schemes and baseline schemes $1$ and $2$ increase with the total number of users $K+J$ due to the ability of these schemes to exploit multiuser diversity. Besides, the proposed suboptimal scheme achieves a similar performance as the proposed optimal scheme, even for relatively large numbers of users. On the other hand, the performance of baseline scheme $3$ does not scale with $K+J$ due to its random subcarrier allocation policy.
Besides, from Figure \ref{fig:secrecy_vs_users}, we observe that the average user secrecy throughput of the proposed schemes grows faster with the number of users than that of baseline schemes $1$ and $2$.
In fact, since baseline scheme 1 adopts fixed beamforming, the information leakage becomes more severe when more users are active. Hence, more power and DoF have to be devoted to AN injection to guarantee communication security which reduces the improvement in the system secrecy throughput.
For baseline scheme 2, in order to limit the MUI, more DoF are employed for generating spatially orthogonal information beamforming vectors compared to the proposed schemes, leaving less DoF for efficiently injecting the AN. However, the proposed schemes exploit the power domain for multiple access which leaves more DoF for user scheduling, beamforming, and AN injection. This leads to a fast improvement in system secrecy throughput as the number of users increases.
Moreover, both the proposed schemes and baseline scheme 2 achieve a lower average user secrecy throughput when more stringent QoS requirements for the DL and UL users are imposed. In fact, for more stringent QoS constraints, the FD BS has to allocate more radio resources to users with poor channel condition to meet the QoS requirements. This reduces the average user secrecy throughput.

\begin{figure}[t]
\centering\vspace*{-6mm}
\begin{minipage}[b]{0.45\linewidth} \hspace*{-1cm}
    \includegraphics[width=3.55in]{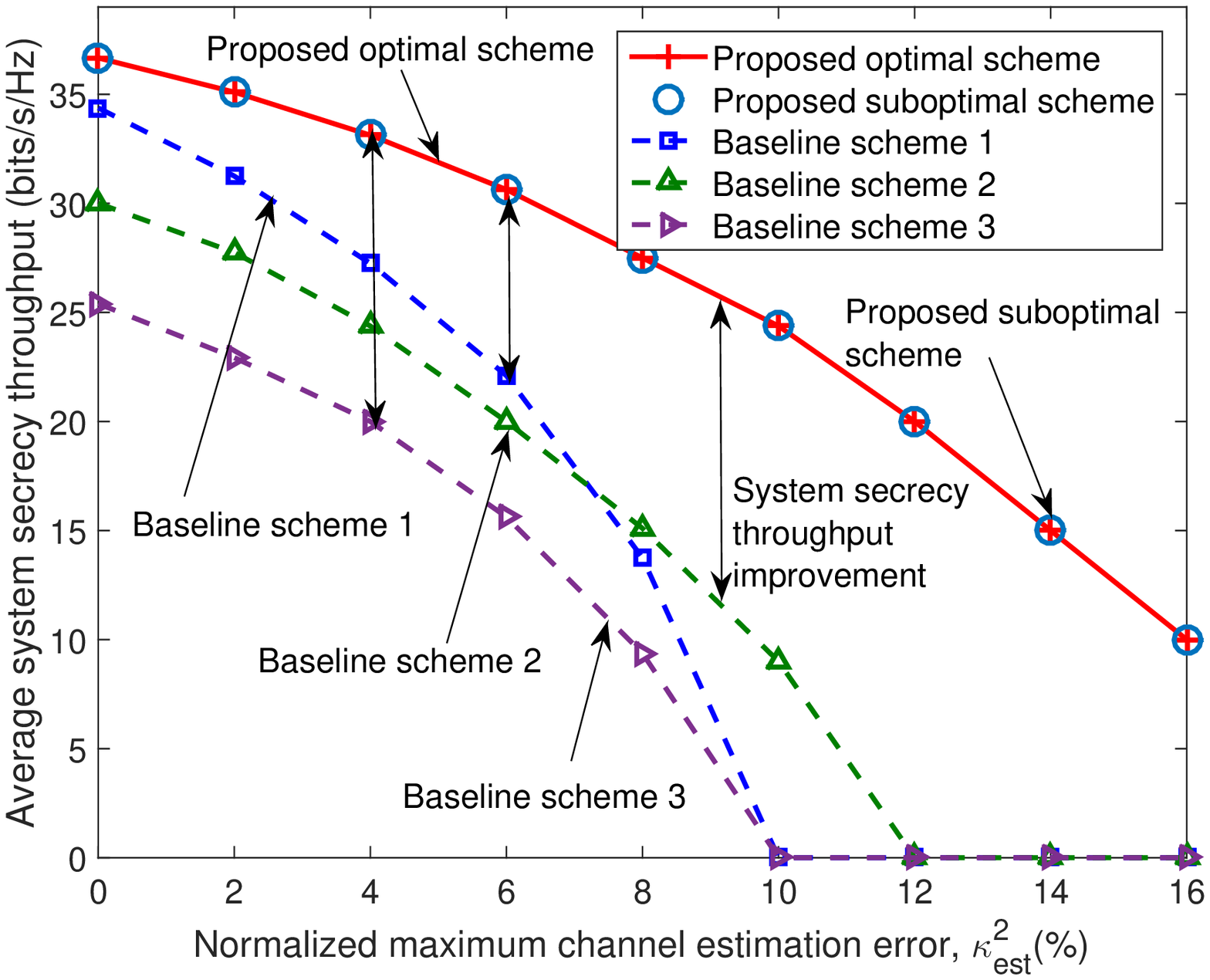}\vspace*{-5mm}
    \caption{Average system secrecy throughput (bits/s/Hz) versus the normalized maximum channel estimation error, $\kappa_{\mathrm{est}}^2$.}
    \label{fig:secrecy_vs_error}
\end{minipage}\hspace*{8mm}
\begin{minipage}[b]{0.45\linewidth} \hspace*{-1cm}
    \includegraphics[width=3.55in]{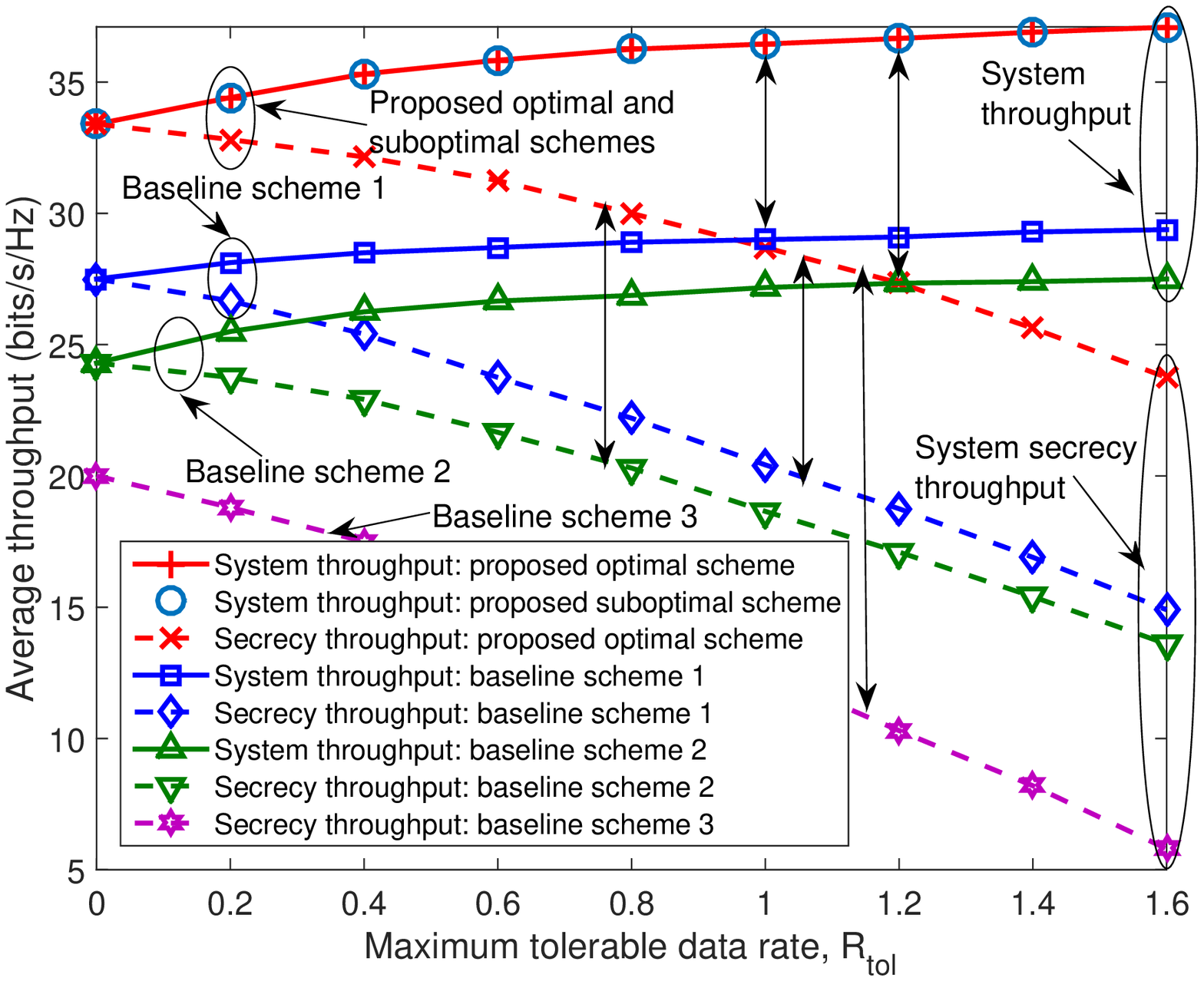}\vspace*{-5mm}
    \caption{Average system throughput and system secrecy throughput (bits/s/Hz) versus maximum tolerable data rate at the equivalent eavesdropper, $R_{\mathrm{tol}}$.}
    \label{fig:secrecy_vs_tolerance}
\end{minipage}\vspace*{-6mm}
\end{figure}

\vspace*{-3mm}
\subsection{Average System Secrecy Throughput versus Maximum Channel Estimation Error}
In Figure \ref{fig:secrecy_vs_error}, we study the average system secrecy throughput versus the normalized maximum channel estimation error, $\kappa_{\mathrm{est}}^2$. As can be observed, the average system secrecy throughput for the proposed schemes and the baseline schemes decrease with increasing $\kappa_{\mathrm{est}}^2$. In fact, the ability of the FD BS to perform precise and efficient beamforming diminishes with increasing imperfectness of the CSI. Therefore, the FD BS has to allocate more power and DoF to the AN to be able to guarantee secure DL and UL transmission. Thus, less power and DoF are available for improving the DL throughput and less DoF can be dedicated to suppressing the SI which has a negative impact on the system secrecy throughput.
Nevertheless, the proposed schemes achieve a significantly higher average system secrecy throughput compared to the baseline schemes.
In particular, the proposed schemes can ensure communication security when $\kappa_{\mathrm{est}}^2$ is smaller than $16\%$ while baseline schemes 1, 2, and 3 can achieve non-zero average system secrecy throughput only when $\kappa_{\mathrm{est}}^2$ is smaller than $10\%$, $12\%$, and $10\%$, respectively.
This implies that the proposed schemes are more robust against channel estimation errors than the baseline schemes.

\vspace*{-3mm}
\subsection{Average System Throughput and System Secrecy Throughput versus Maximum Tolerable Eavesdropping Data Rate}

Figure \ref{fig:secrecy_vs_tolerance} illustrates the average system throughput and the system secrecy throughput versus the maximum tolerable data rate at the equivalent eavesdropper, $R_{\mathrm{tol}}$. As can be observed in Figure \ref{fig:secrecy_vs_tolerance}, the average system throughput increases monotonically with $R_{\mathrm{tol}}$, for both the proposed schemes and baseline schemes 1 and 2.
In fact, a larger $R_{\mathrm{tol}}$ implies a higher information leakage tolerance. Thus, the FD BS can allocate less power to the AN and more power to information transmission, which leads to a higher average system throughput.
However, there is a diminishing return in the improvement of the average system throughput as $R_{\mathrm{tol}}$ increases.
On the other hand, as $R_{\mathrm{tol}}$ increases, the communication security of the considered system decreases.
This is because a higher $R_{\mathrm{tol}}$ implies less stringent requirements on communication security and a smaller value for the lower bound on the average system secrecy throughput. Thus, the proposed schemes and all baseline schemes yield a lower system secrecy throughput but a higher average system throughput for larger values of $R_{\mathrm{tol}}$.
Nevertheless, the proposed schemes achieve a considerably higher average system throughput and system secrecy throughput than the baseline schemes.
In particular, the proposed schemes exploit the power domain which provides additional DoF for resource allocation. Then, the additional DoF are employed for more efficient steering of the transmit beamforming vector and the AN for improved system throughput and reduced information leakage.
In contrast, baseline schemes 1, 2, and 3 adopt suboptimal resource allocation policies which leads to a performance degradation compared to the proposed optimal and suboptimal schemes.

\vspace*{-2mm}
\section{Conclusions}
In this paper, we studied the optimal resource allocation algorithm design for robust and secure communication in FD MISO MC-NOMA systems. An FD BS was employed for serving multiple DL and UL users simultaneously and protecting them from potential eavesdroppers via AN injection. The algorithm design was formulated as a non-convex optimization problem with the objective of maximizing the weighted system throughput while limiting the maximum tolerable information leakage to potential eavesdroppers and ensuring the QoS of the users.
In addition, the imperfectness of the CSI of the eavesdropping channels was taken into account to ensure the robustness of the obtained resource allocation policy.
Exploiting tools from monotonic optimization theory, the problem was solved optimally.
Besides, a suboptimal iterative algorithm with polynomial time computational complexity was developed.
Simulation results revealed that the considered FD MISO MC-NOMA system employing the proposed optimal and suboptimal resource allocation schemes can secure DL and UL transmission simultaneously and achieve a significantly higher performance than traditional FD MISO MC-OMA systems and two other baseline systems. Furthermore, our results confirmed the robustness of the proposed scheme with respect to imperfect CSI and revealed the impact of various system parameters on performance.

\vspace*{-2mm}
\section*{Appendix}
\vspace*{-2mm}
\subsection{Proof of Proposition 1}
First, by applying the basic matrix equality $\det(\mathbf{I}+\mathbf{AB}) = \det(\mathbf{I}+\mathbf{BA})$, we can rewrite constraints ${\mbox{C5}}$ and ${\mbox{C6}}$ as \vspace*{-2mm}
\begin{eqnarray}
\label{eqn:half-det-DL-tol-rate-constraint}
\hspace*{-6.5mm}\overline{\mbox{C5}}\mbox{: }\hspace*{-1.5mm}\det(\mathbf{I}_{M}\hspace*{-1mm} + \hspace*{-1mm} (\mathbf{X}^{[i]})^{-1/2}\mathbf{Q}(\mathbf{X}^{[i]})^{-1/2}) \hspace*{-2.5mm} &\le& \hspace*{-2.5mm} 2^{R_{\mathrm{tol}_{m}}^{[i]\mathrm{DL}}}, \,\,\,
\label{eqn:half-det-UL-tol-rate-constraint}
\overline{\mbox{C6}}\mbox{: }\hspace*{-1.5mm} \det(\mathbf{I}_{M}\hspace*{-1mm} + \hspace*{-1mm} (\mathbf{X}^{[i]})^{-1/2} \mathbf{E}(\mathbf{X}^{[i]})^{-1/2}) \hspace*{-1mm} \le \hspace*{-1mm}2^{R_{\mathrm{tol}_{r}}^{[i]\mathrm{UL}}}\hspace*{-1mm},
\end{eqnarray}
respectively, where $\mathbf{Q}=\mathbf{L}^{[i]H} \tilde{\mathbf{W}}_{m,n,r,t}^{[i]\mathrm{DL}_m}\mathbf{L}^{[i]}$ and $\mathbf{E}=\tilde{P}_{m,n,r,t}^{[i]\mathrm{UL}_r} \mathbf{e}_{r}^{[i]}\mathbf{e}_{r}^{[i]H}$.
Besides, we note that $\det(\mathbf{I}+\mathbf{A}) \ge 1+\Tr(\mathbf{A})$ holds for any semidefinite matrix $\mathbf{A} \succeq \mathbf{0}$ and the equality holds if and only if $\Rank(\mathbf{A}) \le 1$ \cite{li2013spatially}.
Thus, \vspace*{-2mm}
\begin{eqnarray}\label{eqn:DL-QoS-det-trace}
\det(\mathbf{I}_{M} \hspace*{-0mm} +\hspace*{-0mm}(\mathbf{X}^{[i]})^{-1/2} \mathbf{Q}(\mathbf{X}^{[i]})^{-1/2}) \hspace*{-0mm} \ge\hspace*{-0mm} 1 \hspace*{-0mm} + \hspace*{-0mm} \Tr((\mathbf{X}^{[i]})^{-1/2}\mathbf{Q} (\mathbf{X}^{[i]})^{-1/2}),
\end{eqnarray}
always holds since $(\mathbf{X}^{[i]})^{-1/2} \mathbf{Q}(\mathbf{X}^{[i]})^{-1/2} \succeq \mathbf{0}$.
As a result, by combining $\overline{\mbox{C5}}$ and (\ref{eqn:DL-QoS-det-trace}), we have the following implications\vspace*{-2mm}
\begin{eqnarray}\label{eqn:LMI-DL}
\Tr((\mathbf{X}^{[i]})^{-1/2} \mathbf{Q}(\mathbf{X}^{[i]})^{-1/2}) \le \vartheta_{\mathrm{DL}_m}^{[i]}
&\overset{}{\Longrightarrow} & \hspace*{-0mm}\lambda_{\mathrm{max}}((\mathbf{X}^{[i]})^{-1/2} \mathbf{Q}(\mathbf{X}^{[i]})^{-1/2}) \le \vartheta_{\mathrm{DL}_m}^{[i]} \notag\\ [-0mm]
\Longleftrightarrow \hspace*{-0mm}(\mathbf{X}^{[i]})^{-1/2} \mathbf{Q}(\mathbf{X}^{[i]})^{-1/2} \preceq \vartheta_{\mathrm{DL}_m}^{[i]}\mathbf{I}_{N_\mathrm{R}}
&\Longleftrightarrow& \hspace*{-0mm}\mathbf{Q} \preceq \vartheta_{\mathrm{DL}_m}^{[i]}\mathbf{X}^{[i]}.
\end{eqnarray}
Besides, if $\Rank(\tilde{\mathbf{W}}_{m,n,r,t}^{[i]\mathrm{DL}_m}) \le 1$,  we have \vspace*{-2mm}
\begin{eqnarray}
\Rank((\mathbf{X}^{[i]})^{-1/2}\mathbf{Q}(\mathbf{X}^{[i]})^{-1/2}) &\le& \min\Big\{\Rank((\mathbf{X}^{[i]})^{-1/2}\mathbf{L}^{[i]H}),\Rank(\tilde{\mathbf{W}}_{m,n,r,t}^{[i]\mathrm{DL}_m}\mathbf{L}^{[i]}(\mathbf{X}^{[i]})^{-1/2})\Big\} \notag \\[-0mm]
&\le& \Rank(\tilde{\mathbf{W}}_{m,n,r,t}^{[i]\mathrm{DL}_m}\mathbf{L}^{[i]}(\mathbf{X}^{[i]})^{-1/2}) \,\,\, \le 1,
\end{eqnarray}
and the equality in (\ref{eqn:DL-QoS-det-trace}) holds. Moreover, in (\ref{eqn:LMI-DL}),
$\Tr((\mathbf{X}^{[i]})^{-1/2}\mathbf{Q}(\mathbf{X}^{[i]})^{-1/2}) \le \vartheta_{\mathrm{DL}_m}^{[i]}$  is equivalent to
$\lambda_{\mathrm{max}}((\mathbf{X}^{[i]})^{-1/2} \mathbf{Q}(\mathbf{X}^{[i]})^{-1/2}) \le \vartheta_{\mathrm{DL}_m}^{[i]}$.
Therefore, $\mbox{C5}$ and (\ref{eqn:LMI-DL}) are equivalent if $\Rank(\tilde{\mathbf{W}}_{m,n,r,t}^{[i]\mathrm{DL}_m}) \le 1$.
As for constraint ${\mbox{C6}}$, we note that $\Rank((\mathbf{X}^{[i]})^{-1/2}\mathbf{E}(\mathbf{X}^{[i]})^{-1/2}) \le 1$ always holds. Therefore, similar to \eqref{eqn:DL-QoS-det-trace}, we have\vspace*{-2mm}
\begin{eqnarray}
\label{eqn:UL-QoS-det-trace}\det(\mathbf{I}_{M}+(\mathbf{X}^{[i]})^{-1/2}\mathbf{E}(\mathbf{X}^{[i]})^{-1/2}) = 1+\Tr((\mathbf{X}^{[i]})^{-1/2}\mathbf{E}(\mathbf{X}^{[i]})^{-1/2}).
\end{eqnarray}
Then, by combining $\overline{\mbox{C6}}$ and (\ref{eqn:UL-QoS-det-trace}), we have the following implications: \vspace*{-2mm}
\begin{eqnarray}
{\mbox{C6}}\hspace*{-1mm} & \Longleftrightarrow & \hspace*{-2mm} \Tr((\mathbf{X}^{[i]})^{-1/2}\mathbf{E}(\mathbf{X}^{[i]})^{-1/2}) \le \vartheta_{\mathrm{UL}_r}^{[i]}
\Longleftrightarrow  \hspace*{0mm}\lambda_{\mathrm{max}}((\mathbf{X}^{[i]})^{-1/2}\mathbf{E}(\mathbf{X}^{[i]})^{-1/2}) \le \vartheta_{\mathrm{UL}_r}^{[i]}\notag \\[-0mm]
& \Longleftrightarrow & \hspace*{-2mm}(\mathbf{X}^{[i]})^{-1/2}\mathbf{E}(\mathbf{X}^{[i]})^{-1/2} \preceq \vartheta_{\mathrm{UL}_r}^{[i]}\mathbf{I}_{N_\mathrm{R}}
\Longleftrightarrow  \hspace*{0mm}\mathbf{E} \preceq \vartheta_{\mathrm{UL}_r}^{[i]}\mathbf{X}^{[i]}. \quad
\end{eqnarray}

\vspace*{-5mm}
\subsection{Proof of Theorem 1}
The  relaxed SDP problem in (\ref{sdp-max-min-pro}) is jointly convex with respect to the optimization variables and satisfies Slater's constraint qualification \cite{book:convex}. Therefore, strong duality holds and solving the dual problem is equivalent to solving the primal problem \cite{book:convex}. To formulate the dual problem, we first need the Lagrangian function of the primal problem in (\ref{sdp-max-min-pro}) which is given by \vspace*{-2mm}
\begin{eqnarray}
\label{eqn:Lagrangian}\notag{\cal L} \hspace*{-2mm} &=& \hspace*{-2mm} \overset{N_{\mathrm{F}}}{\underset{i=1}{\sum}} \hspace*{-0.3mm} \overset{K}{\underset{m=1}{\sum}} \hspace*{-0.3mm} \overset{K}{\underset{n=1}{\sum}} \hspace*{-0.3mm}
\overset{J}{\underset{r=1}{\sum}} \hspace*{-0.3mm}
\overset{J}{\underset{t=1}{\sum}} \hspace*{-0.3mm} \Big(\gamma \Tr(\tilde{\mathbf{W}}_{m,n,r,t}^{[i]\mathrm{DL}_m})  - \Tr\big(\mathbf{R}_{{\overline{\mathrm{C5}}}_{m,n,r,t}^{[i]\mathrm{DL}_m}} (\tilde{\mathbf{W}}_{m,n,r,t}^{[i]\mathrm{DL}_m},\mathbf{Z}^{[i]},\varrho_{m,n,r,t}^{[i]\mathrm{DL}_m}) \mathbf{D}_{{\overline{\mathrm{C5}}}_{m,n,r,t}^{[i]\mathrm{DL}_m}}\big) \\[-3mm]
\hspace*{-2mm}&-&\hspace*{-2mm}  \Tr(\tilde{\mathbf{W}}_{m,n,r,t}^{[i]\mathrm{DL}_m} \mathbf{Y}_{m,n,r,t}^{[i]\mathrm{DL}_m}) \Big) - \sum_{d=1}^{4D}\theta_d \big[f_d(\tilde{\mathbf{W}},\tilde{\mathbf{p}},\mathbf{Z})-\alpha_x z_d^{(k)}g_d (\tilde{\mathbf{W}},\tilde{\mathbf{p}},\mathbf{Z}) \big] + \Lambda.
\end{eqnarray}
Here, $\Lambda$ denotes the collection of terms that only involve variables that are independent of $\tilde{\mathbf{W}}_{m,n,r,t}^{[i]\mathrm{DL}_m}$. Variables $\gamma$ and $\theta_d$ are the Lagrange multipliers associated with constraints ${\mbox{C2}}$ and ${\mbox{C14}}$, respectively. Matrix $\mathbf{D}_{{\overline{\mathrm{C5}}}_{m,n,r,t}^{[i]\mathrm{DL}_m}} \in {\mathbb{C}^{(N_{\mathrm{T}}+M)\times (N_{\mathrm{T}}+M)}} $ is the Lagrange multiplier matrix for constraint $\overline{\text{C5}}$.
Matrix $\mathbf{Y}_{m,n,r,t}^{[i]\mathrm{DL}_m} \in {\mathbb{C}^{N_{\mathrm{T}}\times N_{\mathrm{T}}}}$ is the Lagrange multiplier matrix for the positive semidefinite constraint ${\text{C12}}$ for matrix $\tilde{\mathbf{W}}_{m,n,r,t}^{[i]\mathrm{DL}_m}$.
Thus, the dual problem for the SDP relaxed problem of (\ref{sdp-max-min-pro}) is given by \vspace*{-4mm}
\begin{eqnarray}\label{eqn:dual-constraint}
{\underset{\theta_d,\gamma \ge 0, \mathbf{Y}_{m,n,r,t}^{[i]\mathrm{DL}_m}, \mathbf{D}_{{\overline{\mathrm{C5}}}_{m,n,r,t}^{[i]\mathrm{DL}_m}} \succeq \mathbf{0}}{\maxo}}  \,\, \underset{\tilde{\mathbf{W}},\tilde{\mathbf{p}},\mathbf{Z},\tau} {\mino} \,\,{\cal L}.
\end{eqnarray}

Next, we reveal the structure of the optimal $\tilde{\mathbf{W}}_{m,n,r,t}^{[i]\mathrm{DL}_m}$ of (\ref{sdp-max-min-pro}) by studying the Karush-Kuhn-Tucker (KKT) conditions.
The KKT conditions for the optimal $\tilde{\mathbf{W}}_{m,n,r,t}^{[i]\mathrm{DL}_m^*}$  are given by:\vspace*{-2mm}
\begin{eqnarray}\label{eqn:KKT-condition}
\mbox{K1: } \theta_d^*, \gamma^* \ge 0, \mathbf{Y}_{m,n,r,t}^{[i]\mathrm{DL}_m^*},\mathbf{D}_{{\overline{\mathrm{C5}}}_{m,n,r,t}^{[i]\mathrm{DL}_m}}^{*} \hspace*{-1mm} \succeq \mathbf{0}, \quad
\mbox{K2: } \mathbf{Y}_{m,n,r,t}^{[i]\mathrm{DL}_m^*} \tilde{\mathbf{W}}_{m,n,r,t}^{[i]\mathrm{DL}_m^*} = \mathbf{0}, \quad
\mbox{K3: } \nabla_{\tilde{\mathbf{W}}_{m,n,r,t}^{[i]\mathrm{DL}_m^*}}{\cal L} =  \mathbf{0},
\end{eqnarray}
where  $\theta_d^*$,  $\gamma^*$, $\mathbf{Y}_{m,n,r,t}^{[i]\mathrm{DL}_m^*}$, and $\mathbf{D}_{{\overline{\mathrm{C5}}}_{m,n,r,t}^{[i]\mathrm{DL}_m}}^{*}$ are the optimal Lagrange multipliers for dual problem (\ref{eqn:dual-constraint}), and $\nabla_{\tilde{\mathbf{W}}_{m,n,r,t}^{[i]\mathrm{DL}_m^*}}{\cal L}$ denotes the gradient of Lagrangian function ${\cal L}$ with respect to matrix $\tilde{\mathbf{W}}_{m,n,r,t}^{[i]\mathrm{DL}_m^*}$. KKT condition K3 in (\ref{eqn:KKT-condition})  can be expressed as \vspace*{-2mm}
\begin{eqnarray}\label{eqn:KKT-gradient-equivalent1}
\mathbf{Y}_{m,n,r,t}^{[i]\mathrm{DL}_m^*} &=& \gamma^*\mathbf{I}_{N_{\mathrm{T}}} -  \bm{\Xi},
\end{eqnarray}
where \vspace*{-4mm}
\begin{eqnarray}
\bm{\Xi}&=&\mathbf{B}_{\mathbf{L}^{[i]}} \mathbf{D}_{{\overline{\mathrm{C5}}}_{m,n,r,t}^{[i]\mathrm{DL}_m}} \mathbf{B}_{\mathbf{L}^{[i]}}^H + \theta_{d`}^* \mathbf{H}_{m}^{[i]} +  \theta_{2D+d`}^* (1- \alpha_x z_{2D+d`}^{(k)})\rho \mathbf{H}_{\mathrm{SI}}^{[i]}\Diag(\mathbf{V}_r^{[i]})\mathbf{H}_{\mathrm{SI}}^{[i]H} \notag \\
&+& \theta_{3D+d`}^* (1- \alpha_x z_{3D+d`}^{(k)})\rho \mathbf{H}_{\mathrm{SI}}^{[i]}\Diag(\mathbf{V}_t^{[i]})\mathbf{H}_{\mathrm{SI}}^{[i]H},
\end{eqnarray}
and $\theta_{d`}^*$, $\theta_{2D+d`}^*$, and $\theta_{3D+d`}^*$ are the Lagrange multipliers corresponding to constraint C14.
First, it can be shown that $\gamma^*>0$ since constraint C2 is active at the optimal solution. Then, we show that  $\bm{\Xi}$ is a positive semidefinite matrix by contradiction. Suppose  $\bm{\Xi}$ is a negative definite matrix, then from  \eqref{eqn:KKT-gradient-equivalent1},  $\mathbf{Y}_{m,n,r,t}^{[i]\mathrm{DL}_m^*}$ becomes a full-rank and positive definite matrix. By KKT condition K2 in \eqref{eqn:KKT-condition}, $\tilde{\mathbf{W}}_{m,n,r,t}^{[i]\mathrm{DL}_m^*}$ has to be the zero matrix which cannot be the optimal solution for $P_{\max}^{\mathrm{DL}}>0$.  Thus, in the following, we focus on the case $\bm{\Xi}\succeq \mathbf{0}$. Since matrix $\mathbf{Y}_{m,n,r,t}^{[i]\mathrm{DL}_m^*} = \gamma^*\mathbf{I}_{N_{\mathrm{T}}} -  \bm{\Xi}$ is positive semidefinite, \vspace*{-10mm}
\begin{eqnarray}
\gamma^* \geq \kappa_{\mathbf{\Xi}}^{\max} \ge 0
\end{eqnarray}
must hold, where $\kappa_{\mathbf{\Xi}}^{\max}$ is the real-valued maximum eigenvalue of matrix $\mathbf{\Xi}$. Considering the KKT condition related to matrix $\tilde{\mathbf{W}}_{m,n,r,t}^{[i]\mathrm{DL}_m^*}$ in \eqref{eqn:KKT-gradient-equivalent1}, we can show that if $\gamma^* \geq \kappa_{\mathbf{\Xi}}^{\max}$, matrix $\mathbf{Y}_{m,n,r,t}^{[i]\mathrm{DL}_m^*}$ becomes a full rank positive definite matrix. However, this yields the solution $\mathbf{W}_{m,n,r,t}^{[i]\mathrm{DL}_m^*} = \mathbf{0}$ which contradicts KKT condition K1 in \eqref{eqn:KKT-condition} as $\gamma^* > 0$ and $P_{\max}^{\mathrm{DL}}>0$. Thus, for the optimal solution, the dual variable $\gamma^*$ has to be equal to the largest eigenvalue of matrix $\bm{\Xi}, $ i.e., $\gamma^* = \kappa_{\mathbf{\Xi}}^{\max}$. Besides, in order to have a bounded optimal dual solution, it follows that the null space of $\mathbf{Y}_{m,n,r,t}^{[i]\mathrm{DL}_m^*}$ is spanned by vector $\mathbf{u}_{\mathbf{\Xi},\max}\in\mathbb{C}^{N_\mathrm{T}\times1}$, i.e., $\mathbf{Y}_{m,n,r,t}^{[i]\mathrm{DL}_m^*}\mathbf{u}_{\mathbf{\Xi},\max} = \mathbf{0}$, where $\mathbf{u}_{\mathbf{\Xi},\max}$ is the unit-norm eigenvector of $\mathbf{\Xi}$ associated with eigenvalue $\kappa_{\mathbf{\Xi}}^{\max}$. As a result, the optimal beamforming matrix $\mathbf{W}_{m,n,r,t}^{[i]\mathrm{DL}_m^*}$ has to be a rank-one matrix and is given by \vspace*{-3mm}
\begin{equation}
\mathbf{W}_{m,n,r,t}^{[i]\mathrm{DL}_m^*} = \nu \mathbf{u}_{\mathbf{\Xi},\max} \mathbf{u}_{\mathbf{\Xi},\max}^H,
\end{equation}
where the value of parameter $\nu$ is such that the DL power consumption satisfies constraint C2. \hfill\qed

\vspace*{-4mm}
\bibliographystyle{IEEEtran}
\bibliography{Robust_MC_FD_NOMA}

\end{document}